# A Practical Guide to Transcranial Ultrasonic Stimulation from the IFCN-endorsed ITRUSST Consortium

Keith R. Murphy[1], Tulika Nandi[2,3], Benjamin Kop[3], Takahiro Osada[4], Maximilian Lueckel[2,16], W. Apoutou N'Djin[5], Kevin A. Caulfield[6], Anton Fomenko[7], Hartwig R Siebner[8,9,10], Yoshikazu Ugawa[11], Lennart Verhagen[3], Sven Bestmann[12], Eleanor Martin[13,14], Kim Butts Pauly[1], Elsa Fouragnan[15], Til Ole Bergmann[2,16]

[1] Department of Radiology, Stanford University, Stanford CA, USA

[2] Neuroimaging Center, Focus Program Translational Neuroscience, Johannes Gutenberg University Medical Center, Mainz, Germany

[3] Donders Institute for Brain, Cognition, and Behavior, Radboud University, Nijmegen Netherlands

[4] Department of Neurophysiology, Juntendo University School of Medicine, Tokyo, Japan

[5] LabTAU, INSERM, Centre Léon Bérard, Université Claude Bernard Lyon 1, F-69003, Lyon, France

[6] Medical University of South Carolina, Department of Psychiatry & Behavioral Sciences, Charleston, SC, USA

[7] Krembil Research Institute, University Health Network, Toronto, Canada

[8] Danish Research Centre for Magnetic Resonance, Centre for Functional and Diagnostic Imaging and Research, Copenhagen University Hospital Amager and Hvidovre, Hvidovre, Denmark

[9] Department of Neurology, Copenhagen University Hospital Bispebjerg and Frederiksberg, Copenhagen, Denmark

[10] Department of Clinical Medicine, University of Copenhagen, Copenhagen, Denmark

[11] Department of Human Neurophysiology, Fukushima Medical University, Fukushima, Japan

[12] Department of Clinical and Movement Neuroscience, UCL Queen Square Institute of Neurology, University College London, UK

[13] Department of Medical Physics and Biomedical Engineering, University College London, London, U.K

[14] Wellcome/EPSRC Centre for Interventional and Surgical Sciences, University College London, London, UK

[15] School of Psychology, Faculty of Health, University of Plymouth, Plymouth, UK

[16] Leibniz Institute for Resilience Research (LIR), Mainz, Germany

**Corresponding author**

Til Ole Bergmann, NeuroImaging Center (NIC), Johannes Gutenberg University Medical Center Mainz, Langenbeckstr. 1, Bldg. 308c, 55131 Mainz, tobergmann@uni-mainz.de

**Highlights**

1. Transcranial ultrasonic stimulation (TUS) is a novel technique for the non-invasive neuromodulation of superficial and deep brain regions with high spatial precision.

2. A subcommittee of the ITRUSST consortium has formulated best-practice recommendations for TUS neuromodulation of the human brain.

3. Recommendations cover critical concepts, required lab equipment, and state-of-the-art experimental procedures.

**Abstract**

Low-intensity Transcranial Ultrasonic Stimulation (TUS) is a non-invasive brain stimulation technique enabling cortical and deep brain targeting with unprecedented spatial accuracy. Given the high rate of adoption by new users with varying levels of expertise and interdisciplinary backgrounds, practical guidelines are needed to ensure state-of-the-art TUS application and reproducible outcomes. Therefore, the International Transcranial Ultrasonic Stimulation Safety and Standards (ITRUSST) consortium has formed a subcommittee, endorsed by the International Federation of Clinical Neurophysiology (IFCN), to develop recommendations for best practice in TUS applications in humans. The practical guide presented here provides a brief introduction into ultrasound physics and sonication parameters. It explains the requirements of TUS lab equipment and transducer selection and discusses experimental design and procedures alongside potential confounds and control conditions. Finally, the guide elaborates on essential steps of application planning for stimulation safety and efficacy, as well as considerations when combining TUS with neuroimaging, electrophysiology, or other brain stimulation techniques. We hope that this practical guide to TUS will assist both novice and experienced users in planning and conducting high-quality studies and provide a solid foundation for further advancements in this promising field.

**Table of Content**

# 1. Introduction

## 1.1. Purpose and structure of this guide

This **Practical TUS guide** has been designed and is maintained by the **Practice Working Group** of the **ITRUSST** (**I**nternational **Tr**anscranial **U**ltrasonic Stimulation **S**afety and **St**andards) consortium, to provide practical guidance on the application of transcranial ultrasonic stimulation (TUS) for neuromodulation in a research setting. It supports researchers at all career stages who are interested in entering the TUS field.

The guide starts with a brief introduction to the principles of ultrasound physics and sonication parameters (**section 1**), to help researchers make informed decisions. Next, we review the **equipment** required in a TUS lab (**section 2**), including ultrasound drive systems, ultrasound transducers, coupling media, acoustic measurement systems, control and acoustic modeling software, and neuronavigation systems, while providing guidance on how to select the best equipment for particular needs. Next, we discuss **experimental design and procedures** (**section 3**) for TUS studies, including different experimental approaches and confounds, the relevance of anatomical target definition, planning of the sonication focus, transducer design/selection and navigation, validation of target exposure, and proof of neural target engagement via the combination with imaging, electrophysiology, and other brain stimulation techniques.

This guide **does not** provide an in-depth review of the existing TUS literature, the underlying neural mechanisms of neuromodulation, or recommendations of specific parameters for stimulation including adherence to regulatory recommended safety guidelines. Where applicable, this guide refers to existing introductory publications and other more detailed resources provided by ITRUSST.

As a **precaution**, new researchers should be aware that the use of TUS for neuromodulation is still in its infancy. Standard practices have yet to be established and there is a need for replication of many core findings. While this represents a great opportunity for new researchers to make fundamental discoveries in the field, some level of experimental variability or troubleshooting is almost guaranteed. For this reason, the content of this guide will remain dynamic, and updates will be made available via the ITRUSST website as new information and practices arise (https://itrusst.com).

## 1.2. Fundamentals of transcranial ultrasonic stimulation

Ultrasound is a mechanical wave with frequencies greater than 20 kHz. In biological systems, these waves can rapidly compress and stretch tissue, depositing energy as they travel. Ultrasound can be focused to target both superficial and deep brain structures with resolution that is limited by wavelength. When applied using appropriate parameters, TUS can be used to alter neuronal activity safely and reversibly and has emerged as a next-generation technique for non-invasive stimulation of the brain. Importantly, TUS in this guide specifically refers to the use of *low-intensity focused ultrasound* (LIFU) for transient neuromodulation; this should be distinguished from the technique of *high-intensity focused ultrasound* (HIFU) which is typically used for surgical ablation (Meng et al., 2021). While there are no explicit ranges defined for these two categories, they are largely distinguished by the capacity of HIFU to ablate tissue through controlled heating. However, it is worth noting that temperature rise can be observed and may contribute to the effects of ultrasound neuromodulation (Darrow et al., 2019); although the relative contribution between thermal and mechanical effects is not fully understood. Low-intensity TUS for direct neuromodulation should also be distinguished from other methodologies like **blood brain barrier opening,** which involves the use of exogenous agents such as microbubbles, or ultrasound neuromodulation through **drug uncaging,** a term which implies the release of pharmacological compounds from encapsulating particles (Aryal et al., 2022; Burgess et al., 2015). These methods may indirectly alter neural activity and may use common ultrasound parameters for focusing or energy delivery. Although the biological mechanisms of TUS neuromodulation are not yet fully understood, there is a large body of evidence suggesting that transient mechanical and thermal perturbation of the cell membrane, altered permeability of mechanically, thermally, or chemically gated ion channels, and changing membrane capacitance may all play a role (Blackmore et al., 2023; Collins and Mesce, 2022; Kubanek et al., 2016; Prieto et al., 2018; Sorum et al., 2021; Yoo et al., 2022).

Direct examination of distinct cell types, as well as analysis of net effects within brain regions, supports both excitatory and inhibitory neuromodulation, and delineating the parameters required for robust neuromodulation is among the top priorities of the field (Murphy et al., 2024). Although effective stimulation parameters may be approximated from existing studies, we have only scratched the surface in examining the great breadth of cell-types, brain regions, and peripheral nerve systems that can be targeted, and stimulation parameters that can be applied. Further complicating the matter, the directionality of response net neural change is sometimes assumed from a given behavior, rather than a direct readout of neural activity. For instance, a behavior may occur when activity of its driving circuit is increased, or when the activity of its inhibitory inputs is decreased. Thus, there is much exploration, tool building, and standardization that lie ahead for newcomers to the field.

### 1.3. Basics of ultrasound physics and interactions

While the physics of acoustics and ultrasound for biomedical applications (Cobbold, 2006) have been characterized in great depth elsewhere (Blackstock and Atchley, 2001), a few principles are critical for understanding experimental requirements and equipment function. These principles will provide a conceptual understanding of how to control ultrasound focus properties and its biological effects.

Ultrasound is a mechanical wave which propagates through a physical medium. While there are several types of waves resulting from mechanical oscillation, the term "ultrasound waves" usually refers to *compression waves*. A compression wave begins with a source of displacement in which an object or material moves in space and compresses the adjacent matter. This compression radiates outward from the source and is followed by a period of low pressure or *rarefaction*, as the particles return back to their origin. Although these particles absorb at least some energy from these waves resulting in a net displacement for the duration of ultrasound pulsing, they have a very limited shift away from their original location over time. Compared to the compressional phase, the negative pressure experienced during the rarefaction phase is often discussed in the context of cavitation, as the negative pressure draws gas from solution and can expand or create bubbles in tissue. This cavitation may contribute to the neuromodulation through biomechanical effects exhibited on the membrane (Lemaire et al., 2021, 2019; Plaksin et al., 2016).

A second type of ultrasound waves are *shear waves* which travel orthogonal to the motion of the medium associated with compression waves. Because these waves require some ability to resist shear force, or shear strength, to propagate, they are almost entirely absent in fluids and soft tissue. However, bone has excellent shear strength allowing shear waves to propagate at high speeds compared to other tissues (Taljanovic et al., 2017). While shear waves have been historically neglected in the context of TUS, compression waves reaching the skull-soft tissue boundary can generate shear waves and vice versa (White et al., 2006). The presence and magnitude of these shear waves is a result of incidental compression waves but can also be interchange back into intracranial compression waves (White et al., 2006). Thus, they will inevitably play some role in the in-situ pressure field and resulting biological effects seen in ultrasound neuromodulation paradigms. The actions and governing equations of these two wave types vary substantially; for simplicity, any general reference to "waves" can be thought of as compression waves in the context of this guide.

Among the many physical features of ultrasound waves, their amplitude, frequency, and relative phase are the most relevant for neuromodulation. Since the interrelationships between all of these features are extensive, we will emphasize the most critical relationships within each feature. A list of key sonication parameters and equations is summarized in **Table 1**.

*Acoustic pressure* is the amount of force applied per unit area and is typically thought of as a local pressure deviation relative to the ambient surround pressure. It is commonly represented with the symbol *p,* and is given in Megapascal (MPa) units.

***Phase*** refers to the location or timing of a point within a sound wave cycle. When two ultrasound waves reach a common point in space, the difference in phase between them (i.e., relative phase) determines their interference. If the waves are in-phase, they will constructively interfere and their amplitudes are added; if the waves are out of phase, they will destructively interfere and their amplitudes are subtracted. It is commonly represented with the symbol ***ϕ,*** and is given in degree (°) or radian (rad) units.

***Amplitude*** is used to describe the extent of the pressure variation over the acoustic cycle. Ultrasound waves often take the form of a sinusoidally varying pressure signal, where the amplitude describes the height of the sine wave. For low amplitude waves with a single frequency, the pressure amplitude is equal to the peak positive and negative pressures, and half of the peak-to-peak pressure. The amplitude is thus the temporal peak acoustic pressure commonly represented with the symbol ***$p_0$***(Preston, 2012)***,*** and is given in Megapascal (MPa) units.

***Frequency*** is the number of wave cycles per second and is determined by the rate at which the ultrasound source oscillates and influences many aspects of tissue interaction. It is commonly represented with the symbol ***f,*** and is given in Megahertz (MHz) units.

***Wavelength*** is calculated as the speed divided by the frequency. It is the spatial distance between two points at the same phase, for example, the spatial distance between two successive high-pressure points is the wavelength. It is commonly represented with the symbol ***λ,*** and is given in millimeter (mm) units.

***Sound speed*** is the distance traveled per unit of time by the ultrasound wave and is inversely related to the compressibility of the medium. It is commonly represented with the symbol c and is given in meter per second ($m \cdot s^{-1}$). The relationship among the frequency, wavelength, and speed of sound is expressed as $c = f \cdot \lambda$.

***Acoustic impedance*** refers to the degree of resistance when ultrasound passes through a medium. It is calculated by the product of density of the medium and speed of ultrasound and is commonly represented with the symbol ***Z*** (*see Table 1 for example usage*). For example, low density, compressible materials such as foam have a very low acoustic impedance, while dense, rigid materials, such as steel, have a very high acoustic impedance.

***Absorption*** of ultrasound refers to acoustic amplitude or energy lost and deposited to the medium in the form of heat. The absorption coefficient is approximately proportional to the frequency of the ultrasound. It is commonly represented with the symbols ***α.*** It is given in Neper per meter ($Np \cdot m^{-1}$) unit for amplitude absorption or in Decibel per centimeter ($dB \cdot cm^{-1}$) unit for energy absorption.

***Attenuation*** of ultrasound is the reduction of ultrasound energy as it passes through a material. This is primarily caused by absorption and scattering of the ultrasound beam. It is commonly represented with the symbols ***A.*** It is given in Neper per meter ($Np \cdot m^{-1}$) unit for amplitude attenuation or in Decibel per centimeter ($dB \cdot cm^{-1}$) unit for energy attenuation.

***Reflection*** occurs when ultrasound travels between two materials with different acoustic impedances. A fraction of the ultrasound bounces back in reflection, whereas another fraction passes into the second medium. Reflections are undesirable in TUS because of lower energy transmission and the predictability of wave motion within heterogeneous media. Nevertheless, some amount of reflection is unavoidable given the differences in acoustic impedance posed by the skull. Reflections have an angle effect where the reflected energy increases as the angle of incidence on the interface is increased from the normal to the interface. There is generally little that can be done about reflections beyond use of coupling material at the skin surface and applying ultrasound perpendicular to the skull surface.

***Scattering*** of ultrasound is a type of reflection where small segments of a non-smooth surface cause a variety of reflections in directions that cannot be estimated from the shape of the larger structure.

***Refraction*** occurs when the incident beam is not perpendicular to the interface between the two media. In this case, the transmitted beam will be at a different angle from the incident beam. Refraction is described by *Snell's law*. Applying ultrasound perpendicular to the skull surface will minimize refraction.

Among the many interactions between these quantities, one of the most important is between frequency and attenuation, since targeting beyond the highly attenuating skull is a primary goal of TUS. Generally speaking, the higher the frequency, the greater the ultrasound attenuation over a given distance. This is primarily due to higher energy absorption (**Figure 1**). In practice, lower frequencies will penetrate more easily through the skull due to the frequency dependence of absorption, and the smaller scale of inhomogeneities and scattering internal skull structures compared to the longer wavelengths at lower frequencies, resulting in lower absorption, and less distortion or aberration of the field. However, less energy will be deposited at the target because of the lower absorption coefficient. Another relevant relationship is the interplay between *frequency* and *focal size*, specifically the constructive interference pattern of multiple ultrasound waves. In this relationship, higher frequency results in a smaller focal size. As an intuitive example, one can envision perfect superposition of all incoming wave peaks at a single point in space. In this case, the width of the additive waveform can only be as small as a single constituent wave. Thus, the smaller the individual waves, the more compact the ideal focus.

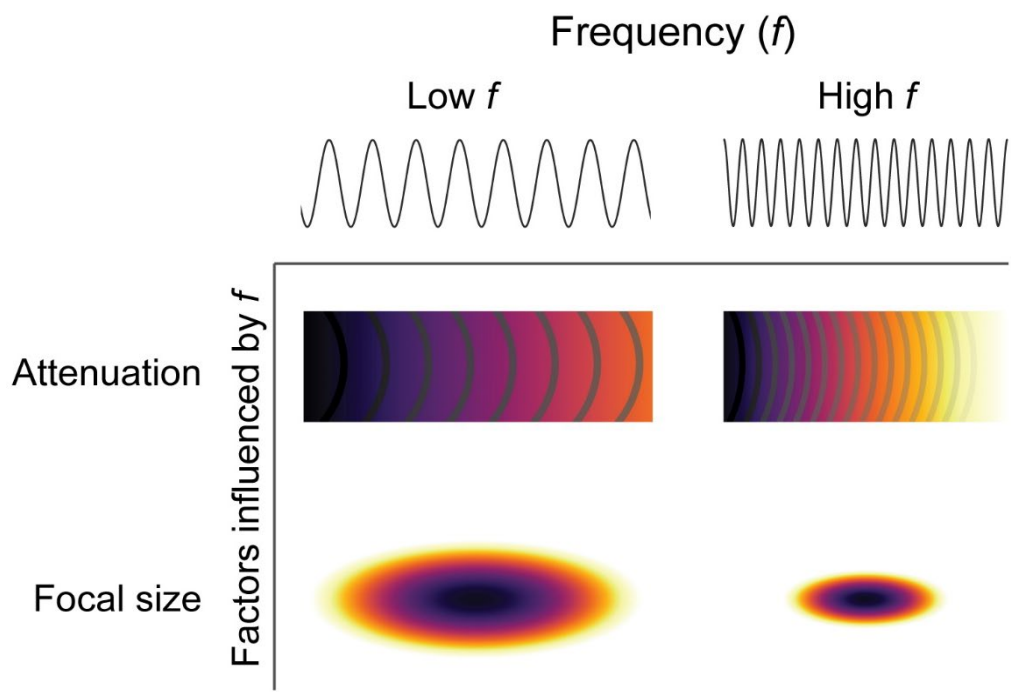

**Figure 1. Illustration of frequency dependence of attenuation and focal size.** Darker colors indicate areas of higher pressures where lighter colors indicate lower pressures.

Another important, but slightly more complex, interaction is the magnitude of reflection at a tissue interface. At any given interface between materials of differing *acoustic impedance*, part of the wave energy is transmitted across the boundary, and part is reflected. The fraction of the incident wave energy that is reflected at the boundary is directly related to the difference in acoustic impedance between the two materials, or tissue types. For example, muscle and fat have an acoustic impedance of 1.71 and 1.34 MRayls, respectively, while bone has an impedance of 7.8 MRayls. Thus, a transmission from muscle to fat with a low impedance difference will be much greater than transmission from muscle to bone, which has a very high impedance difference.

### 1.4. Sonication parameters

Similar to other transcranial neuromodulatory techniques such as ***transcranial magnetic stimulation (TMS)***, ***transcranial electrical stimulation (tES)***, or ***deep brain electrical stimulation (DBS)***, there are a number of stimulation parameters that define a specific TUS protocol. Variations in parameter space can impact both the magnitude and direction of the acute neuronal response to ultrasound as well as its prolonged effects (Kim et al., 2014; King et al., 2013; Kubanek et al., 2018; Manuel et al., 2020; Munoz et al., 2022; Murphy et al., 2024; Murphy and de Lecea, 2023). However, the mechanisms by which these parameters differentially modify biomechanical and/or neurophysiological properties to alter neuronal response remains largely unknown. First and foremost, neuromodulatory TUS can be applied either continuously or in a pulsed fashion. For human applications seeking to induce biomechanical effects with limited thermal impact, pulsed protocols are more commonly used to distribute the total energy delivered over a period of time while limiting the peak temperature rise (Darmani et al., 2021). Through the process of *perfusion*, blood and glymphatic flow in the brain are constantly absorbing and distributing heat to colder areas of the body. Thus, pulsing protocols with sufficiently long intervals may allow for greater blood or cerebrospinal fluid perfusion, which can minimize undesired focal heating of skin, bone, and brain. Pulsed protocols may also be used to avoid excessive depolarization or hyperpolarization of the membrane, allowing the neurons to return to resting membrane potential between pulses. While this rationale is common to electrical and optogenetic stimulation paradigms, it is not yet apparent in TUS studies. Nevertheless, consistent documentation and terminology will allow cross-study comparison and further understanding of the relationship between pulsing protocols and neuromodulation.

A *continuous wave* (CW) is simply an uninterrupted pulse often described by its length, amplitude and fundamental frequency. In contrast, *pulsed protocols* can be thought of as a series of hierarchical repeating patterns which include the characteristics of continuous waves (Martin et al., 2024a). Ramping is the smooth increase and decrease in waveform amplitude applied at the beginning and end of a pulse. Ramps are often applied to pulses used in neuromodulation to reduce audibility so should be carefully described due to the potential impact of auditory confounds and biological effects on tissue (Johnstone et al., 2021; Mohammadjavadi et al., 2019). Although the commonly referenced features of a pulse repetition protocol include frequency and duty cycle, **Figure 2** depicts the extensive set of parameters for a simple TUS pulsing scheme and **Table 1** defines the key parameters typically used to describe them (Preston, 2012). Studies often report an incomplete set of these parameters, affecting reproducibility and comparison to other studies. In some instances, parameters can be derived from a combination of those provided (see formulas in **Table 1)** (Blackmore et al., 2019) or recovered by contacting the authors. Since certain terms, such as "burst," "sonication," and "stimulus," lack consistent definition, careful consideration of terminology will enable better comparison and understanding of diverse studies. For instance, the term "burst" is often used for what we will describe as pulse duration; we recommend the latter for consistency. Since clear and consistent parameter reporting enables the research community to assess the efficacy and safety of published protocols, the ITRUSST Working Group on Standardized Reporting also offers reporting guidance (Martin et al., 2024a). The Radboud FUS Initiative also provides a resource for calculating and deriving features from basic waveform descriptors (https://www.socsci.ru.nl/fusinitiative/tuscalculator/).

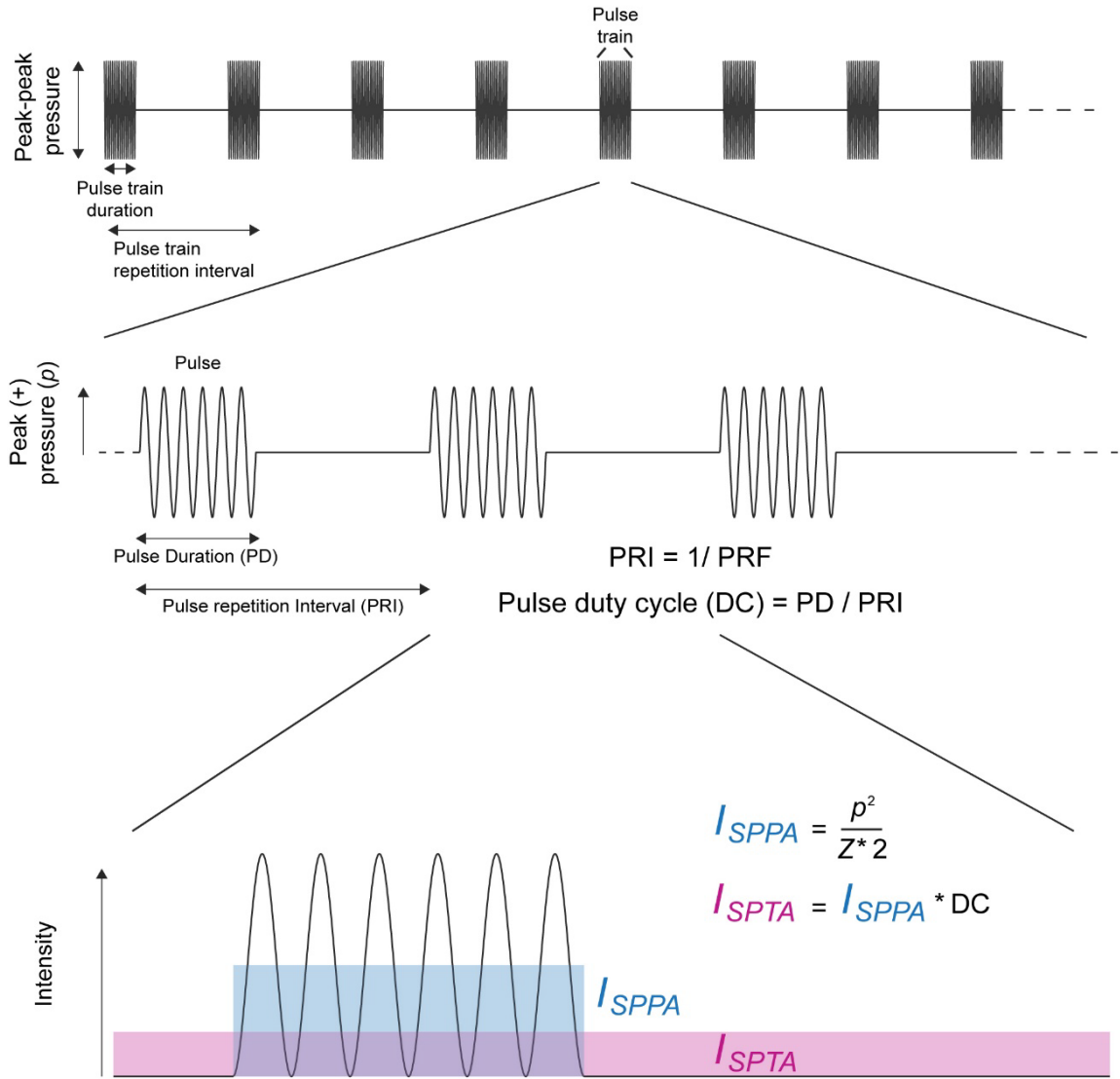

**Figure 2. Visual representation of key TUS parameters in a simple pulsing scheme.** See **Table 1** for a more complete list and explanations of the respective terms.

**Table 1.** List of key sonication parameters. Z = Acoustic Impedance

| Parameter | Abbrev. (Unit) | Description | Formula |
|---|---|---|---|
| Ultrasound operating frequency | $f_0$ (kHz or MHz) | The frequency of ultrasonic pressure waves. | |
| Pulse duration | PD (ms) | The shortest continuous period of sonication. When applying continuous TUS, pulse duration is equal to the pulse train duration. | |
| Pulse repetition frequency | PRF (Hz) | The frequency of pulse delivery within the pulse train duration | 1/ PRI |
| Pulse repetition interval (or Inter- | PRI (ms) | The length of time between the onset of two pulses within a train of pulses. | 1/ PRF |
| pulse interval) | | | |
| Pulse train duration | (s) | The length of time a train of pulses is delivered. | |
| Pulse train repetition interval | PTRI (s) | The length of time between the onset of pulse trains. | |
| Duty cycle | DC (%) | The percentage of time sonication is delivered during the pulse repetition interval. The duty cycle is only defined for rectangular pulses. (or pulse train duration) | PD/PRI |
| Instantaneous pressure | $p$ (MPa) | The instantaneous pressure at any given point of an ultrasound wave. | |
| Instantaneous intensity (commonly referred to as intensity) | $I(t)$ | The time varying intensity, the intensity at each point intime during a pulse. | $p^2$/ density * speed of sound (= $p^2/Z$) |
| Peak positive pressure | $p+$, $p_c$ (MPa) | The peak (+) pressure found at the peak of an ultrasound wave. Note: peak (-) and (+) pressure are equal when low pressures are used. (c) refer to compression. | |
| Peak negative pressure | $p-$, $p_r$ (MPa) | The peak (-) pressure found at the trough of an ultrasound wave. Note: peak (-) and (+) pressure are equal when low pressures are used. (r) refers to rarefaction. | |
| Spatial peak pressure | $p_{sp}$ | The amplitude of the pressure during the steady part of the pulse at the location of the spatial peak. | |
| Pulse intensity integral | PII | Integration of the instantaneous intensity over the duration of the pulse. The pressure term for a pulse with a square envelope simply calculates the average pressure within the pulse during oscillation ($p/2$). | $\int^{PD} I_i \cdot dt$<br>*or ~ p / 2 for a pulse with a square envelope, where p is pressure amplitude* |
| Spatial peak pulse average intensity | $I_{SPPA}$ (W/cm$^2$) | The intensity of the pulse at the spatial peak (focal peak) time averaged over the pulse duration. | *For arbitrary pulse shapes (e.g. ramped pulses):*<br>$I_{sppa} = \frac{1}{PD\,Z}\int_0^{PD} p_{sp}(t)^2 dt$<br>For pulses with a square envelope:<br>$I_{sppa} = \frac{p_{sp}^2}{2Z}$ |

| Spatial peak temporal average intensity | I$_{SPTA}$ (mW/cm$^2$) | The intensity of the ultrasonic stimulus at the spatial peak time averaged over a defined period of time (i.e., across the stimulus duration). The temporal window must be defined. | I$_{SPPA}$ * DC (duty cycle) |
|---|---|---|---|
| Spatial peak temporal peak intensity | I$_{SPTP}$ (W/cm$^2$) | The temporal peak of the instantaneous intensity of the ultrasonic stimulus at the spatial peak. | $\max\left(\frac{p_{sp}^2(t)}{z}\right)$ |
| Mechanical Index | MI | A unitless index for assessing the likelihood of cavitational bioeffects. Where p$_{r.3}$ is the pressure derated using $\alpha$ = 0.3 dB cm$^{-1}$ MHz$^{-1}$ | $MI = \frac{p_{r,.3}}{\sqrt{f}}$ |
| Transcranial Mechanical Index | MI$_{tc}$ | A unitless index. $p_{r,\alpha}$ is an estimate of in-situ rarefactional pressure, to assess the likelihood of cavitation bioeffects in transcranial applications (tc), derating is performed using properties of the skull and soft tissues. | $MI_{tc} = \frac{p_{r,\alpha}}{\sqrt{f}}$ |
| Thermal Index | TI | A unitless index for assessing the likelihood of thermal bioeffects. | *power at focus / power required to increase temperature 1°C* |
| Cranial Thermal Index | TI$_C$ | Unitless index for assessing temperature rise in the skull bone. It is the power transmitted by the transducer, divided by 40* the diameter of the beam on the skull. | |

## 2. TUS Equipment

### 2.1. Ultrasound systems

Compared to ultrasound imaging systems, which can transmit and receive ultrasound waveforms, ultrasound neuromodulation devices often contain only components for transmitting ultrasound. There are many configurations for these transmit systems with both essential and non-essential components described below. The main components of an ultrasound system are a **signal generator**, **signal amplifier**, and a **transducer**; which outputs the acoustic signal (**Figure 3**). The system may also include several additional components to improve the control or characteristics of waveform outputs such as electrical impedance matching networks. Transducer construction is optimized for the application by adjusting matching layers, backing layers and thickness of the piezo electric layer, and by the addition of features such as acoustic lenses, and suitable coupling agents. While many turnkey systems have these components built-in and hidden from the user, a fundamental understanding of each will allow for better communication when procuring or designing custom transducers or control systems. It will also allow users to take full advantage of system capabilities. Here, we'll examine the functionality and example configurations of each of these parts.

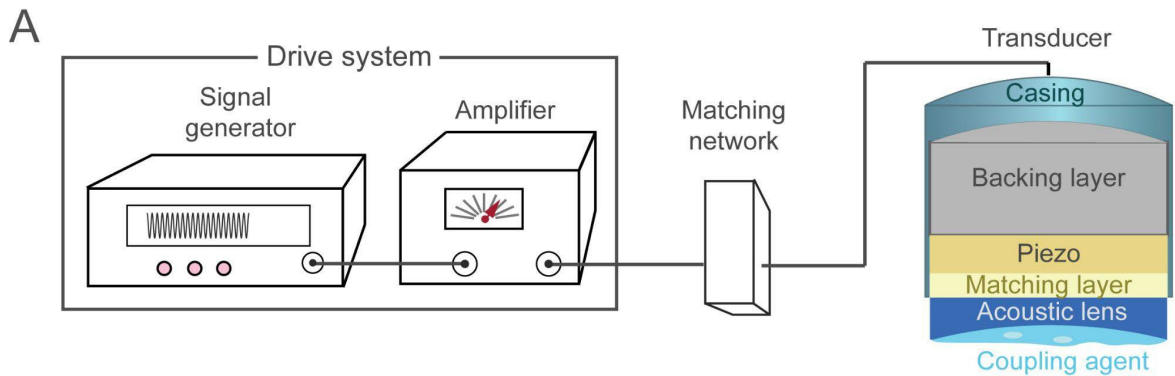

**Figure 3. Schematic representation of the individual components of an example TUS system**, consisting of a signal generator and RF power amplifier serving as the drive system, an electrical impedance matching network, and a transducer.

The *transducer* is the component which converts electrical energy to mechanical vibrations and thus ultrasound waves. The electrical energy originates from a *signal generator* in the form of an oscillating voltage. This oscillation can often be characterized by a sine wave function but may have added complexities such as ramping or pulsing. A transducer will usually be driven by a signal containing a sine wave at a certain frequency which may be gated or pulsed at a given PRF, perhaps with a ramped envelope to produce signal waveforms required for TUS. Although function generators create the desired waveform shape, most cannot generate peak voltages required to generate the pressure amplitudes required for TUS neuromodulation. Therefore, an *amplifier* is usually placed in series between signal generator and transducer to achieve the desired output amplitude. In alternative methods, power may be stored and transmitted by gating electronics in a pattern matching the desired waveform.

A signal generator and amplifier together constitute the ultrasound *drive system*. Coupling of the transducer to the driving system is typically achieved via coaxial cables (one per transducer element) containing inner (+) and outer (-) wiring layers. A *matching network* is an electrical circuit which may be inserted between driving system and transducer, and which can reduce electrical reflections from the transducer. Similar to the concept of acoustic impedance mismatch described earlier, these reflections occur because many transducers have a very different electrical impedance to the cabling and drive systems, which typically have a 50-Ohm electrical impedance. These reflections can reduce the transfer of voltage and power, leading to inefficient pressure generation. Transducers are often provided with electrical matching networks which reduce the impedance mismatch to ensure efficiency with standard 50 Ω cables and driving circuits. For custom transducers, electrical matching is an important consideration. To observe the efficiency of energy transfer to the transducer, a *power meter* may be placed in-line between the amplifier and the matching network. However, measured electrical power from the driving system to the transducer should be related to a measured acoustic pressure output from the transducer as discussed in Section 2.4.

The combined operation of these signal chain components allows conversion of electrical into acoustic energy by the transducer, resulting in physical surface displacements which transmit the acoustic wave into the medium. In addition to the commonly used coupling gels, a variety of additional *coupling media* may be used, such as built-in solid-encapsulated water coupling, gel pads, silicon pads, or a water bladder/balloon coupling system (see section 2.3.). TUS systems can be assembled from individual components or purchased as complete turnkey ready systems off the shelf (cf. **Supplementary Table S1** for a list of commercially available systems).

## 2.2. Ultrasound transducer characteristics

TUS transducers span a broad range of aperture diameter, radius of curvature, materials, constructions, and electrical interfaces. They can exist as a single element or multiple elements acting in concert with varying geometric configurations; for a generic drawing see **Figure 4**. Together, these features ultimately determine the shape, size, intensity, and steering range of the ultrasound focus. Since all of these aspects have a range of options, the sheer number of combinations that can be used for a single purpose is staggering, and selection of the appropriate transducer can be a daunting task. Here, we provide important information about the design principles and general characteristics of ultrasound transducers and their impact on the resulting sonication focus. In section 3.3, we will build on these principles when discussing the rationale for selecting or designing the appropriate transducer for a particular anatomical target; users are well advised to read that section before purchasing a transducer for their new TUS lab. Researchers may seek a single transducer for a broad range of targets, including small, large, deep, and superficial brain structures. Importantly, optimal transducer choice depends strongly on the location, size, and shape of the anatomical target. Since a single transducer cannot typically be optimized for multiple targets of interest, it is best to start with the single most important or representative target to simplify the selection process. Compared to single-element transducers, multi-element transducers allow axial or even lateral electronic steering of the focus. Although steering can cover a larger range of targets, it can come at the expense of a well-shaped focus. Similar to cameras with optical zoom lenses, they allow for a dynamic focus, but with less 'sharpness' at some focal depths relative to lenses with fixed focal length. Generally, the process of transducer selection also benefits from the use of targeting software and acoustic simulations (see **section 2.8** and (Aubry et al., 2022) for an overview).

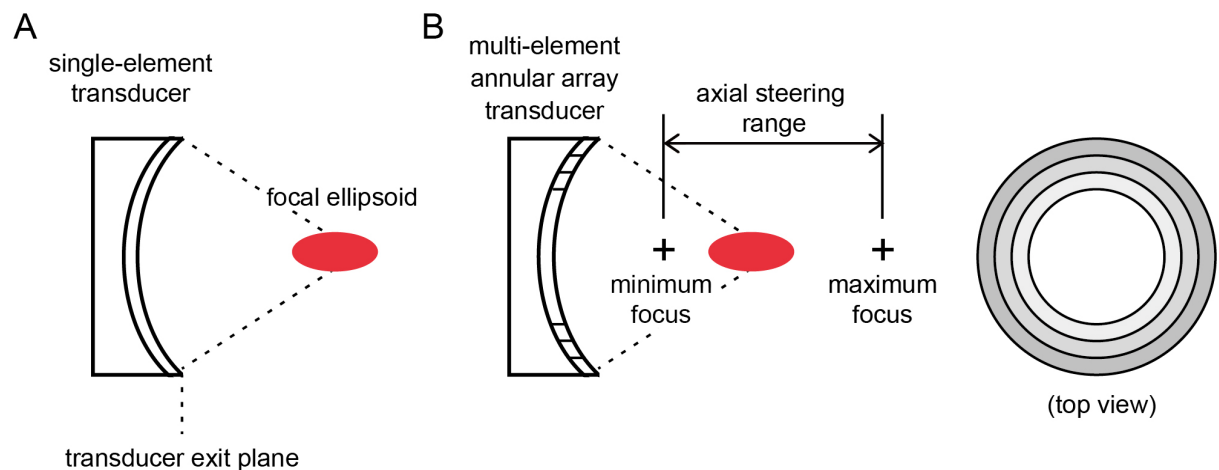

**Figure 4. Schematic drawing of a transducer and ultrasound beam.** (**A**) For a single-element transducer, the position of the focus is fixed relative to the transducer exit plane. In the case of commonly used spherically focusing transducers, this position is determined by the radius of curvature and aperture diameter of the transducer. In general, the focal size is determined by the aperture size, focal distance, and the wavelength of the ultrasound. The shape of a typical ultrasound focus is an ellipsoid. (**B**) In a multi-element annular array transducer, the position of the focus can be adjusted by controlling excitation phase of each element, in this example to move it over the axial steering range. A four-element annular array transducer is depicted as an example.

The TUS focus is visualized by examining the spatial profile of the acoustic pressure along the beam axis, has a characteristic high-pressure region. The length and width of this high-pressure focal region is often used to describe the size of the focus, using different pressure or intensity thresholds (i.e., fraction of the spatial maximum pressure or intensity) to make the calculation. The **full width half maximum** (FWHM) is described along the axis orthogonal to the transducer face (the **axial focal length**) and parallel to the face or surface plane (the **lateral focal width**). The same values for the focal size can be calculated from either the pressure or intensity profiles in terms of dB, using the appropriate fractional threshold. The FWHM of the intensity profile defines the -3 dB region (**Figure 5**), in which intensity exceeds 50% of the maximum values. Notably, the same region contains pressures exceeding 70% of the maximum value. Correspondingly, the -6 dB region of the intensity profile is defined as exceeding 25% of the maximum intensity and 50% of the maximum pressure (corresponding to the FWHM of the pressure profile).

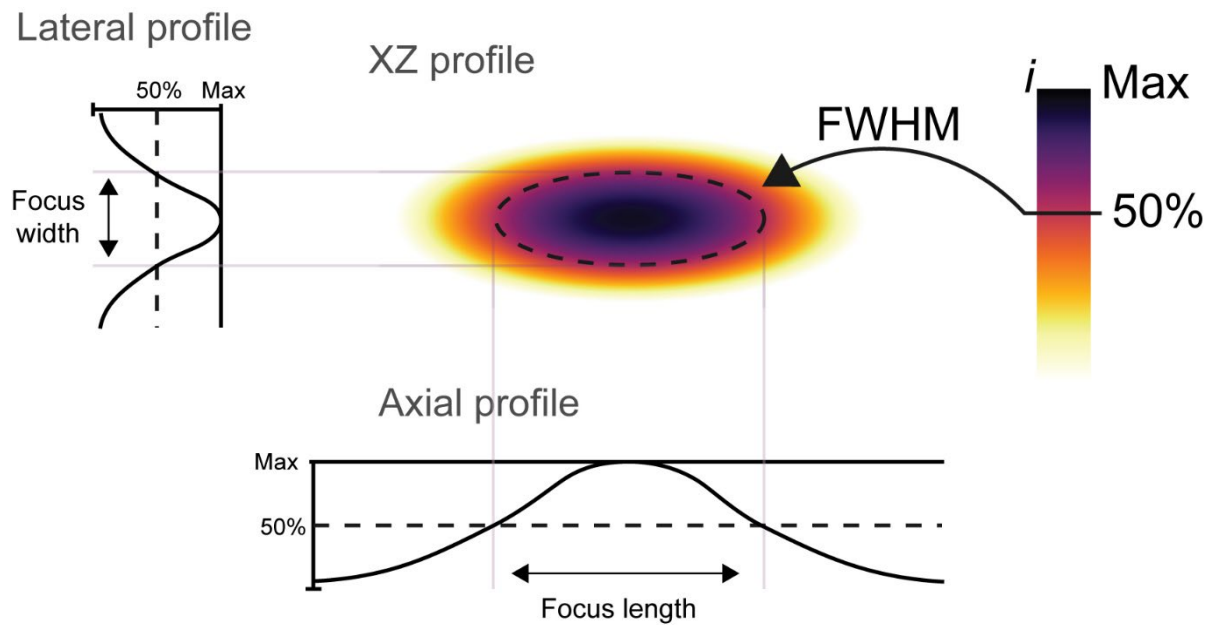

**Figure 5. Beam profile plots showing the 50% intensity (*i*) cutoff often used to describe the Full Width Half-Maximum (FWHM) boundary**, shown here as a dashed oval. This is also called the -3dB contour. Note that this does not indicate the area of assumed effective neuromodulation but only provides a spatial characterization of the beam focus.

There is a common misconception that boundary descriptors of ultrasound pressure fields are necessarily equivalent to the area of biological effect. For instance, it should not be assumed that the 50% intensity boundary will give 50% biological effect. These are relative boundaries, and as absolute thresholds for effect and dose-response relationships are not yet well defined or understood, the region of effect cannot be precisely determined from the in-situ exposure parameters However it is likely that TUS neuromodulation has some form of exposure dependence linking the spatial distribution of the in-situ exposure parameters with biological effect. **Figure 6** illustrates an example, in which the portion of the ultrasound field, which reaches 50% of the maximum biological effect (**ED$_{50}$**), is substantially smaller than the FWHM. In this case, researchers assuming FWHM is equivalent to ED$_{50}$ would have over-estimated effect size. Importantly, Figure 6 is a hypothetical scenario, and the ED$_{50}$ across brain regions, cell types, and biological states remains largely unknown. It is also possible that response curves are non-sigmoidal or even non-monotonic.

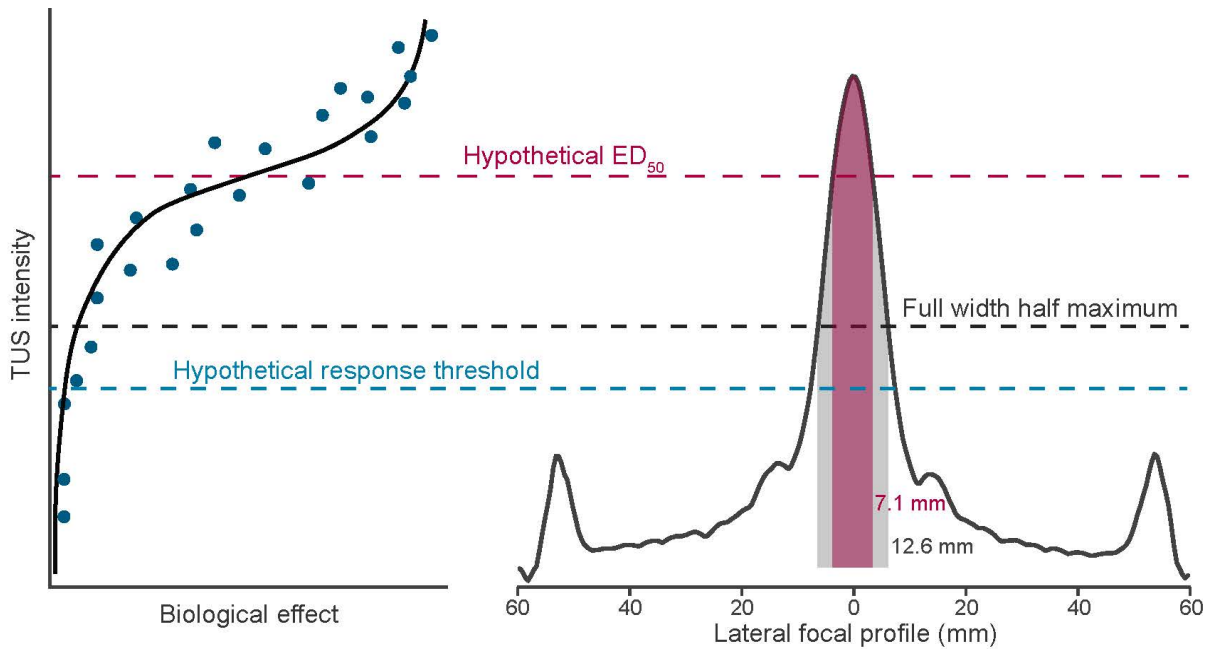

**Figure 6. Hypothetical comparison of ED$_{50}$ TUS intensity compared to the field full width half maximum.** Note that the dose response curve axes are inverted for illustrative purposes.

In the case of **single-element transducers**, the focal length and width primarily depend on the transducer operating frequency and geometry, namely its **aperture diameter** and radius of curvature (**Figure 7**). With everything else remaining constant, reducing the **radius of curvature** will reduce both the **length and width of the focus**. It will also bring the focus closer to the surface of the transducer, that is decrease **focal distance**. Reducing the aperture diameter will increase both the focal length and width, so long as the radius of curvature remains constant. For intuitive quick estimates of focal position, you can simply imagine a circle formed by continuation of the transducer aperture and place a point at the center of that imaginary circle, as shown by the dashed lines in **Figure 7**.

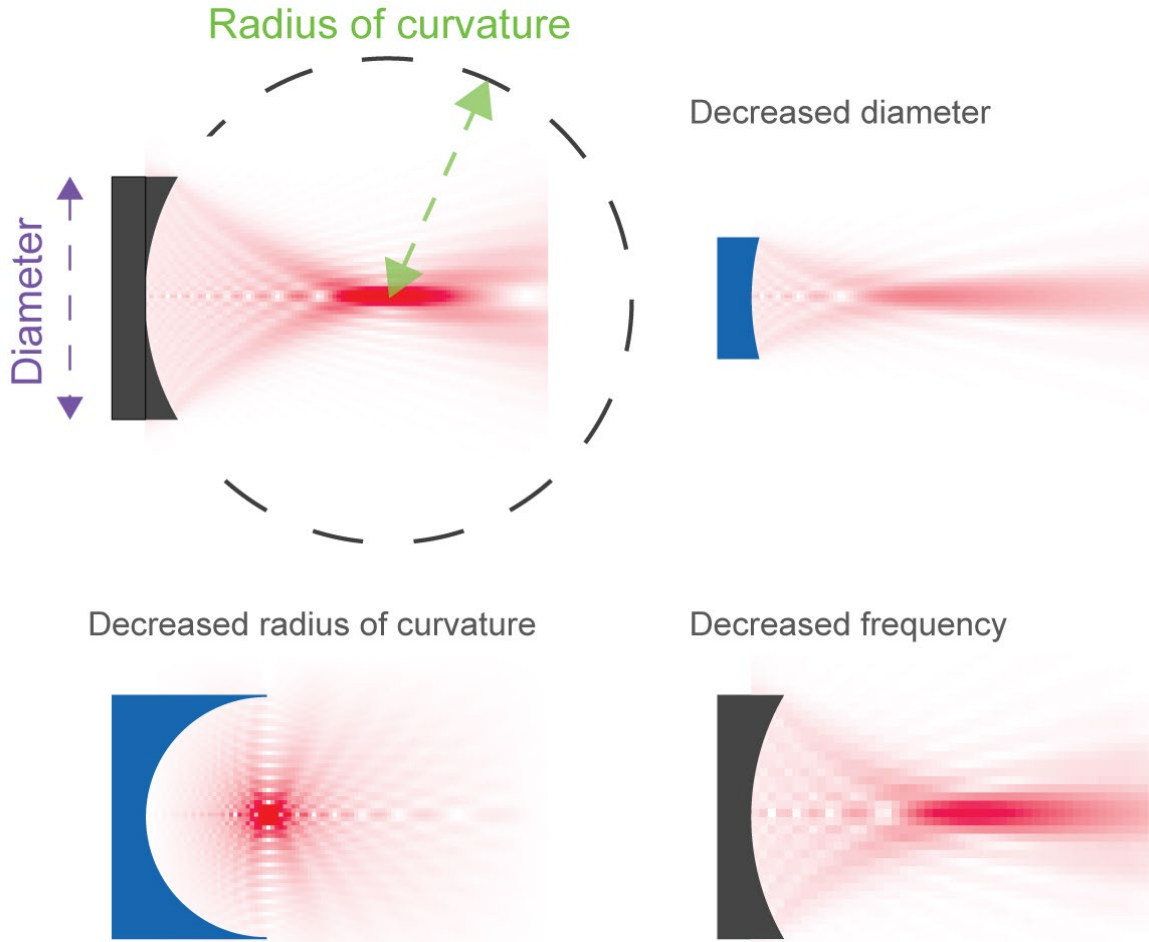

**Figure 7. k-Wave simulations illustrating changes in focal size, shape and position** due to alteration of radius of curvature, aperture diameter, or frequency for a single element spherically focusing transducer.

Another key parameter influencing focal size is the **operating frequency** of the transducer. In principle, an ultrasound transducer can be excited at any frequency. However, the thickness and material of a transducer are carefully designed to maximize pressure output at a given frequency. Deviation from this frequency will reduce transmit efficiency or conversion of electrical energy into focal energy. As described earlier (section 1.3), the operating frequency contributes to the beam profile (e.g., its focal width and length), with higher frequencies resulting in smaller foci. In addition, attenuation is dependent on frequency. To balance transmission across the skull, focal size, and mechanical effects arising at very low frequencies, neuromodulation transducers will typically operate between 250-1000 kHz. The length and width of the focus can be estimated using the following simple relationships between aperture diameter, D, focal distance, F and wavelength, λ:

$$-6\ dB\ focal\ width\ = 1.41\frac{\lambda F}{D},\ -6\ dB\ focal\ length: = 9.7\lambda\left(\frac{F}{D}\right)^2.$$

Axial and lateral profiles through the focus can also be easily calculated using analytical solutions such as the O'Neill solution for the pressure generated by a spherically focusing source (Martin et al., 2016; O'Neil, 1949). Alternatively acoustic simulation software can be used to explore the fields generated by different sources.

In addition to single-element ultrasound transducers, **ultrasound arrays** (a.k.a. **multi-element transducers** or **phased arrays**) can also be used for TUS. At a high level, ultrasound arrays are a series of individual ultrasound transducer elements organized to work together in creating an ultrasound focus. These elements can be arranged as a normally spaced grid, a series of concentric rings, or even random patterns. Although curvature can still be applied to an array, what makes them unique from a single-element transducer is their ability to focus ultrasound waves by changing their emission timing. Ideal timing ensures that all emitted wave fronts reach the focal target at the same time; this can be achieved reasonably well using a simple technique called ray-tracing. In acoustics, ray tracing implies calculating the time it takes a wave to travel from one point in space to another, requiring the ultrasound source and target coordinates, and the speed of sound across the coordinate space. By example, take three ultrasound wave sources at different positions relative to a focal target, within an acoustically homogenous field, such as water (**Figure 8A**). A line can then be drawn between the target and source and the average speed of sound can be calculated along that line, where the distance divided by speed of sound gives time of flight (TOF). Starting with the furthest element with the longest TOF (TOF$_{long}$), each subsequent element wave output should be delayed by the difference in time TOF relative to TOF$_{long}$, ensuring that the waves will reach that point at the same time (**Figure 8A**). Since the phase within each wave cycle drives constructive interference, rather than the cycle itself, signal inputs can also be delayed by fractions of a cycle; this feature is commonly known as a phase delay. Phase delays can be extracted from the full delay by the remainder, or modulo (mod) of the delay divided by the wave period (**Figure 8A**). The resulting drive signal and an example wave emission pattern can be seen in **Figure 8B and C**. In addition to ray-tracing, more sophisticated methods can be used to determine delay times needed to focus, such as time-reversal focusing and machine learning-based approaches. Irrespective of calculation method, these delays can be used within acoustic simulation software to produce estimate pressure fields which can then be used in hardware and experimental design. When performed properly, this timing difference allows for targeted constructive interference patterns (**Figure 8**).

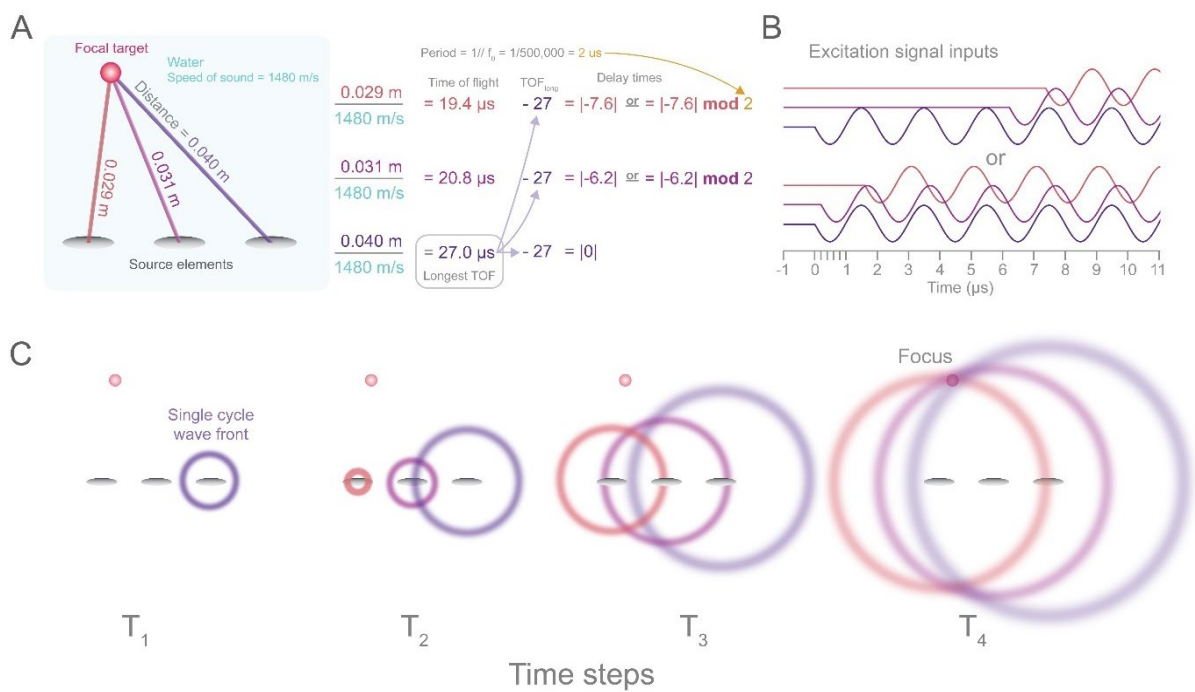

**Figure 8. Ray trace methodology for ultrasound focusing A)** Example ray tracing from three ultrasound source elements to a focal target in a homogenous medium. Calculation of phase delays is shown to the right and can be derived from ray length and speed of sound. **B)** Input signal to transducer elements based on phase delays calculated. **C)** Time course illustration of single cycle pulses from each element with appropriate delays. Constructive interference of the wave fronts at the focal target can be seen at time point 4 (T$_4$).

While arrays offer flexible steering of focal location, they also come with added design and operation complexity. While the relationships discussed in the context of single-element transducers still apply here, such as the dependence of focal size on operating frequency and aperture diameter, ***steering range*** is an additional critical feature of phased array transducers. Designing multi-element transducers can be quite complicated, given the great variety of features associated with their geometry. While detailed phased array design methodology is outside the scope of this guide, some primary considerations are described herein.

In a regularly distributed phased array, element geometry is described by both pitch and kerf (**Figure 9A**). Pitch is the distance between the centers of individual elements, while kerf is the space between the edges of these elements. By varying the size, position, number, and spatial pattern of elements, researchers can find a balance between design complexity and steering and focusing capabilities. One of the most important considerations is the distribution of energy at the focus compared to areas outside the focus. Generally speaking, the further the beam is steered from an orthogonal projection or the geometric focus, the more likely it is that constructive interference will occur off-target. This energy distribution is not uniform and, for regularly arranged arrays, it can result in the incidental formation of off-target interference patterns known as grating lobes. Grating lobes are caused by the periodic arrangement of elements in an array transducer, especially when the element spacing (pitch) is larger than half the wavelength of the ultrasound. This spacing violates the Nyquist criterion, leading to additional regions of constructive interference at angles away from the main beam due to aliasing effects. These grating lobes can be as intense as the main lobe under certain conditions causing significant off-target effects.

In addition to grating lobes, side lobes are another form of undesired acoustic energy distribution. Side lobes are caused by the diffraction of ultrasound waves as they emanate from the finite-sized aperture of the transducer elements. When ultrasound waves emanate from an element, they spread out due to diffraction, creating a wavefront that propagates through the medium. This diffraction leads to interference patterns, resulting in regions of constructive and destructive interference. By carefully designing the element geometry and spatial pattern and simulating pressure fields, researchers can minimize the formation of both grating lobes and side lobes, thereby improving the accuracy and effectiveness of focused ultrasound applications.

A common way to assess and display this balance between off and on-target energy is to simulate the field with the focus steered to different locations over some spatial region and extract the pressure amplitude at the focus as well as the maximum pressure in any region outside the focus. The focal amplitude at the different positions should also be compared to that at the geometric focus or central position. A contour map can then be derived showing the boundaries within which the focus maintains some level of pressure/intensity relative to the maximum possible output. For instance, **Figure 9B** shows that the pressure obtained when steering the focus to ~8 cm axial, ±2 cm lateral (green lines) is still within -1dB (89%) of the peak output obtained at the natural focus, or the point at which focusing is maximized along the orthogonal projection. In contrast, a focus targeted at ~13 cm axial, ±4 cm lateral (red lines) creates a focus that is only ~50% of the peak output. This also generates a stronger off-target grating lobe. As this plot implies, the greater the distance from the natural focus, the lower the focal pressure; a simulated example set of various degrees of steering can be seen in **Figure 9C**. The effects of under-sampling a regular array can be observed in **Figure 9D**, where a prominent grating lobe forms adjacent to the intended focus. This grating lobe can be substantially mitigated through various strategies such as full element sampling, random spatial patterning, or smaller element size.

With these principles at hand, it is advisable to define a worst-case scenario in advance of exploring array configurations. What is the furthest away from the natural focus you expect to target? What is the ratio of pressure between on-target vs. off-target that is considered tolerable? And perhaps, most importantly, what is the allocated budget for the array? Many steering issues are simply solved by reducing element size and increasing count. However, since each element likely requires its own drive source, this pursuit can quickly become very complicated and expensive.

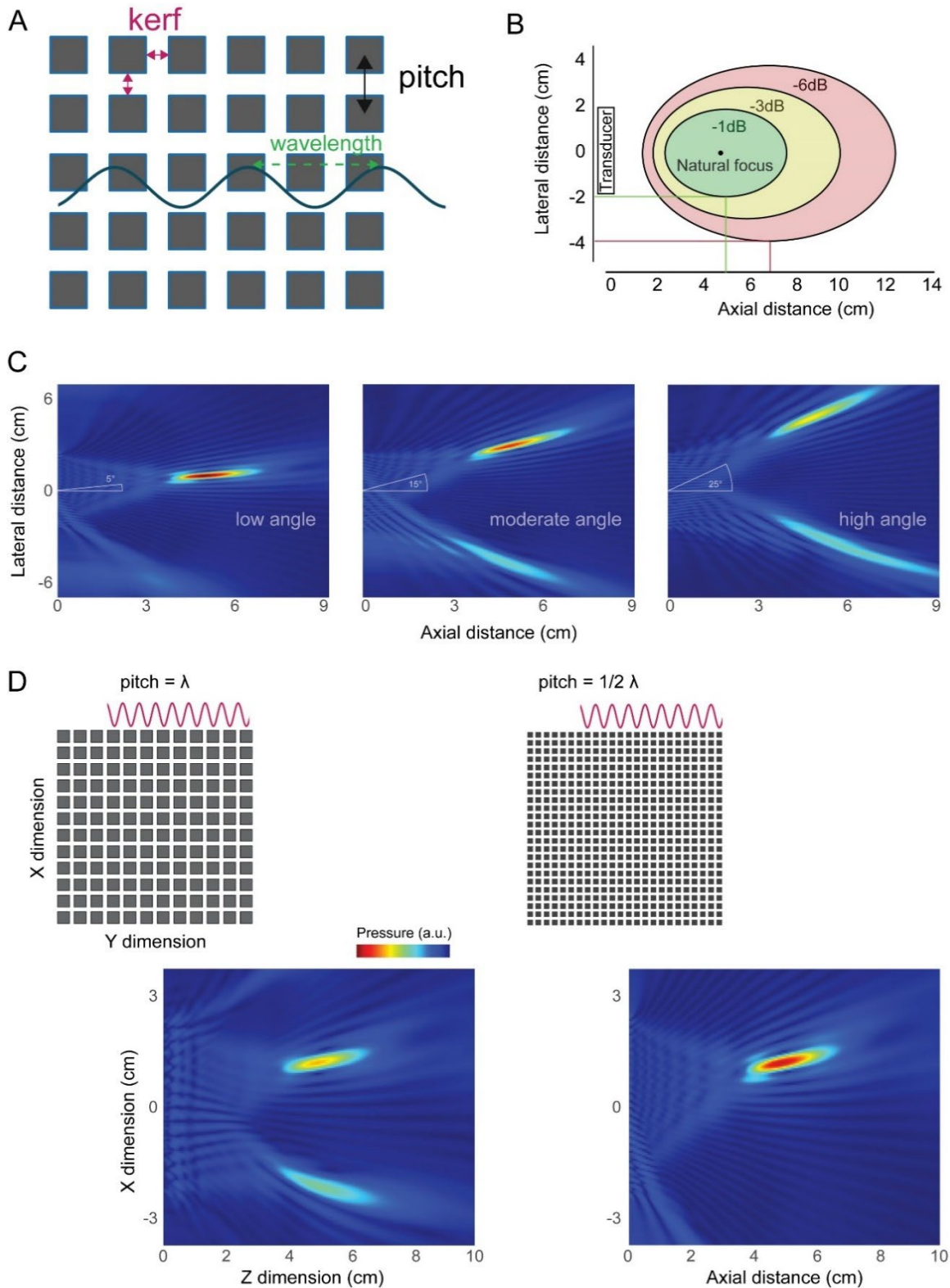

**Figure 9. Characteristics of multi-element phased arrays. A**) Sample geometry of a planar 2D phased ultrasound array. Kerf indicates distance between element edges and pitch indicates the distance between element centers. **B**) Contour map comparing focal peak pressure relative to the natural focus peak pressure at a given steering location. **C**) Simulations of a 64 element 2D array focal profile when steering at various angles. **D**) A comparison of a steered focus with an under sampled (left) and correctly sampled (right) 2D array.

As an alternative to grid organized element arrays, elements can also be organized into concentric rings of increasing radius, constituting so-called *annular arrays* (**Figure 10A**). Ultrasound phase can be varied across each ring to move the focus along the axial, but not lateral, domain. Similar to the two or three-dimensional steering, the focal geometry is subject to change with different focal locations. As shown in **Figure 10**, increasing the distance from the transducer face elongates the beam. In doing so, the energy is also spread along the orthogonal axis, lowering the peak intensity of the focus. In some cases, manufacturers will compensate for this effect by increasing system voltage with greater distance foci; however, this will necessarily increase the total amount of energy delivered through the skull and into the brain.

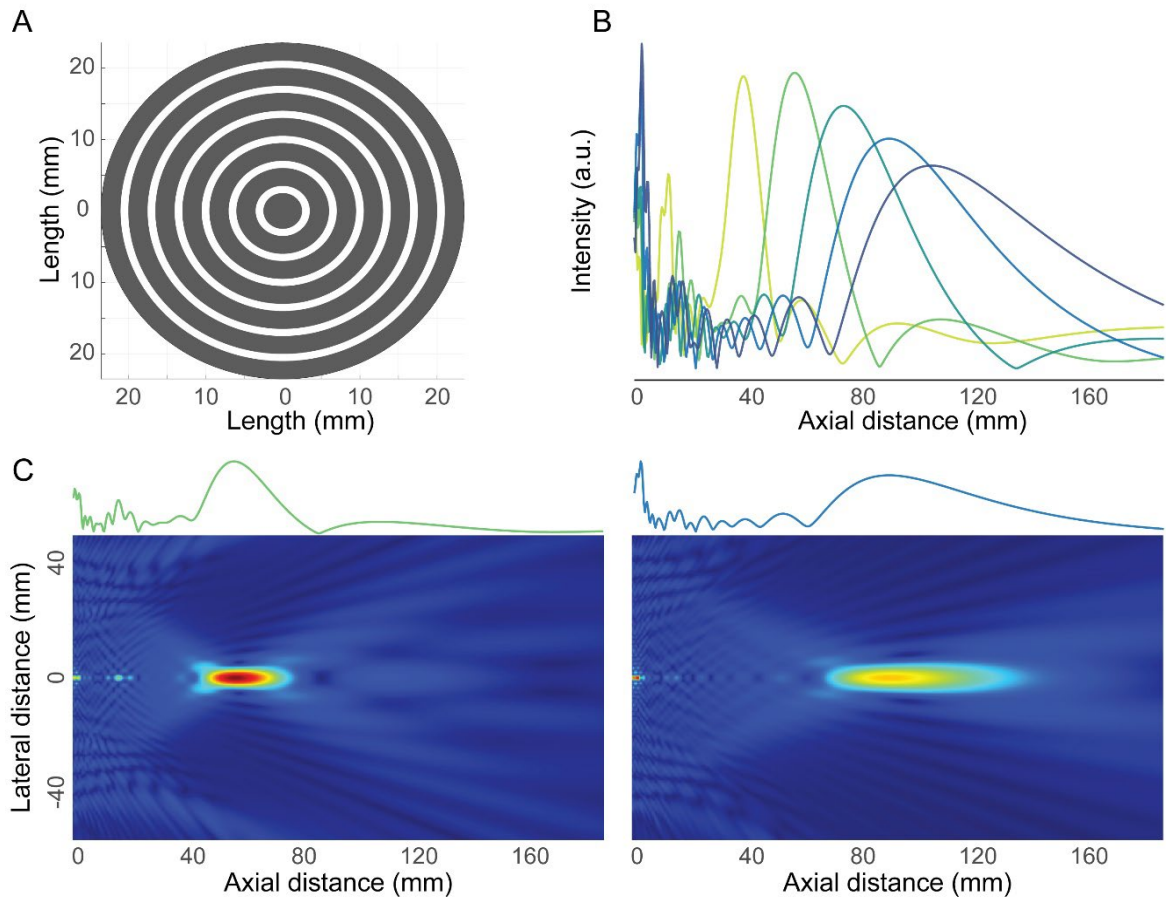

Figure 10. Characteristics of annular phased arrays. **A**) A sample multi-element concentric ring transducer configuration **B**) Axial pressure profiles of the transducer shown in (A) when focused at varying depths. **C**) Focal profiles at different depths along the axial steering range.

Although many types and arrangements of therapeutic transducers exist, not all transducers are suitable for TUS. For instance, *diagnostic ultrasound imaging* transducers are tuned to operate at much higher frequencies (~1-20 MHz) and are unable to transmit sufficient amounts of energy through the human skull. There are many companies designing and building ultrasound transducers, including those optimized for TUS for neuromodulation in humans (**Supplementary Table S2**). A large number of commercial transducers are readily available, and most companies can adapt or design new transducers according to the customer's requirements.

### 2.3. Coupling media and systems

### 2.3.1. Relevance of acoustic coupling

Air has significantly different acoustic impedance and speed of sound relative to ultrasound transducer materials, skin, and underlying soft tissues. As reviewed in section 1.3, differences in acoustic impedance between two media result in a proportional acoustic reflection and a reduction of power delivery to the intended target (Duck, 2007), as shown in **Figure 11A**. As a result, air pockets or layers formed between a transducer and the subject's skin will cause reflections and scattering of ultrasound waves and reduced ultrasound transmission to the target brain region (**Figure 11C**). Given the stark difference in impedance between air and aqueous materials such as soft tissue, the reflection will be greater than 99%. These strong reflections may damage the transducer through heating or mechanical interference. Thus, limiting the presence of these air pockets is of paramount importance. The presence of hair is likely the most common source of air pockets, particularly for cranial applications. In fact, patients undergoing HIFU surgical ablation procedures must have their hair removed to avoid these issues. Another issue that becomes more prominent with increasing transducer aperture is a contour mismatch between the cranial bone and transducer face where local peaks and valleys in the skull don't allow the transducer to make flat contact. Skull constraints may also require that the transducer be tilted at a surface to reach a target, causing an expanding gap along the transducer plane. Fortunately, these issues can be partially corrected using acoustic coupling agents (**Figure 11B**). Note, however, that increasing transducer tilt relative to the skull bone surface also increases the amount of energy reflected.

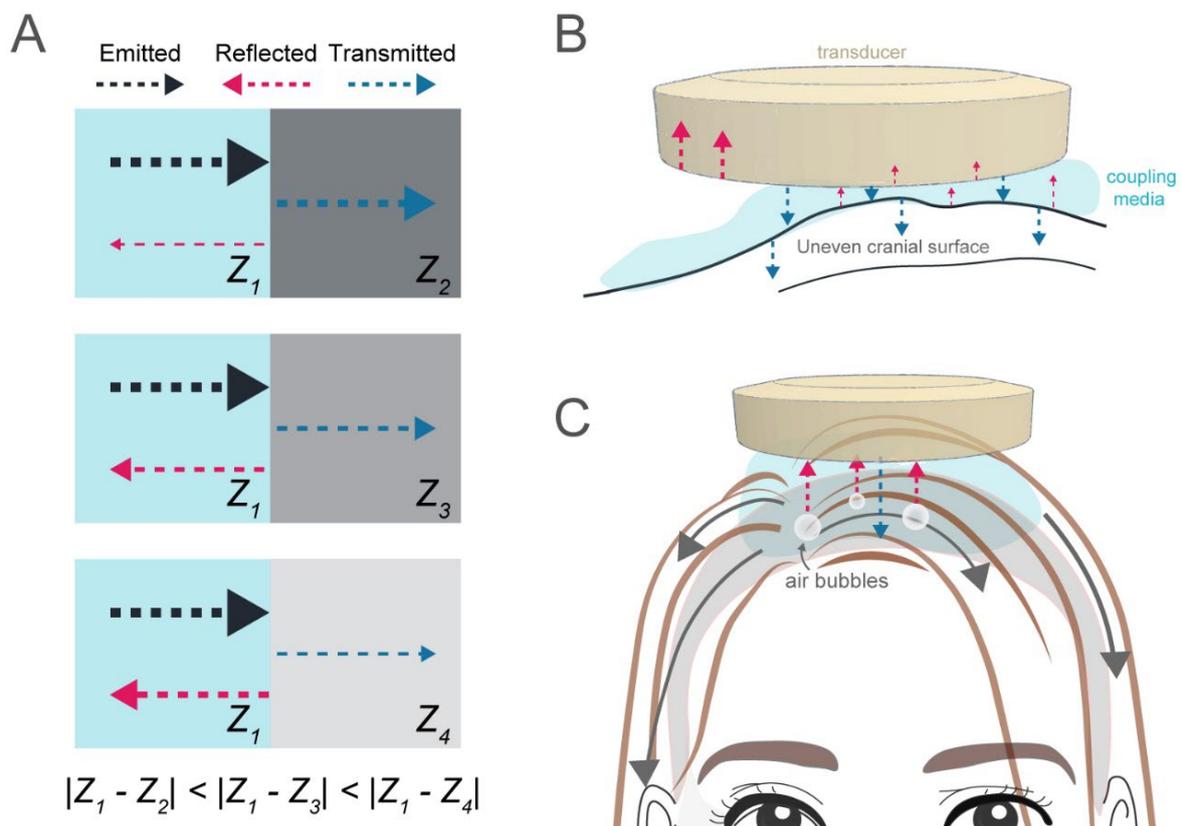

**Figure 11. Effects of insufficient coupling A**) The absolute difference in acoustic impedances between two media determines the amount of reflection at the mediums' boundary. $Z$ denotes the acoustic impedance of a specific medium; the numbers 1-4 represent examples of different mediums. The lighter the color, the greater the impedance of the medium compared to $Z_1$. **B**) A contour mismatch between transducer face and scalp/skull bone can result in considerable air gaps that have to be filled with coupling agents. **C**) Gel should be combed into hair along the natural parting (gray arrows) to remove any air pockets which will cause near total reflection of ultrasound waves.

### 2.3.2. Fluid coupling agents

Generally speaking, acoustic coupling agents are semi-viscous homogenous materials with acoustic impedance close to that of skin. A wide variety of materials can be used as acoustic coupling agents but the most commonly used is aqueous **ultrasound gel**. Biocompatible moisturizers, hair conditioners or products, silicone agents, and oils may also be used, which readily flows into small gaps. However, the specific transmission and biocompatibility should be examined or empirically validated prior to use.

During the process of application, it is critical that the coupling agent fills all voids where air bubbles might reside. Although this process is not standardized, there are numerous steps researchers can take to ensure homogenous coupling agents. For example, ultrasound gel may contain a great deal of air bubbles formed during the handling of the bottle or containing vessel. While the impacts and typical variation in bubble formations are not well understood from a quality control standpoint, we recommend examining common laboratory gel samples before using. If the gel contains visible bubbles, degassing the gel can ensure consistent coupling. **Figure 12** shows the result of degassing ultrasound gel through centrifugation using a 50 mL polycarbonate syringe spun for 5+ minutes at 3,000+ rotations per minute. Centrifuged gel can be stored normally following degassing. Prior to use, the gel should be visually inspected to ensure that it remains free of any bubbles or contaminant. Importantly, further studies are needed to aid quantification of air bubble content in gel and its impact on transmission, as most regularly handled gel dispensers may not meaningfully benefit from degassing.

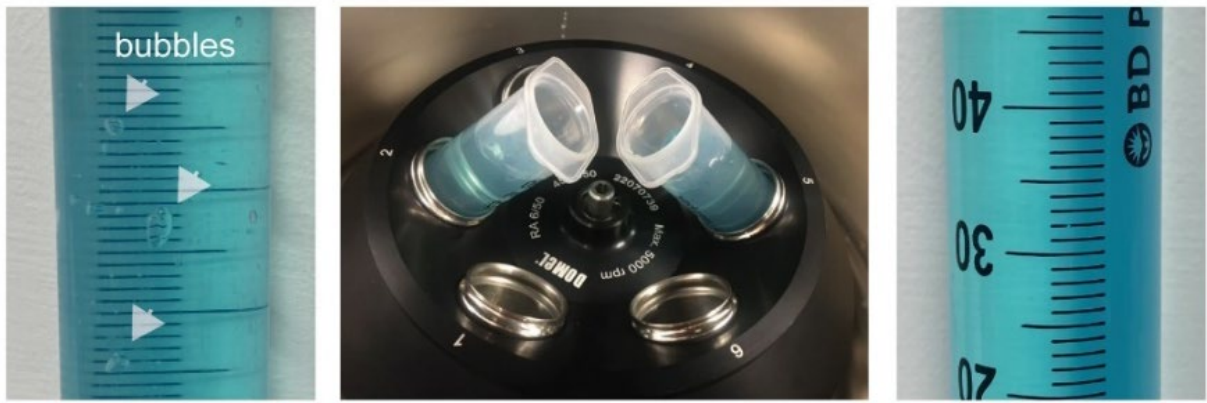

Figure 12. **Effect of a degassing procedure on aqueous gel**. Example of aqueous gel prior to (left), and after (right) centrifugation. Alternatively, the syringes can be placed in a vacuum chamber for degassing.

While there are many ways to apply gel, an exemplary method that we recommend is described here. Participants should be asked to arrive with clean hair. While hair cleanliness or cleaning product use has not yet been studied in the context of ultrasound transmission, clean hair would be expected to have less tangles and debris serving as nuclei for air pocket formation. Prior to gel application, researchers should comb the subject's hair gently to remove entanglement. The tip of a rattail comb can be used to create a central parting (precise line) at the location where the transducer will be placed on the head (**Figure 13**). This parting allows experimenters to see a narrow line of skin clearly where the parting has been done. Ideally, the parting should follow the natural direction of the individual's hair to reduce resistance and subsequent hair recoil to its natural position. Next, researchers should apply a line of gel along the axis of the skin line just created. The gel should be spread outwards parallel to the direction of the hair and the part line. The hair can then be gently combed lateral to the part line. This should be repeated every 2mm with parallel lines. Researchers should take into consideration different hair types. For thicker hair, these sections may need parting lines that are smaller than 2 mm. After each section is completed, the hair should be folded over, much like turning the pages of a book, and repeated on the other side. After covering the desired area with this method, the entire area should be gently flattened. Then, apply additional gel on top to ensure that the hair is thoroughly soaked. It is important to note that detecting bubbles through tactile sensation or auditory cues is one of the most effective methods for their removal. Indeed, the researcher can place their ear close to the surface of the head and listen for sound cues when smoothing the gel with their finger; bubbles sound like a small crackling sound. These bubbles may also be felt by the finger. When a crackling sound is no longer heard, the bubbles are likely removed. Additional gel may be applied to the surface of the prepared hair to ensure coupling. While this method is generally applicable, certain hair types or styles may require adjustments to ensure sufficient gel application. Investigators should use their discretion to confirm that the hair is adequately saturated with gel for effective TUS application, taking care to adapt the method as needed. Depending on the protocol being used, researchers may clamp the transducer to the head with their hands or with some mechanical system (see **section 3.3.1.4**).

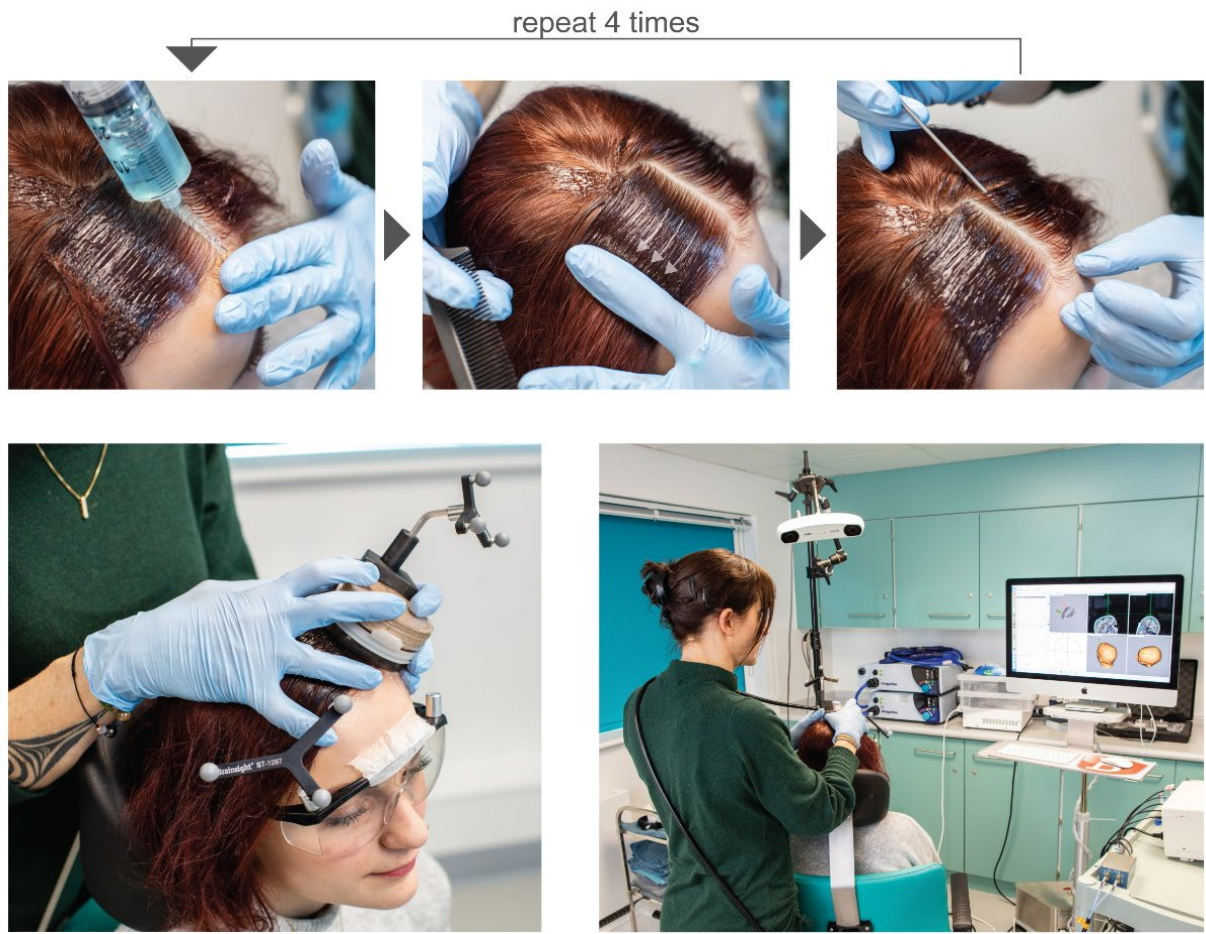

**Figure 13. Gel coupling of ultrasound transducer on hair.** Degassed gel is applied following the parting line of the subject's hair (top left). Applying finger pressure in sweeping motion to saturate hair with gel and press out air bubbles (top middle). Example of additional hair parting following the natural direction of the hair (top right). Bottom panels show an example of a fully coupled subject during a study.

### 2.3.3. Coupling pads and balloons

In some instances, a large gap may be present between the transducer and subjects head to improve comfort or reduce temperature sensations related to the probe temperature. If the gap is too large to contain ultrasound gel for an extended period, compressible *hydrogel or cryogel pads* can be used to fill the gap (Lee et al., 2014). These pads have acoustic transmission similar to ultrasound gel with far more structural rigidity. These pads are hypoallergenic and easily stored for long term projects. They can be cut to a desired size and shape from a larger block using cutting wires or thin and sharp blades. Alternatively, custom made pads from commercially available de-gassed high-viscosity gel polymer matrices have been evaluated and reported to have good coupling properties while holding shape at room temperature and during sonication (Strohman et al., 2023). In practice, the pads should be coupled to the transducer with ultrasound gel and a clamping mechanism. Immersing the transducer, gel and pad in water while clamping can help ensure the coupling is free of air bubbles. Another alternative is to use pliable dry materials such as *silicone pads* which closely matches the acoustic impedance of skin and can easily mold to the surface of a subjects cranial surface (Norman, 2017). To bridge even larger gaps with high degrees of flexibility, *coupling balloon* systems can be used. These balloons are highly flexible pouches which can be filled with water or aqueous solutions and can help to account for larger contour mismatches or larger transducer angulation. They can also be used to deliberately increase the distance of the transducer from the target as an alternative to electrical axial steering. To create a water balloon, either a stretchable membrane is attached directly to the transducer housing sealed with a silicone ring to create a half-balloon in between transducer face and membrane, or a *full balloon* is directly attached to the transducer face. Degassed water is then filled into these balloons to create a flexible but generally convex surface. The required pressure is obtained by using a simple syringe or a dedicated pump, typically injecting water via one tube and letting out air bubbles via another. In principle, using two tubes also a fluid circulation system for active cooling can be created. Note that any of these additional materials must also be coupled with aqueous gel or other fluid coupling media to avoid air bubbles being trapped in between surfaces.

### 2.4. Acoustic measurements and equipment calibration

Many transducers and drive systems are supplied with a calibration report created by the equipment manufacturer, which should provide the relationship between drive settings (output level and focal position if applicable) and the position and amplitude of the focal pressure. Measurements should be made to establish this relationship where the information is not provided, for example for custom transducers. Since transducers are often stress-tested (a.k.a. "burned in") prior to delivery, the decline in transducer performance over time should be limited. Nevertheless, it is recommended that researchers conduct regular checks to ensure that the acoustic output of the transducer remains stable over time to ensure repeatable acoustic energy delivery to subjects throughout a study. Otherwise, damaged cables, connectors, or transducer elements may go unnoticed, leading to corrupted transducer output and electronic beam steering and flawed study results. Measurements performed as part of this process must be quantitative and related to acoustic output quantities such as acoustic pressure, intensity, or acoustic power. This may involve direct measurement of these quantities or secondary measurements such as transmitter electrical power, driving voltage, or transducer electrical impedance. Qualitative observations such as audible noise emitted from a transducer are not adequate for ensuring that the transducer output is stable or operating at a safe and effective level. Depending on equipment availability, accessibility, and training of lab members, a variety of checks can be performed at a regular interval or prior to each subject's exposure. While there are no explicit timelines established for testing, we recommend incorporating at least one of the following tests from **Table 2**. Importantly, the researcher must decide a range of tolerance in performance deemed acceptable (i.e. ± 10% or other) for subject testing and ideally reported. Guidance on performing calibration and routine testing is also provided by the ITRUSST Equipment Working Group and the latest recommendations can be found on the ITRUSST website. Here, we provide an overview of basic measurement methods and hardware (https://itrusst.github.io/documentation/equipment/).

**Table 2.** List of equipment tests.

| Test | Description | Considerations |
|---|---|---|
| Focal pressure characterization | Using a calibrated hydrophone aligned to the focus of field, the peak focal pressure can be measured. | Requires specialist equipment and can be time-consuming. The placement of transducer with respect to the hydrophone must be highly repeatable to ensure measurement repeatability. The hydrophone performance should be routinely checked at an interval recommended by the manufacturer. |
| Acoustic radiation force balance measurement | A radiation force balance with an appropriate absorber can be used to measure total acoustic output from the radiation force exerted on an absorbing or reflecting target. | Requires specialist equipment, sensitive to vibrations. The absorber material must be suitable for the frequency range of interest and of an appropriate size to capture the field. |
| Electrical power, voltage or impedance measurement | A power meter can be placed in line between the amplifier and transducer to measure the dissipated electrical power at the output level settings used in studies. Drive voltage and current can also be measured separately. | Simple measurement. May require separate measurement for each element. |

A *hydrophone* is an acoustic receiver which generates an electrical voltage in response to an incident pressure field, providing the time varying waveform. Pressure can then be derived using a calibration curve relating voltage to pressure. As suggested by their name, hydrophones operate and are calibrated in water. There are various types of hydrophones with special characteristics including piezoelectric membrane, probe (e.g., capsule and needle), and optical type hydrophones. A list of applicable hydrophone manufacturers is provided in **Supplementary Table S3**. Like the TUS transducer, it is difficult to comment on which type is best suited for any laboratory's particular needs. For measurement of focused fields used in TUS, probe type hydrophones are well suited as they have a small surface area, reducing the potential for reflections or standing waves between the transducer and hydrophone. They are also less costly and can be less delicate. In any case deciding on a hydrophone requires consideration of three main features. The size of a hydrophone relates to the size of the sensitive element over which the pressure signal is collected. Sensitivity increases with area, but with a corresponding increase in spatial averaging, which can cause significant underestimation of peak pressures when the hydrophone sensitive area begins to approach the size of the focal region. Hydrophone size is also directly related to its *directivity,* which describes its sensitivity to pressure as a function of angle of incidence. This can be very important since for some hydrophones, depending on effective element size and the frequency of the ultrasound field, sensitivity can fall sharply with increasing angle of incidence (**Figure 14A, B**). The orientation of the hydrophone during pressure measurement should match the orientation used during the pressure calibration. *Frequency response*, or the sensitivity to waves at different frequencies, may also be important since hydrophone response can be non-uniform as a function of frequency. In general, needle and fiber optic hydrophones have the most non-uniform frequency responses. This can introduce uncertainty when measuring nonlinear or broadband fields (such as HIFU fields or optically generated fields), which is not a concern in TUS applications. However, for measurement of TUS fields, it is important to obtain a calibration at the frequency of interest rather than extrapolating from neighboring frequencies. Hydrophone calibrations can be obtained from the manufacturer or from an alternative calibration provider such as a national measurement lab. The frequency range and step size should be specified when obtaining the calibration in either case. Finally, most hydrophones are extremely delicate in and should be handled very carefully. They are highly susceptible to damage by high acoustic pressures and cavitation, so care should be taken to maintain safe conditions by ensuring water purity and avoiding high focal pressures, especially at lower frequencies.

A hydrophone system often consists of the hydrophone sensitive element, preamplifier (often submersible), cable, and power supply, which then connects to the sampling oscilloscope via a BNC cable. The hydrophone system should always be used under the configuration in which it was calibrated– with the same combination of hydrophone, preamplifier, power supply, angle connectors, attenuators, or other signal amplifiers for example. The preamplifier usually serves to match the high impedance hydrophone element to the 50 Ω cables and electronics in the rest of the signal chain, in addition to amplifying the signal. Some hydrophones may not have a preamplifier and may have a high impedance output and typically have a short cable. Termination impedance at the oscilloscope must match the output impedance of the hydrophone system and the characteristic impedance of the connecting BNC cables where used. Similar to acoustic reflections from boundaries between materials of different acoustic impedance, electrical reflections can occur where there is a change in electrical impedance. Improperly terminated connections will affect the measurement of voltages at the oscilloscope resulting in incorrect conversion of voltage to pressure when applying the hydrophone calibration. Many oscilloscopes have a choice of either 1 MΩ or 50 Ω impedance for each channel. Alternatively, a 50 Ω shunt or terminator can be connected between the cable and channel input. Hydrophone systems which have a 50 Ω output impedance should be terminated in 50 Ω at the oscilloscope. Hydrophones with a high output impedance should be terminated in 1 M Ω at the oscilloscope. Note that RG58 BNC cables have a characteristic impedance of 50 Ω, and that extending a high impedance cable by connecting an extra length of BNC cable will result in electrical reflections and affect the overall sensitivity of the signal chain.

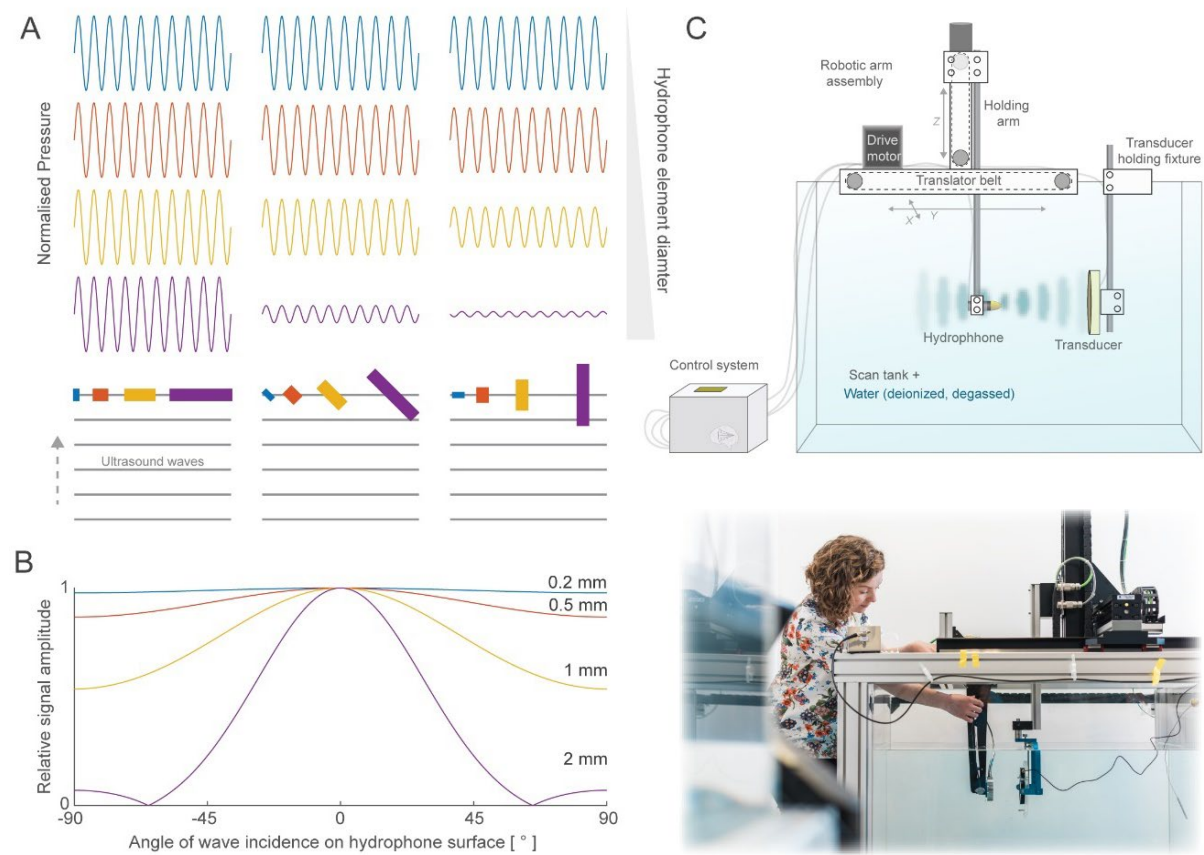

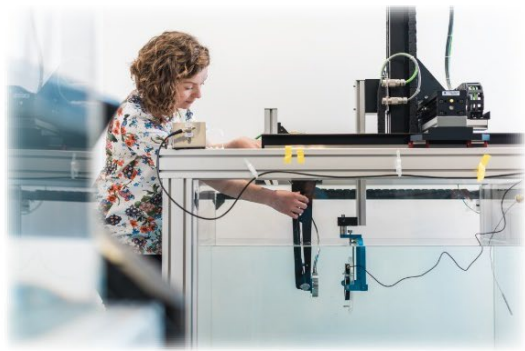

**Figure 14. Hydrophone characteristics and scanning tank setup. A)** Illustration of the example of sinusoidal signal output from hydrophones of increasing diameter (colored rectangles) positioned with increasing angle of incidence relative to the direction of wave propagation. **B)** A hydrophone directivity plot for the hydrophones showing the relative signal amplitude as a function of ultrasound wave incidence with the transducer surface. **C)** An illustration and photograph of a hydrophone scanning tank.

*Hydrophone scanning tanks* are mechanical systems that manipulate a hydrophone in the pressure field generated by an ultrasound transducer radiating into water to facilitate pressure measurements (**Figure 14C**). Briefly, a motor-mounted mechanical arm steps the position of a hydrophone through water along several axes. Voltage waveforms are acquired from the hydrophone at each position and are assembled to create a spatial pressure profile. In some cases, these systems may also come with water filtration and degassing systems to prevent biological growth and cavitation at lower frequencies. Several manufacturers make off-the-shelf hydrophone scanning tanks which consist of a water tank with a mount for holding the source transducer and a hydrophone mount attached to a motorized translation stage for positioning the hydrophone (**Supplementary Table S3**). An open-source instruction for building a low-budget version (Clinard et al., 2022) can be found on this website: https://osf.io/ys7jr/. While this equipment is nearly essential for characterizing an ultrasound field, it is not necessarily required for assessing transducer consistency post calibration. If a hydrophone can be precisely placed within an ultrasound focus repeatably between measurements, consistency checks can be made at a single point. While this method can be performed simply without a scanning system, it assumes that the larger field inherits any degradation or change measured at a single point and that the positioning of the measurement is highly reliable.

While scan tank hydrophone measurements by the lab or manufacturer prior to every use would be an ideal scenario for quality control, this would likely be too time intensive for most laboratories. Instead, power meters can be used to assess the power draw of a system during a hydrophone characterization. The transmitted and/or reflected powers recorded can then be used as a reference point since changes in power draw likely reflect changes in the system. We recommend assessing the variance in this measurement while employing fixed protocols which will be used in the study, where the transducer is placed in water or some acoustically transmissive media. Prior to each experimental use, the system power draw can be rapidly assessed. If the power draw values are outside the range of initial measurements, the system should be characterized again using scan tank hydrophone measurements. The acceptable range should be set at the discretion of the investigator to ensure deviation does not result in safety concerns and should be described carefully in the methods.

### 2.5. TUS control software

For some commercial turnkey ultrasound neuromodulation systems, the embedded control systems are largely hidden from the user and allow only the selection and start of predefined protocols, while others allow to change at least a subset of the stimulation parameters (within certain safety boundaries) via a graphical user interface or even remotely via an *application programming interface (API)* as well as external triggering of the sonication. While such interfaces lower the complexity and safety concerns the user has to deal with, they also lower the customizability and make it more difficult to design novel protocols or to make ad-hoc modifications in real-time. Custom systems provide more flexibility in driving the system, e.g., allowing to program custom waveforms for individual channels, and for less complex systems, such as single-element transducers, a function generator can principally be controlled directly via a custom software. Function generators may contain firmware for communication through a general-purpose interface bus (GPIB), a protocol that allows external command-based control over the function generator's operation including the ability to change the output waveform shape and voltage, modulation inputs, and the exact timing of when the device output is on. However, such customized solutions require the user to implement all relevant safety measures by themselves. As is true of all medical devices using computerized systems, the possibility for unexpected conditions or errors in hardware or software may pose safety risks to human subjects. For instance, an applied voltage may be excessively high producing higher amplitude ultrasound than expected or a pulse could run indefinitely leading to excessive temperature rise. Extensive safety testing and documentation should also be performed by the study researchers to ensure no harm is brought on to the participants.

## 2.6. Equipment requirements for combining TUS with strong electromagnetic fields

When combining TUS transducers in close spatial proximity with devices inducing an electromagnetic (EM) field, such as MRI or TMS coils, it is important to rule out device interference in both directions. The primary concern is operational integrity and sonication behavior of the TUS transducer while subject to large EM fields, such as MRI bores and as part of an online TUS-TMS combined stimulation(Fomenko et al., 2020; Legon et al., 2018b; Zeng et al., 2021). It is also important to consider the potential for added neuromodulatory effects arising with mechanical and electromagnetic field interactions (Webb et al., 2023). Prior to conducting experiments in combination of TUS with MRI or TMS, researchers should perform a thorough investigation of the material specifications of the transducer. This can usually be obtained from the manufacturer, and special note of any ferromagnetic materials within the transducer body, housing, or cables should be made. Transducers classified as "MR-safe" or "MR-compatible" by the manufacturer should operate to specification when in proximity to an active TMS coil, since the focal magnetic field strength of a TMS coil is often similar to that of a static magnetic field used in MRI imaging (1.5-2.0 Tesla)(Pawar et al., 2008). An example of proper operation of a custom non-ferromagnetic TUS transducer in proximity to a figure-8 TMS coil can be found in works by Legon et. al (Legon et al., 2018b). When using transducers not intended for transcranial neuromodulation applications, great caution should be taken before the use with TMS/MRI, as significant damage to the device may occur from in the presence of ferromagnetic components (Ai et al., 2018a, 2016).

### 2.6.1. Combining TUS with MRI

Several concerns arise when using TUS equipment in an MRI room. First, consideration must be given to emergency procedures and ensuring that the TUS setup does not impede the evacuation of the participant. Second, standard transducers which contain lead zirconate titanate piezoelectric material or cable connectors made from nickel can produce susceptibility artifacts. These artifacts may have a substantial effect on the signal emanating from cortical areas below the transducer and may limit the brain areas that can be effectively studied. Although it is difficult to entirely block these artifacts, they can be minimized by use of MR-compatible transducers and/or careful shimming and optimization of MRI parameters and pulse sequences. Third, some MR configurations operate with a very high specific absorption rate (*SAR*), particularly those with specific RF coils or use of high static magnetic field strengths. These configurations may raise the temperature of tissue in contact with the transducer in addition to the TUS-induced temperature rise, and must be carefully monitored and, ideally, tested on phantoms beforehand. Third, appropriately sized transducers and/or MR head-coils must be chosen so the transducer can be comfortably and accurately placed within the restricted MR environment. There should also be adequate space for transducer fixation materials, foam padding, and optical navigation trackers in case MR-compatible neuronavigation is available. Finally, either cables of sufficient length that are run through low-pass filter plates are required to connect the transducer to a drive system outside the MR room, or a sufficiently MR-compatible drive system has to be located in the MR room connected to a control system outside the MR room via fiberoptic cables. Cable traps made from ferrite beads are often used on cables run through the scanner bore to prevent radio frequency (RF) interference by blocking unwanted RF currents and thus signal distortions; whether such measures are required needs to be determined by signal quality measurements for a given setup. Despite the technical challenges, concurrent TMS-MRI studies have successfully been performed in humans at both 3T and 7T (Ai et al., 2018b)– see section 3.5.3 for examples.

### 2.6.2. Combining TUS with TMS

When performing TUS and TMS at the same time (see section 3.5.4 for examples), it is important to consider whether the presence of a transducer underneath a TMS coil might influence the magnitude or waveform of the temporally and spatially overlapping TMS-EM field. The obvious impact is the increased scalp-coil distance created by insertion of a TUS transducer. The increased distance will reduce the magnetic field reaching the brain and thus the electric field induced in the target region, which may limit the ability of TMS to fully depolarize neurons and trigger action potentials. However, if the transducer is sufficiently thin to minimize the resulting increase in coil-cortex-distance, and the TMS coil is sufficiently sized and powered to compensate for the decay of magnetic field strength with increased coil-cortex-distance, adequately sized **motor-evoked-potentials (MEPs)** (i.e., up to 1.0 mV) can be achieved. Notably, the volume of the elicited EM field in the brain will be substantially larger in this scenario due to the required increase in coil size and stimulation current. Care should be taken to maintain both the orientation and position of TMS coil and TUS transducer relative to the target and to each other as well as coupling medium integrity to avoid the introduction of systematic confounds. However, ensuring stable positioning of the transducer and the TMS coil can be challenging, dedicated transducer/coil holders (e.g., fixing one firmly to the other) and neuronavigation are mandatory, and multiple experimenters may be required. Similarly, if a TMS sham/control/baseline condition is to be performed, the unpowered transducer (or a respective placeholder) and any coupling media should be present to control for scalp-coil distance, which may otherwise confound the MEP amplitude.

If the TUS transducer is custom made, and the presence of ferromagnetic materials in the housing or cables is uncertain, a validation of the TMS-induced EM field in the presence of the transducer can be performed. This will increase confidence that any changes in MEP amplitude are likely from TUS neuromodulation itself, rather than an equipment-related interference or reduction of TMS induced fields. Several control conditions may be compared during testing, such as the TMS coil with a dummy transducer made from wood/plastic to assess the effect of added distance or a TMS coil with an unpowered/powered transducer in-situ to test assess interference effects with transducer materials and the sonication itself, respectively. Direct or indirect measurements of the induced TMS field at the intended focus can easily be made with a simple circuit and either a multimeter, or a data-acquisition interface.

The indirect measurement involves wrapping a thin, insulated conductive wire into a small (preferably 1 cm or less in diameter) coil. The wrapping can be aided by coiling around a pencil or a suitably-sized cannula/rod. The terminal ends of the wrapped wire can then be connected to an Ammeter, or a Multimeter, or any device capable of measuring current, preferably with a high sampling rate. The coil can then be carefully positioned at the desired location (i.e., near the TUS transducer and at the epicenter of the induced TMS magnetic field), and measurements of induced current can be made during TMS and/or TUS activation. Of note, the magnetic field is a 3D vector, so positioning of the coil in the three spatial axes should be done if a proper 3D assessment is desired. A more direct and precise measurement of the EM field can be made by purchasing an inexpensive transistor-type chip called a **Hall effect sensor**. These are miniaturized sensors designed to convert any static or dynamic magnetic field applied to their face into a time-varying voltage, proportional to the magnetic field. No manual coiling of wire or calibration is needed, and only careful positioning in 3D space of the sensing face is required (Mancino et al., 2020).

### 2.7. MRI and/or CT imaging for acoustic modelling

During experiment planning, to simulate the intracranial acoustic field (section 2.8), it is important to estimate the skull geometry and bone density for each subject due to the significant effect of the skull on the propagation of the ultrasound field. **Two steps** are ideally performed for the stimulation planning: (i) acquire **CT images** or **Ultrashort echo time (UTE) MRI** for each subject to estimate skull characteristics and (ii) acquire **structural (e.g., T1-weighted) MRI images** for each subject to identify the region to stimulate. CT Hounsfield Units have been correlated with the speed of sound and attenuation. Alternatively, UTE MR images have been demonstrated as a reliable option for bone density estimates, which is particularly important in reducing the burden of accessing CT scanners, requiring subjects to undergo two scans, and the radiation exposure inherent to CT acquisition. UTE images can be inverted and scaled to provide CT-like images or **pseudo-CT images**, which correlate well with density profiles across the skull (Wiesinger et al., 2018) (**Figure 15**). UTE variants can be found on most, if not all, of the available MR scanners. As shown in **Figure 15**, cranial bone can have dramatically different representation across scan types, where density and speed of sound information may be more difficult to obtain from T1 images as compared to UTE. Open-source MATLAB toolboxes exist for converting the PETRA UTE sequence images to pseudo-CTs (https://github.com/ucl-bug/petra-to-ct). However, the T1 scan is superior for anatomical contrast of brain tissue which is useful for identifying sonication targets. Thus, a combination of a T1-weighted and CT or UTE scans are typically used for neuromodulation studies. As an alternative to predefined scan sequence data, pseudo-CT images can be generated from T1-weighted MRIs based on deep learning algorithms (Siti N. Yaakub et al., 2023). While these implementations hold great promise, deeply trained models are hardly generalizable to new data and thus will not contain any idiosyncratic information about individual skulls. Thus, we recommend careful review new approaches and compare them to manual skull segmentations and established toolboxes such as SimNIBS (simnibs.org) (Thielscher et al., 2015) or ITK-SNAP (itksnap.org).

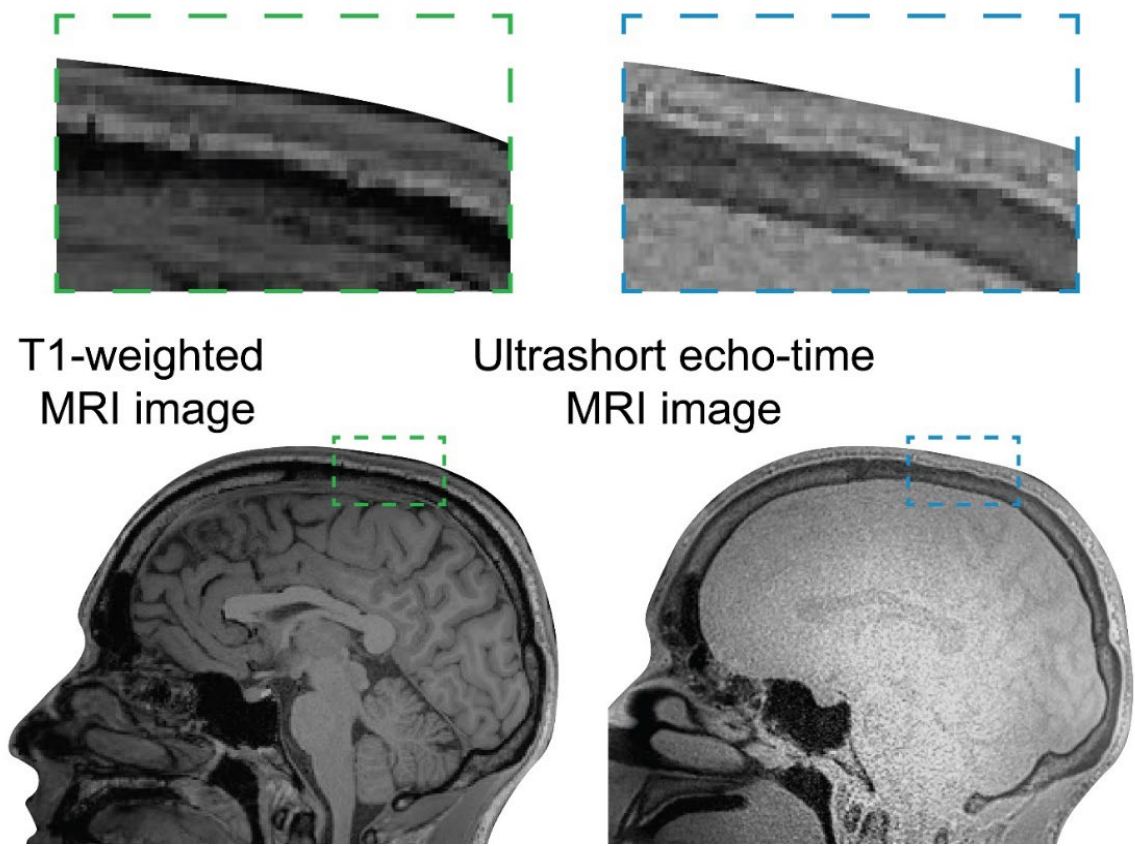

**Figure 15. Comparison of T1-weighted and UTE sequences.** Sample T1-weighted MRI scan (left) compared to an ultrashort echo time (UTE) MRI scan (right). Note the variation in signal given by the bone between scans.

### 2.8. Acoustic simulation models and software

***Numerical modeling*** of ultrasound propagation within biological tissues can be used to estimate in-situ exposure and resulting thermal (Constans et al., 2018) and/or mechanical effects (Ozenne et al., 2020) induced by TUS. One purpose of these models is to derive the in-situ focal position, shape, size, and pressure for a given **transducer** position, orientation, and/or electronic steering parameters. While this section is not intended to be a detailed instruction set, it will provide an understanding of fundamental steps required to implement modeling in TUS planning (**Figure 16**).

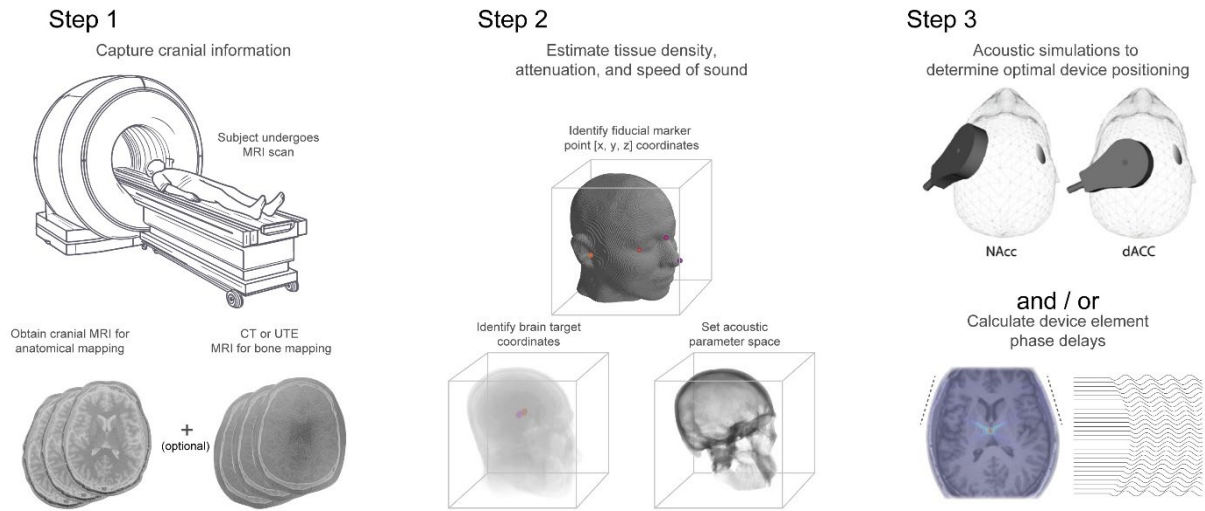

**Figure 16. Stepwise use of acoustic simulations in the experimental workflow.** Steps for incorporating acoustic simulations into TUS planning begin with the acquisition of anatomical MRI for soft tissue maps as well as either CT or MR (e.g., UTE) scans for cranial bone density imaging (Step 1). Next, acoustic space mapping of cranial features is performed, including brain target location, facial registration points (such as the nasion, canthi, tragi, as shown), and skull property estimates (Step 2). Finally, acoustic simulations of TUS devices can be conducted prior to stimulation to determine the correct transducer placement relative to the intended brain targets or the element phasing required to steer a focus to a particular target (Step 3). The top panel of Step 3 illustrates the optimal positioning of a fixed-focus transducer relative to the nucleus accumbens (NAcc) and dorsal anterior cingulate cortex (dACC). The bottom panel shows a 2D cross-sectional plane of two multi-element arrays placed over the temporal window, alongside a sample signal input from one of the transducer arrays.

The skull represents a challenging, but manageable hurdle in TUS due to its high ultrasound attenuation coefficient and high sound speed compared to skin and brain. The high sound speed coupled with spatially varying skull thickness and internal scattering distorts or aberrates the ultrasound field, affecting focusing, targeting accuracy and specificity. The high attenuation caused by both absorption and scattering in the skull prevents a large part of the ultrasound energy from reaching the brain target. This absorption can also cause cranial heating that can diffuse to the skin and brain. Simulating the propagation of ultrasound through the skull can help to evaluate these effects in individual subjects to plan and evaluate targeting of specific brain structures with a given transducer. Modeling acoustic pressure fields based on CT-/MR-derived bone and brain information in combination with neuronavigation tools can enable the user to recreate a planned transducer position, suitable for targeting the desired brain region with human participants. Post-hoc simulations of TUS experiments can be used to evaluate the applied exposure for reporting and safety assessment, to aid correlation of experimental results with TUS exposure, and to refine hypotheses based on experimental observations (Legon et al., 2023, 2022; Strohman et al., 2024). Personalized planning simulations can also be used to scale the transducer output to achieve consistent exposure across participants. When multi element array transducers are used, simulations can be used to derive the ultrasound element phases required to correct for skull aberrations and focus on the target (**Figure 17**). Alternatively, acoustic lenses can be printed for single-element transducers to perform this aberration correction (Hu et al., 2022; Jimenez-Gambin et al., 2022; Maimbourg et al., 2020, 2018; Sallam et al., 2024, 2021).

In addition to simulation of fundamental acoustic parameters, secondary effects such as temperature rise resulting from the deposition of acoustic energy may also be simulated. This is performed using a heat source calculated from the intracranial acoustic intensity distribution and the acoustic absorption coefficient as an input to the Penne's bioheat equation (Leung et al., 2019), or comparable equations. Although the modeling process can be achieved in many ways, a step-by-step procedural example is outlined in Figure 16 and further details can be found in the literature (Angla et al., 2023; Treeby, 2019).

**General principles of numerical modeling:** Broadly, acoustic simulation involves estimation of the ultrasound pressure at each voxel or grid point in space over consecutive time points, based on solutions of the governing equations with inputs which define the behavior of the source and describe the propagation medium. Acoustic properties such as density, speed of sound, and attenuation, must be defined throughout the simulation domain, with particular attention to the acoustic properties and geometry of the skull. In the context of ultrasound propagation, the properties of soft tissues such as white and grey matter of the brain are very similar and usually described by a single brain value. The skull geometry and acoustic properties may be defined in different ways. Perhaps the most comprehensive is voxel wise mapping of density and sound speed from medical images as described in Sec. 2.7, although clinical CT resolution is not sufficient to image the skull microstructure resulting in spatial averaging of properties within voxels. Alternatively, the skull can be represented by a three-layer model describing the outer dense layer of cortical bone, a spongy middle layer of trabecular bone also known as the diploe, and an inner layer of dense cortical bone (**Figure 17**). Two different sets of constant acoustic properties are then applied to the two types of bone within the three layers. In the simplest case, a whole skull mask can be used with average properties of skull bone applied within all voxels inside the mask. The anatomical accuracy of the skull representation can have a large impact on the accuracy of the model, with higher errors in pressure amplitude and focal position observed with simplified models (Hynynen and Jones, 2016). It is important to note that no matter which method is used, there is always uncertainty in the acoustic properties derived using these methods (Leung et al., 2021, 2019). Attenuation coefficient is often defined by a single value over the skull, which is often taken from literature values obtained from small sets of skull fragments (Gimeno et al., 2019; Pinton et al., 2012), as there is currently no well-established mapping from image values to attenuation coefficient. The uncertainty on skull attenuation or absorption coefficients is therefore considerable and can propagate to large errors in simulated pressure in some cases. Tools and methodologies have been made available which can aid new investigators in assigning acoustic properties to skull (Cain et al., 2022; Miscouridou et al., 2022; Pinton et al., 2012; Siti N. Yaakub et al., 2023).

In addition to the acoustic properties, inputs relating to blood perfusion may be included in the model to obtain a more realistic estimate of temperature changes (Leung et al., 2019). Importantly, more parameters are needed to include perfusion in simulations and may not be included by default.

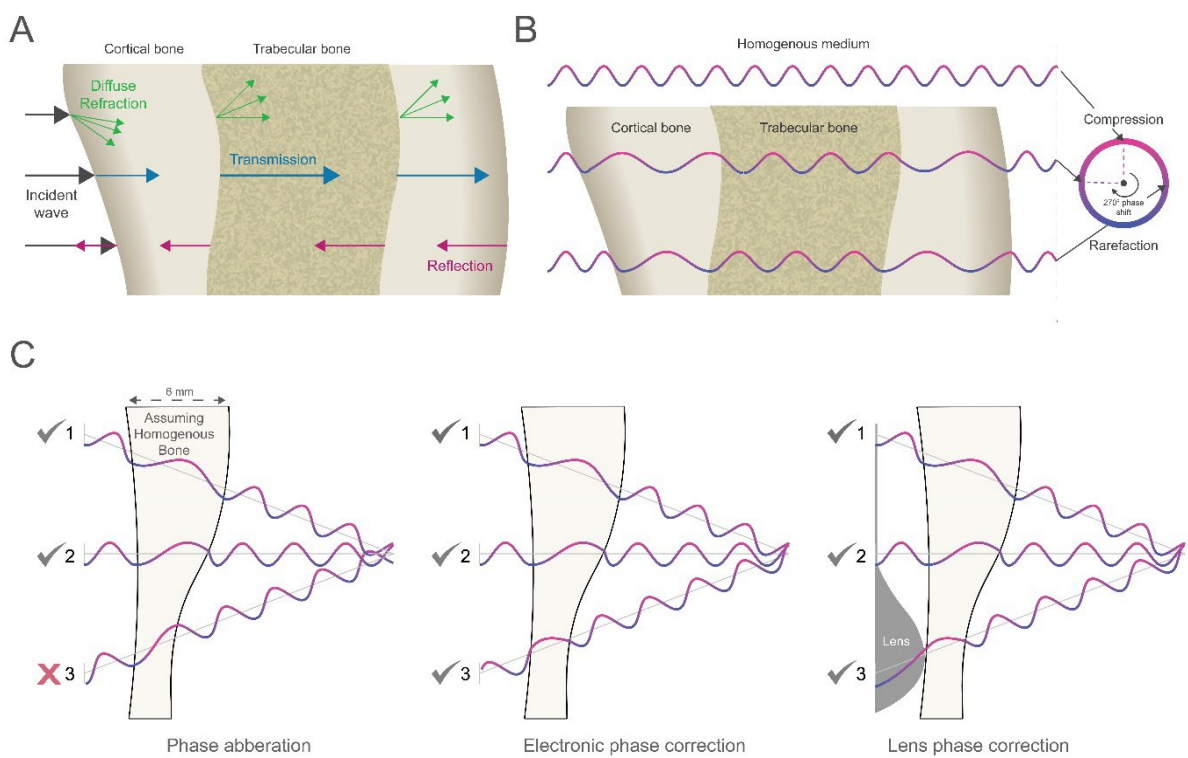

**Figure 17. Bone aberration and correction of ultrasound phase. A**) Various modes of travel for waveforms passing into a bone interface **B**) Phase shifting occurs between a wave with homogenous speed and one with dynamic speed across a bone interface. Note each wave begins with the same phase and ends with a different phase despite equivalent distance traveled. **C**) Bone aberration or shifting of phase results in out of phase wave fronts at an intended focus; the red **X** indicates the wavefront which is out of phase. Electronic phase correction is achieved by changing the timing or phase of the wave source emissions to ensure constructive phase interference at the focus. Similarly, an acoustic lens can correct the phase during wave travel.

After defining the propagation medium, a description of the acoustic source is introduced. This includes information defining the geometry of the source, including the size, shape, and position of all transducer elements, as well as the operating conditions or source signal which includes operating frequency, and amplitude and phase of the signals. To obtain realistic estimates of the field generated by a given source, it is important to define the source in simulation with a description which accurately recreates the measured field of the real device. The use of the nominal parameters of the transducer with an idealized transducer model can result in significant differences from the field generated by the actual device. Real transducers often differ from ideal transducers (e.g., a simple uniformly oscillating piston source) as their construction can give rise to surface waves and mechanical and electrical cross talk between elements, all of which contribute to the generated field (Martin et al., 2021, 2016).

In addition to careful definition of the acoustic source and propagation medium, further careful consideration of the parameters used to define and run the model is needed to achieve accurate results. Numerical models are defined in a given spatiotemporal coordinate system *(x,y,z,t)*. As a general rule, the grid size must be less than half a wavelength to satisfy the Nyquist frequency, however numerical solvers require finer sampling in order to minimize numerical errors. Spatial sampling may need to be increased where there are large absolute changes in the acoustic properties of the medium. For pseudo-spectral methods, a minimum of three points per wavelength (PPW) are required to minimize errors in simulations in water. However, when boundaries between soft tissue and bone are modelled, 6-9 PPW are required to correctly represent wave behavior at the boundary (Robertson et al., 2017); the shortest estimated wavelength within the simulation space should be considered for this calculation. Temporal sampling should be sufficiently granular that a wave peak does not travel over too large of a distance compared to the spatial step size in a single time step; this can result in simulation instability and unrealistic operation. The step size can be usefully described by the Courant-Friedrichs Lewy (CFL) number, which is the ratio of the distance the wave travels in one time step to the spatial step size: $c_0 \Delta t / \Delta x$. For weakly homogenous media (e.g., soft tissues), a value of 0.3 can be a good balance between accuracy and computational requirements. Where there are bone and soft tissues, a smaller CFL number will likely be required for accurate and stable simulations. The required spatial and temporal sampling varies with governing equations and numerical methods, so it is important to demonstrate convergence with model inputs before running simulations, for example, by repeating simulations with increased spatial sampling until there is no change in model output (e.g., focal pressure amplitude and position). Previous work has described discretization parameters used across a range of simulation software (Aubry et al., 2022) and provides an overview of ultrasound field simulation in the body.

The ultrasound pressure field *P(x,y,z,t)* emitted by the modeled transducer and propagated in the modeled tissue structures can then be estimated. There are different approaches using different governing equations or solutions to the wave equations. The approach chosen should capture the relevant physics. In the case of TUS, models should be able to simulate propagation through heterogeneous absorbing media. The spatial distribution of acoustic intensity *I(x,y,z,t)* can be calculated simply under the plane wave assumption from the pressure distribution and tissue impedance. In order to simulate any **thermal effects directly induced** by TUS, the **thermal energy** deposited by ultrasound in the tissue, **or the volume rate of heat deposition**, *Q(x,y,z,t)*, can be calculated from the pressure field or acoustic intensity and the absorption coefficient in the tissue of interest. The Bio Heat Transfer Equation (BHTE) is then used to estimate the spatial/temporal evolution of the **target tissue temperature**, *T(x,y,z,t)*, induced by repetitive pulsed ultrasound in the perfused tissue. The evolution of tissue temperature over time is directly related to the thermal diffusion in space, the effect of tissue blood perfusion, and the contribution of ultrasound-induced heat deposition. Modeling blood perfusion maintained at a constant body temperature allows to estimate the cooling effect of tissues during a pulsed US protocol, including (i) a slowing down of ultrasound-induced temperature increases during active ultrasound pulses and (ii) an acceleration of temperature decreases between active ultrasound pulses. Moreover, the **mechanical effects directly induced** by the ultrasound field on the tissues can be evaluated by calculating the acoustic radiation force field, *$F(x,y,z,t)$* from either the pressure and particle velocity or from the acoustic intensity. Also, **mechanical effects indirectly induced** by the ultrasound field may be explored by considering the probability of cavitation microbubble creation, e.g., when reaching a certain level of negative pressure in the brain tissue. **Supplementary Table S4**. provides a list of available software packages for numerical simulations of acoustic fields. It is worth noting that different acoustic simulation models with identical inputs agree well when model parameters are correctly optimized (Aubry et al., 2022). In general, researchers should choose a tool suitable for modelling the physics of TUS fields which is compatible with their access to computational resources.

### 2.9.    Neuronavigation

Due to the heterogeneity of individual skull and brain morphology, accurate and precise targeting of specific brain structures is not easily accomplished without neuronavigation. Previous work has used localization based on MEPs or visually-evoked phosphenes, or other, scalp-based markers such as EEG electrode positions, to guide transducer placement on the scalp. However, such approaches do not meet the high requirements for precisely steering the small focus of the sonication beam, necessitating some form of MR-informed neuronavigation. The accurate and precise identification of brain structures is particularly important in TUS. Due to the narrow focus of the sonication beam, even slight deviations of the transducer positioning at the scalp level can cause pronounced targeting errors, an effect that is exacerbated for deep target regions of interest. For example, at an 8 cm depth, a small deviation in transducer angle of 5 degrees would lead to a lateral deviation of 0.7 cm. Therefore, targeting deeper brain structures inherently requires a higher transducer placement accuracy compared to more proximal targets, but for the accurate targeting of both MR-guided neuronavigation should be considered state-of-the-art.

*Frameless stereotactic neuronavigation* enables the targeting of specific brain structures, by co-registering an individual's head, brain scan, and neuromodulation device (e.g., TUS, TMS). In principle, the goal of neuronavigation is to target the intended brain structure accurately, track the location of the brain stimulation device relative to the brain target of interest, and to use neuronavigation to keep the device on target with real-time feedback. While there are several commercial and non-commercial devices (**Supplementary Table S5**), the approach is generally the same and can be conceptualized as a two-part process (**Figure 18**). First, the user will begin with **planning** the upcoming neuronavigation session and does not require the presence of the participant. A high-resolution T1-weighted structural MRI scan is loaded into the neuronavigation software, 3D reconstructions of scalp and brain are created, and the researcher identifies and sets anatomical landmarks in the MRI, which are later used to co-register the MRI to the actual head. Typical landmarks, which need to be accurately and reproducibly identified in both the MRI and the participants head, include the nasion, the outer canthi of the eyes, and/or the tragi of the ears. Then, the target coordinates or volume needs to be marked in the native space of the individual brain as well as the optimal transducer position which includes center coordinates, pitch and yaw, and rotation if the transducer elements are configurated asymmetrically. The optimal transducer position may be calculated and imported from acoustic simulations software. For details and a discussion of the challenges and solutions with this iterative approach, see section 3.3. The use of MNI templates instead as a surrogate for each individuals MRI is inadvisable since the high spatial precision of TUS exceeds the likely positional difference between average and individual mapping of any given target structure.

The second step involves the spatial co-registration of the participant's head to the MRI scan loaded in the navigation system. Co-registration requires several tools, including a stereo infrared camera, reflective fiducial spheres that can be optically tracked by the navigation system, a calibration block/plate with fiducials, a pointer tool with fiducials, a subject tracker with fiducials, and a transducer tracker with fiducials (**Figure 18**). The subject tracker is firmly attached to the participant's head (e.g., via adhesive tape, headband, glasses, or dental piece), and the predefined anatomical landmarks are identified with the pointer on the head surface, allowing the system to co-register virtual and actual head surface via a method of least squares. Since the matching of a few landmarks provides limited co-registration accuracy, some navigation systems provide an additional step where numerous random points are acquired from the scalp surface by the pointer tool and matched to the 3D rendered scalp surface. The accuracy of this registration will directly impact the error in estimated device positioning accuracy; which may be on the order of millimeters (Nieminen et al., 2022; Schönfeldt-Lecuona et al., 2005). The transducer tracker is attached to the transducer using tape or 3D-printed mechanical fixtures and is placed within the cranial mapped space by using the calibration block/plate or the pointer tool. As a **warning**, ultrasound transducers may contain brittle epoxy housings which can crack when dropped or removing items attached using strong adhesives. Once the fiducial is attached to the transducer, the researcher can track the real-time location of the transducer relative to the participant's head and anatomical target inside the brain. During use, the neuronavigation software provides online visual feedback regarding any offset between the intended and actual transducer position, allowing the researcher to adjust the transducer position in real time. Depending on the overall stimulation duration, common approaches include holding the transducer by hand, mounting it to a mechanical arm or boom, or positioning the transducer automatically via a robotic arm. Following the initial positioning, the transducer position can be tracked in real-time and automatically saved for each sonication. This stored tracking information allows researchers to monitor any deviation from the target to be considered during analyses.

When simultaneously combining TUS and MRI, fiducials linked to the head and transducer may also provide accurate anatomical information relative to the transducer position. Using a multi-element array may allow for single placement while steering the beam to the target without any mechanical intervention. With a fixed single focus, or in the absence of sufficient steering freedom, iterative repositioning and scanning are required. Some MR-informed, camera-based neuronavigation systems can operate either in the lab or the scanner after initial structural images have been obtained. Other systems are designed to operate purely under continuous MR-guidance. Several manufacturers of neuronavigation systems originally developed for TMS coil navigation have adapted their systems to accommodate TUS transducers. This adaptation requires the implementation of respective virtual transducer models and the modification of trackers and calibration procedures to suit ultrasound transducer geometries.

Importantly, neuronavigation only informs the user about the transducer position, but does not specify the true distribution of the sonication beam, target exposure, or neural target engagement. Target exposure refers to the physical distribution of the acoustic field within the target structure, and its quantification requires simulated (section 2.8), or empirically measured (for MR-ARFI, see section 3.4.1) acoustic fields and their physical effects (for MR-thermometry, see section 3.4.2). Neural target engagement refers to whether the neural activity in that target structure is actually modulated by the biomechanical effects of TUS. The extent of modulation may depend on the intensity, temporal pattern of the stimulation (section 1.4), and the stimulation threshold or activity state of the target structure (cf. section 2.2 and **Figure 6**). Thus, neuronavigation is necessary but not sufficient to accurately target a brain structure with TUS. However, it is likely that future neuronavigation software will incorporate real-time estimations of target exposure from TUS (Park et al., 2023) as this is already occurring in TMS electric field estimates in real-time in multiple neuronavigation software packages.

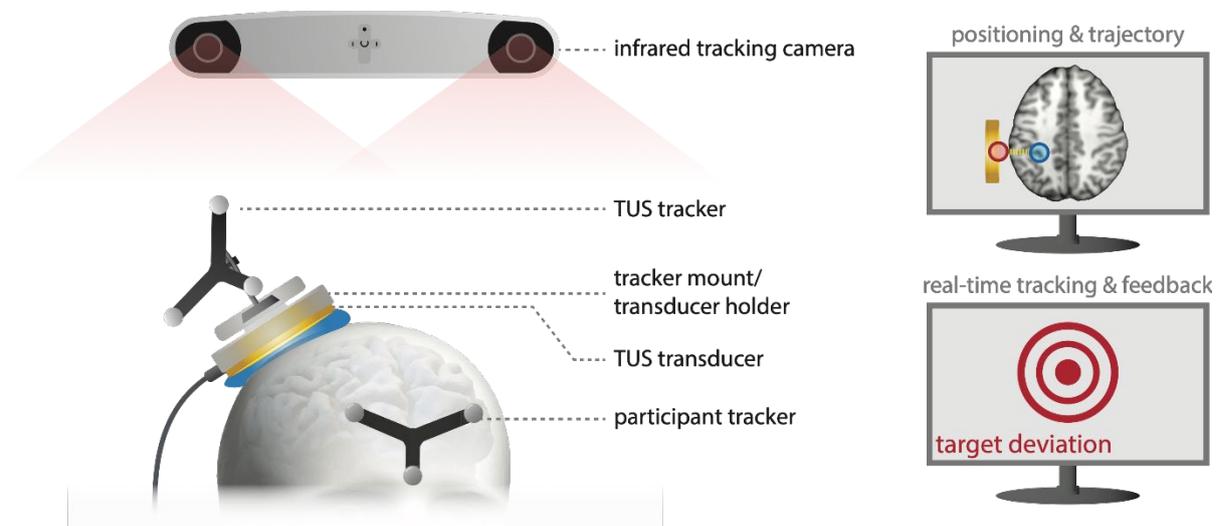

**Figure 18. Schematic of neuronavigation.** An infrared tracking camera captures the position of trackers with reflective spheres. The TUS tracker fiducials are attached to the transducer with a tracker mount. The participant tracker fiducials are typically attached to the participant's forehead. Neuronavigation software then facilitates TUS positioning and trajectory, and provides real-time tracking, feedback, and storage of position during sonication.

## 3. Procedures & Experimental Design

In this section, we will provide advice on how to design and conduct a valid TUS experiment, including the selection of the general approach, the **experimental design** to **control confounding factors**, the **application planning** of spatial targeting of the ultrasound focus, pulsing pattern, and stimulation intensity, the as **empirical validation** of successful **target exposure** using MRI techniques and the proof of **neural target engagement**.

### 3.1. Online vs. offline approaches

TUS, like other NIBS techniques, can be implemented using either an "online" or "offline" approach. The "online" method involves administering stimulation while simultaneously recording the outcome measure of interest. This approach typically aims to observe immediate or short-lived effects of TUS on neuroimaging, physiological, or behavioral metrics. Here, stimulation and measurement are mostly concurrent. Conversely, the "offline" method refers to first administering TUS, and later recording the outcome measure of interest. This approach investigates the subsequent and potentially long-lasting effects of TUS, where stimulation and measurement are consecutive.

*Offline experimental designs*" will capture **after-effects**, possibly due to "**early-phase plasticity effects**", emerging during and persisting beyond the period of stimulation. These effects, which often involve changes in neuronal excitability, can also be quantified by a change in behavioral outcome or a measure of neuronal activity after the end of the stimulation protocol. While it is unclear whether these after-effects are due to synaptic plasticity or changes in functional brain networks, there is some evidence in animal and human works suggesting plasticity through NMDA receptor regulation (Mesik et al., 2024; Shamli Oghli et al., 2023). Offline effects are typically compared to both a baseline period before the protocol and a control condition. The choice of online vs. offline TUS protocols for a planned TUS experiment thus primarily depends on the experimental question at hand and the researcher's neurophysiological assumptions but may also be constrained by certain technical aspects of the combination of TUS with the outcome measures of choice. Note that repeated measures of online effects may possibly be affected by a concurrent build-up of offline effects with repeated stimulation. Shortening the inter-pulse or treatment interval may increase this buildup. Furthermore, higher order aspects of the experimental timeline, such as the trial structure, the inter-stimulus interval or trial duration (absolute duration as well as jitter), number of trials, blocked vs. intermingled application of TUS conditions, as well as the occurrence and duration of breaks, will also affect both the expression of online and the emergence of offline effects. Importantly, there is still extensive TUS parameter space left to be explored. While longer repetitive protocols may be associated with longer-lasting effects, researchers should carefully consider the potential for online and/or offline effects of a given protocol regardless of the intended study design.

### 3.2. Confounding factors, countermeasures, and control conditions

Substantiated causal inferences based on TUS-induced modulation of neural activity and/or behavior critically depends on the exclusion of confounding factors. These factors include electrical and mechanical interference between TUS and the outcome variable, experimenter biases, non-neuronal effects, peripheral co-stimulation, and more. To allow valid conclusions and facilitate reproducibility, these confounds must be eliminated or experimentally controlled to distinguish direct neuromodulatory effects from indirect confounding effects. Similar to alternative NIBS techniques (Bergmann and Hartwigsen, 2020; Siebner et al., 2022), TUS is often associated with unintended peripheral stimulation including auditory and/or somatosensory confounds at the site of stimulation (**Figure 19A**)(Johnstone et al., 2021).

Although TUS operates outside the human audible frequency range, its administration at certain pulse repetition frequencies (PRF), such as 1-1000 Hz, which can create noticeable sounds. These sounds are often described as a 'beep', 'clicking', or 'tapping' sound and arise from complex harmonics due to steep gradients in commonly used rectangular pulse envelopes (i.e., for each pulse, the amplitude envelope of the ultrasound wave typically jumps immediately to full intensity at the beginning and back to zero at the end without any ramping). These sounds are primarily transmitted to the inner ear via bone conduction and, to a lesser extent, via airborne sound waves. In addition, continuous ultrasonic stimulation may also be perceived as the highest audible frequency for a given participant (Gavrilov and Tsirulnikov, 2012). Such audible components of pulsed TUS are known to produce confounding activation of the auditory cortex in rodents (Airan and Butts Pauly, 2018; Guo et al., 2018; Sato et al., 2018) and can produce auditory event related potentials in humans. However, this sound may be attenuated by **smoothing**, or ramping up and down, of the TUS pulse amplitude (Mohammadjavadi et al., 2019) (**Figure 19B**). Notably, a sufficient ramp duration is required for effective minimization of the auditory confound, rendering this approach less effective for protocols with very short pulse durations, such as the commonly applied 1000 Hz PRF protocol with a pulse duration of 0.1-0.3 ms. In parallel to attenuating auditory confound salience, it's impact can be controlled through concurrent administration of an **auditory masking stimulus**, e.g., via in-ear or bone conduction headphones (Braun et al., 2020; Kop et al., 2024). While masking has been shown to effectively reduce TUS detection rates (Braun et al., 2020); Kop et al., 2024), it remains possible that audible differences between experimental conditions persist (Kop et al., 2024). Insufficient auditory masking may arise from complex harmonics of the modulation frequency and transducer location and skull morphology contributing to perceived location. It is even possible that the participant perceives the sound source as opposite the transducer. Recent evidence showed that use of a multitone random mask, or Auditory Mondrian, consisting of multiple overlapping mini-pulsed-sine tones with random shifts may better mask the complex sounds heard during TUS application (Liang et al., 2023). Perhaps counterintuitively, blocking the ear canal with in-ear headphones or ear plugs for passive attenuation of airborne sound will increase the perceived loudness of bone conducted sound, and masking sound might benefit from application via bone conduction headphones. Notably, masking attempts for unsmoothed pulsed protocols (e.g., at 1 kHz PRF) have proven difficult and may not reliably work in all participants (Johnstone et al., 2021) due to the complex harmonics as well as the complex perceived location of the sound source. It is therefore good practice to systematically assess and report the blinding efficacy of administered masking stimuli. Further, researchers might consider individually titrating auditory mask parameters to optimize masking efficacy per participant, though care must be taken to avoid unblinding participants to the auditory confound(s) during this process.

TUS administration has also occasionally been reported to produce **tactile somatosensation** at the stimulation site, which may be related to transducer vibration or direct peripheral neuromodulation, but this has not yet been systematically investigated. It is, however, well-known that TUS can stimulate peripheral nerves and touch receptors (Riis and Kubanek, 2022), eliciting a range of sensations from innocuously tactile to aversively nociceptive. While sufficient experimental control of this confound may be achieved otherwise, it has yet to be determined whether topical anesthesia (e.g., lidocaine) can attenuate some of these effects as successfully used in tES (van Boekholdt et al., 2021). It is further unclear whether TUS may also produce peripheral modulation of visual or vestibular pathways.

Furthermore, as TUS can produce considerable thermal effects, a *heating of the skin* by even less than 2°C may be noticeable by the participant (Kim et al., 2017). Depending on the experimental design and sonication protocol, participants may not be able to attribute the sensation of a warm skin to single sonication trials, as the thermal effects need time to accumulate. It may, however, unblind the participant to whether or not sonication was applied (or whether different TUS protocols with different thermal effects were applied) if varied across runs or sessions. Active cooling of the coupling media (section 2.3.3) may offer a possibility to counteract such effects to some degree, but no systematic investigation of this aspect has been published so far.

Beside attempts to remove/attenuate peripheral co-stimulation confounds (which are at least less pronounced than in electromagnetic NIBS techniques such as TMS or tES), appropriate control conditions that mimic such confounding factors without producing the same neural effects in the brain should be introduced (**Figure 19C**). A common procedure (well-established for TMS or tES) is so-called **Sham** stimulation, where in this case the transducer is present on the scalp but not effectively stimulating the brain. While sometimes the mere presence of the transducer is considered a sufficient Sham setup, typically the sonication is applied to mimic at least the airborne (but not the bone conducted) aspects of the acoustic confound. With this method, it is recommended that the transducer would be coupled and then removed from the head to replicate the procedural sensations of coupling, while still nullifying ultrasound transmission. Note that still some coupling medium and absorbing materials need to remain on the transducer surface when detaching it from the skin to avoid transducer damage from hard reflections at the transducer-air interface (section 2.3). Making a Sham sonication ineffective has been achieved by different approaches, such as flipping the transducer by 90° or 180° relative to the scalp surface(Fomenko et al., 2020; Legon et al., 2018a, 2014), decoupling it from the scalp(Liu et al., 2021; Yu et al., 2021), or shielding the brain from the sonication by the (double-blind) use of sound transmitting coupling pads vs. sound absorbing sham pads (Schafer and Schafer, 2022). Together with *auditory masking*, sham stimulation may partially control for auditory co-stimulation (more easily for the airborne than the bone conducted part), but rather not for somatosensory or other (unknown) peripheral confounds. In any case, these sham approaches only allow to vary active and sham stimulation conditions across runs or sessions but not trials, as they require the physical manipulation of the transducer position/orientation.

Alternatively, and considered a gold-standard for other NIBS techniques (Bergmann and Hartwigsen, 2020) an **active or inactive control site** can be chosen (**Figure 19C**), by applying an active sonication protocol to another brain region or e.g., a ventricle, producing different ("active control") or no ("inactive control") neural effects, respectively. Assuming (or better empirically verifying) comparable levels of peripheral-co stimulation confounds for the different transducer positions and/or brain targets, this approach represents a much stronger control than sham stimulation. Depending on the steering possibilities of the transducer (section 2.2), electronic refocusing may even allow to interleave experimental and control targets in a trial-by-trial manner; otherwise, the physical transducer position needs to be varied across runs or sessions. When using phased arrays with many elements, it is further possible to defocus the beam to distribute the intracranial intensity across a larger area, reducing it to ineffective levels. This has been explicitly performed by shuffling phases used for target focusing (Fan et al., 2024), or defocusing can also be achieved create plane waves through targeting imaginary far away targets, effectively matching the phase of all elements (Riis et al., 2023), or by randomizing the phases applied to the elements (Martin et. al. 2024). Actively designing phases to optimally reduce local foci across the brain may also be a control strategy. It might also be possible to modify crucial stimulation parameters to produce ineffective control protocols without neuromodulation capabilities while preserving peripheral confounds. Strongly reduced duty cycles (e.g., 10-fold less than active condition) may reduce or alter neuromodulatory effectiveness while recreating parts of the auditory/somatosensory confound, but systematic investigations of this method are still lacking. However, this approach may also eliminate associated skull and near field tissue heating and does not control for peripheral heat sensation which could result in experimental confound. It is also unknown how low the duty cycle must be to eliminate biological effects. In addition, this may result in differential heat produced by the transducer, which should be carefully monitored. While there are few reports of active control conditions in humans (Butler et al., 2022; Kop et al., 2024; Legon et al., 2018b; Zeng et al., 2022) and non-human primates (Folloni et al., 2019; Verhagen et al., 2019), no evaluation of inactive controls or ineffective protocols has been published so far.

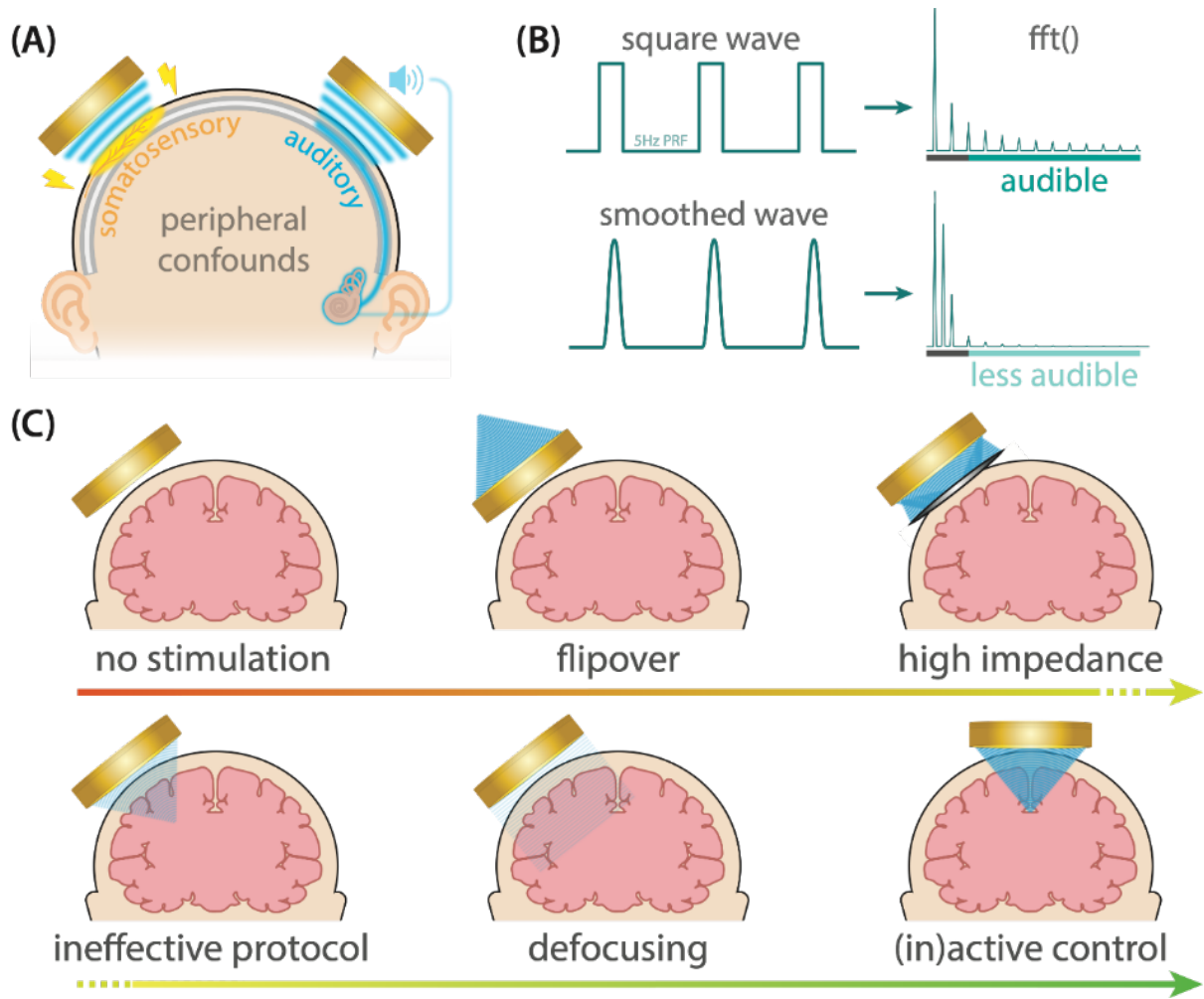

**Figure 19. Peripheral confounds and their control.** (**A**) TUS can be associated with a somatosensory confound likely caused by peripheral stimulation of the scalp, and an auditory confound that is primarily transmitted via bone conduction. (**B**) The main source of the auditory confound is the pulse repetition frequency. With a square modulation wave, the sharp edges evoke complex harmonics such that even for 5 Hz (below hearing range) there can be sufficient power in audible harmonics for TUS to be heard. If the modulation wave is smoothed over a sufficient duration, the auditory confound can be minimized. (**C**) Different methods of sham and control conditions, in order of likelihood of effective blinding and inferential power (arrow with color gradient from red to green). Importantly, flipover and high impedance methods may damage transducers from excessive heating resulting from high levels of reflection.

### 3.3. Application planning

The planning of TUS application for a given experiment and participant is a highly complex problem with many interdependent steps, none of which can be considered in isolation. However, each step will be discussed separately to emphasize their conceptual distinction. When planning an experiment, considering all aspects in combination, potentially in an iterative process, may help achieve optimal target exposure and neuromodulatory outcomes. The key decisions that need to be made during application planning pertain to (i) the spatial targeting of the sonication beam in the brain, (ii) the stimulation intensity both inside and outside the target brain structure, and (iii) the temporal pulsing pattern of the sonication protocol. All of these decisions can and should be accompanied by respective numerical simulations, modelling the acoustic pressure field distribution (for targeting) and focal intensities (for intensity selection) as well as the associated thermal effects (when choosing the temporal pulsing pattern). After the planning phase, the transducer has to be physically navigated into the correct position with the help of MR-informed neuronavigation, and ideally both target exposure and neural target engagement will be evaluated. **Figure 20** provides an overview of the respective workflow of experimental planning and execution.

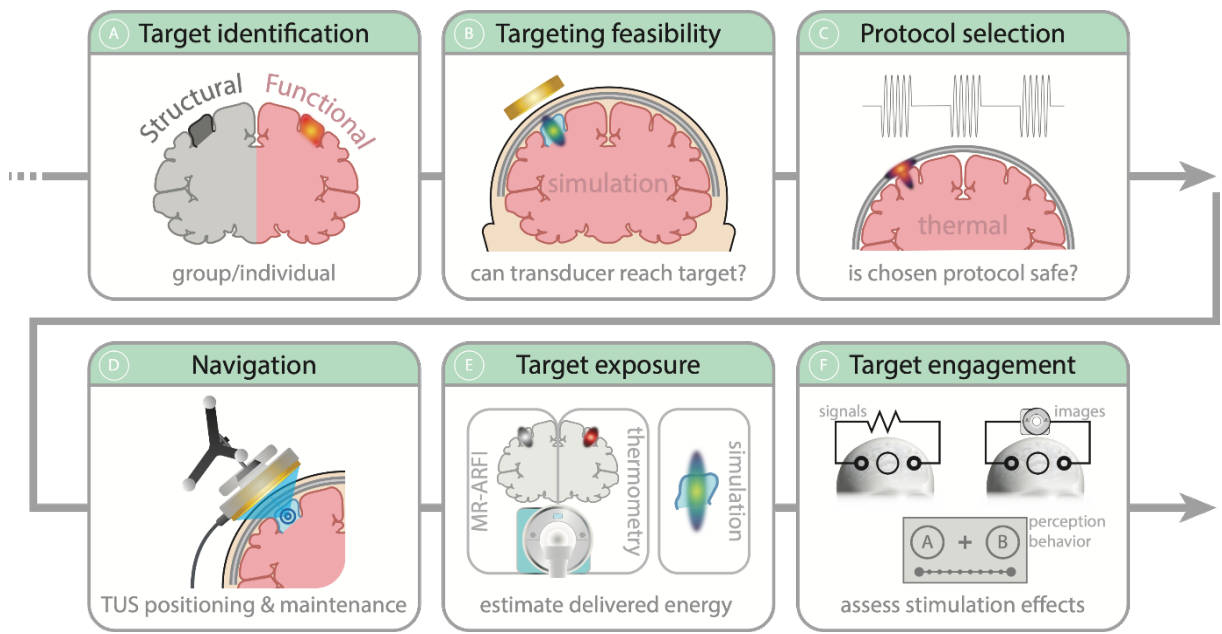

Figure 20. Example workflow of experimental planning and execution. **A**) Identification of the target region on group and/or individual level, typically based on structural (e.g., landmarks, automatic parcellation, coordinates in standard space) or functional (e.g., fMRI localizer, resting state) data. **B**) Assessment of the transducer's ability to reach the target, typically via acoustic simulations (supported by hydrophone measurements). **C**) Selection of stimulation parameters for neuromodulation, including an assessment of both biomechanical (e.g., peak pressure, $I_{sppa}$, MI) and thermal safety (e.g., $I_{spta}$, TIC, simulated thermal rise). **D**) Individual level neuronavigation for precise transducer placement and maintenance of position throughout an experiment. **E**) Verification of target exposure through MR-ARFI and potentially through MR-thermometry, or estimation of target exposure through post-hoc acoustic simulations. **F**) Evidence of target engagement via TUS effects on the outcome measure (e.g., electrical activity, BOLD response, perceptual effects, and/or behavior).

### 3.3.1. Spatial Targeting

Targeting of the sonication beam onto the anatomical structure of interest comprises multiple procedural steps that are often conflated or used interchangeably but are important to be conceptually distinguished. All of the following steps have to be conducted, although not necessarily in that order and likely in an iterative manner: (1) **Anatomical Target Definition**, (2) **Individual Target Identification**, (3) **Transducer Design/Selection, Positioning, and Acoustic Simulations**, and (4) **Physical Transducer Navigation**. For most aspects, solutions with different levels of accuracy are available, each associated with different technical requirements and demands for the investigator. As rule of thumb, the smaller the target structure and the sonication focus, the more accurate the targeting procedure needs to be.

### 3.3.1.1. Anatomical Target Definition

Prior to TUS targeting, the anatomical target of interest needs to be well characterized. While this is true for any brain stimulation technique, this is particularly important for TUS, given the superior spatial specificity of focused ultrasound. Exploiting a key advantage of TUS, many researchers may not only want to target larger superficial or deep cortical structures but also small subcortical nuclei. It is therefore of critical importance to carefully evaluate not only the location but also the size and shape of the target brain region to ensure the selected transducer can produce a sonication focus with the desired spatial exposure of the target (**Figure 21**), while at the same time avoiding or reducing the off-target stimulation of neighboring brain structures. Since not all relevant brain targets are sufficiently defined based on macroscopic brain structure alone (as available in e.g., MR-based template brain atlases and structural connectivity maps), functional characterization based on individual or group-level task-related BOLD fMRI contrasts may also be required.

### 3.3.1.2. Individual Target identification

Having established the principal anatomical target, its position needs to be identified for each individual subject brain. This typically requires at least a high-resolution T1-weighted structural MR image but may also require additional MR-sequences facilitating the identification of specific brain structures through methods such as tractography. The target may now be found by visual identification of anatomical landmarks and labeling on the individual's brain. As mentioned earlier, segmentation can be performed using labeling tools available in ITK-SNAP or similar software. Alternatively, a brain atlas containing standard brain space (e.g., MNI) can be transformed onto the native brain space of the individual subject with the help of segmentation and de-normalization procedures built into standard brain imaging software packages (e.g., ANTs, SPM, or FSL). In addition, depending on the target specifics and given the considerable inter-individual variability of some structure-function relationships, fMRI-based activation or functional connectivity maps or TMS-based (and E-field guided) mapping procedures (aka *functional localizers*) may also be of assistance.

### 3.3.1.3. Transducer Selection, Positioning, and Acoustic Simulations

As discussed in section 2.2, different transducers will generate sonication beams with foci of different shape and size and depth, dependent on the transducer type, its fundamental frequency, and electronic steering capabilities. Therefore, the transducer needs to be selected carefully to be capable of producing the desired sonication focus on the target structure. Note that for small targets the elongated focus of a single transducer is typically several centimeters and may extend considerably beyond the target structure causing undesired off-target stimulation effects (**Figure 21C, middle**). On the other hand, the comparably small focal width is typically several millimeters, and may only stimulate a small portion of a functionally or anatomically defined target. Also, the focus length may change with focal depth when using electronic axial steering, and the skull bone may introduce further distortions of its shape (see below and **Figure 21C top**).

A brain target can be reached from several locations on the head, and with different types of transducer configurations. In a first simple and inaccurate approach, bone-related distortions may be neglected altogether. In this case, a centered orthonormal projection can be made from the transducer exit plane to the target brain region. The intersection of the focus with the target will approximate the required focal distance and the length required to reach and cover the brain area. This method relies only on the spatial beam characteristics derived from calculations or, ideally, water-based hydrophone measurements. Together this method will allow selection of an appropriate transducer type and principal position during early study planning. Researchers should consider avoiding sinuses which will cause near total reflection and be aware that unintended co-stimulation of brain stem structures may theoretically modulate heart rate, breathing, and other vital functions. Note, that in the study sample there may be large inter-individual variation in head and brain size and shape as well as the precise position, shape, and orientation of the target structure, and that the selected transducer must be able to accommodate a certain range of focal distances, e.g., by adding additional physical distance through use of coupling cones for single element transducers or using electronic steering with phased arrays containing multiple elements. Differences in head shape may also require individual adaptations of transducer position and angulation.

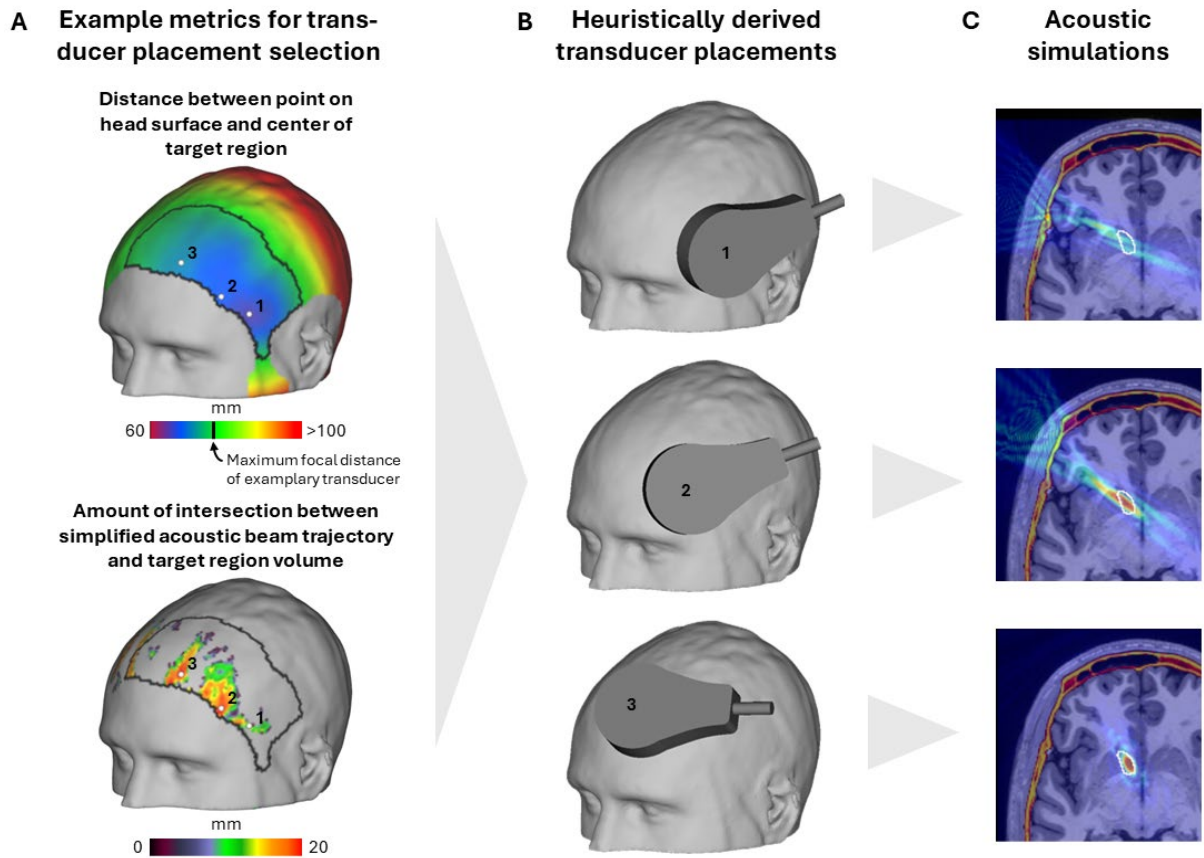

**Figure 21**. **Heuristic planning of ultrasound transducer placement. A**) Color coding of the distance between a reconstructed cranial surface to an intended target. **B**) Synthetic placement of the transducer at 3 distinct locations across the cranial surface for subsequent acoustic simulation in **C**. The simulation results show the amount of field overlap with the intended structure. Position 3 represents the optimal choice where focal intensity is high within the structure and low in surrounding tissue.

Having selected a transducer capable of producing the desired focus and an approximate transducer position, the exact position on the scalp then needs to be determined and adjusted for each individual. This adjustment can be multifactorial and, since there is no existing algorithm for selecting ideal position, consultation with neuroanatomical or acoustic simulation experts may be required. Local bone thickness and surface curvature or smoothness should be considered, as highly distorted or curved bone can lead to distorted and/or off target focusing (**Figure 21C, top**), and thick bone may lead to excessive ultrasound attenuation and skull heating. Similarly, excessive distance to the target may amplify angulation errors in the focus caused by acoustic refraction. The location of undesired off-target structures and whether it has a high chance of falling within the sonication focus should also be considered. While it is ideal for the transducer to fall orthogonal to the skull entry plane to minimize reflections, the necessary tilt of the transducer and how to achieve such tilt is an important practical consideration. While hair coverage should not deter location selection, the absence of hair facilitates easy experimental setup, whereas thick hair not only creates challenges for transducer coupling but also increases transducer-target distance. The presence of air under the transducer, e.g., when beam trajectories pass through the forehead sinus, ear canals, and jaw/mouth, should be explicitly avoided, and trajectories through the cheekbone may pass through air cavities and should be assessed carefully. These numerous considerations are non-exhaustive and not easily accounted for without the help of MR-guidance along with acoustic simulation and navigation software, which allow researchers to simulate and compare focal overlay of multiple scenarios. **Figure 22** illustrates the flow of a possible decision pipeline to determine the optimal transducer position. The workflow begins with general subject-agnostic planning, followed by patient-specific strategies of targeting.

Lückel et al. have recently developed a tool for individualized planning of TUS transducer placement (Lückel et al., 2024) (**Figure 21**). This tool leverages 3D reconstructions of an individual's head and skull, derived from structural MRI scans, to compute and visualize several metrics that are highly dependent on individual head and brain anatomy. These metrics include, among others, the distance between each point on the head surface and the (center of) the target brain region, which helps to identify restricted areas on the scalp where the target region lies within the limited focal distance of the transducer at hand (**Figure 21A, top**). Another useful metric is the extent of intersection between the target region and an idealized ultrasound beam trajectory (i.e. a straight line) perpendicular to each point on the head surface (**Figure 21A, bottom**). The tool automatically identifies "avoidance areas" on the head surface where transducers can or must not be placed such as sinuses, and around the ears (grey areas **Figure 21A, top**). Potential transducer positions can then be interactively and intuitively selected and exported for further use with acoustic simulation software (for validation of the heuristically derived transducer placements) as well as neuronavigation software (for precise device placement according to the planned position). This tool is designed to be used prior to running acoustic simulations, to easily identify the most promising transducer positions for a given individual, target brain region, and transducer. Importantly, however, it *does not replace* proper acoustic simulations (since aspects like aberrations of the ultrasound beam by the skull bone or temperature rise in bone and brain tissue are not accounted for). Similarly, Atkinson-Clement et al have developed an open-source software which takes a scalp and target mask file and minimizes the volume between the transducer and the scalp to account for partial offsets due to curvature in the skull (**Figure 11B**) (Atkinson-Clement and Kaiser, 2024). The method aims to enhance targeting precision and decrease the risk of off-target effects.

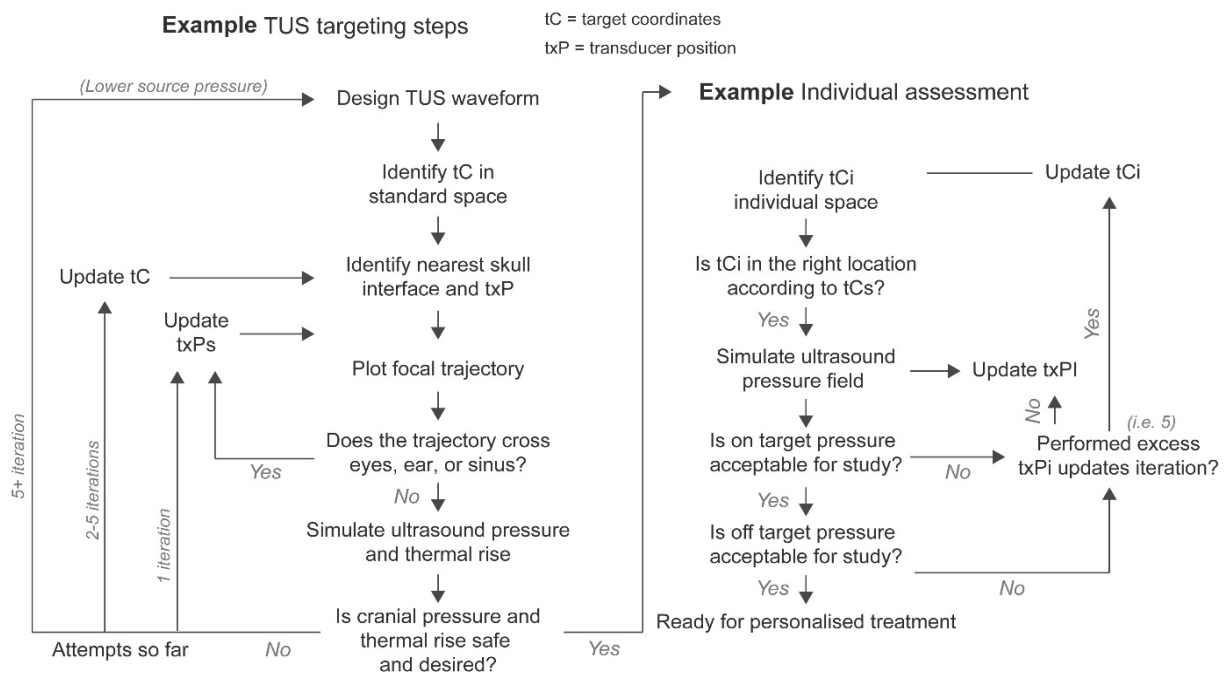

**Figure 22**. **Example decision strategy for positioning a TUS transducer**. Abbreviations: tC = traget coordinates, tC/s = target coordinate/s, txP = transducer position, tCi= target coordinates in individual space, txPi = transducer position in individual space

### 3.3.1.4. Physical Transducer Navigation

Having determined the optimal transducer position with the help of virtual scenarios in planning software, the transducer still needs to be physically moved to and maintained at the desired location and angulation. As described in section 2.9, there are different methods to precisely navigate the transducer, either by means of an MR-based frameless stereotactic neuronavigation system, or via online MR-guidance for MR-compatible transducers equipped with MR-fiducials. Accurate placement of all navigation trackers and exact co-registration of the transducer-tracker to the actual transducer as well as of the individual MRI and the subject's head surface are key to gain the high positioning accuracy (~2-3 mm) expected of state-of-the-art neuronavigation devices. Manufacturer's instructions should be followed carefully, and co-registration processes validated afterwards empirically each time by comparing the physical and virtual positions of the navigation pointer instrument for multiple sites on the scalp surface to detect co-registration errors. Small inaccuracies in the angulation of the transducer can result in comparably large localization offsets for the sonication focus. For instance, a transducer angulation error of only 3° would result in a ~3 mm lateral shift of the focus in 6 cm depth in free water, which is sufficient to move a 500 kHz beam with 3 mm focal width completely off target. In-situ, this relationship is further complicated by skull reflections and aberrations, which should be minimized by positioning the transducer surface in parallel to the skull bone (more important than the scalp) surface. In cases where transducer tilt relative to the skull bone is necessary for placing the focus on the target volume, e.g., when using a transducer without lateral steering capabilities, acoustic simulations which account for these aberrations should be used to determine the actual focal target rather than assuming the focus will behave similar to free field. It is often challenging, or even impossible to reach the planned transducer position on the head, since available acoustic simulation software can sometimes suggest physically impossible transducer positions. Although currently lacking, future developments in acoustic simulation and planning software may allow modeling of coupling media, and even the physical interaction between the transducer and compressible elements such as gel pads, hair, and skin. Thus, it is possible that a well-planned transducer position cannot be achieved in practice and a near-by alternative position must be considered.

In most experiments, targeting will be an iterative process that requires some prior knowledge and experience on what transducer positions may be practically feasible. Where this is too time-consuming, post-hoc acoustic simulations can be performed with the inherent risk that the chosen alternative position has missed the target location.

In addition to identifying the transducer position and orientation, experimenters must find effective means of maintaining the physical transducer position throughout the entire sonication session. This requires either (i) stable transducer holding devices and head fixation, (ii) means for flexible transducer repositioning to compensate head movements, or (iii) methods for firmly fixing the transducer directly to the head. It is common practice to manually hold the transducer against the subject's head, while keeping careful watch of its real-time trajectory on the neuronavigation monitor. With skilled personnel and a steady hand, this approach may provide sufficient accuracy for short periods of sonication. However, for longer applications, spanning tens of minutes to hours, either for a single application (Riis et al., 2023) or throughout an experimental session with many different conditions and TUS being applied in short trials (Lee et al., 2015; Legon et al., 2014; Nandi et al., 2023), physical support is recommended. If head fixation is not desired, robotic arms in combination with neuronavigation systems provide an alternative to automatically compensate small head movements, keeping the transducer in an exact position throughout the entire experiment. Alternatively, many researchers have invented strategies for affixing the transducer directly to the head with textile caps, adjustable helmets, rubber or Velcro straps, and 3D printed elements for transducer angulation (B. Badran et al., 2020; Park et al., 2022). Recent efforts have further designed contoured subject-specific helmets or face masks, which register the position of transducers to the head (Park et al., 2022). In any case, the actual transducer position at the time of each sonication should be saved in the navigation system. If this is not possible, this information should at least be stored before and after the experiment. With this data, respective acoustic simulations should be conducted to verify sufficient on-target stimulation during the experiment. Note that achieving the exact physical transducer position is most important for single-element transducers or annular arrays without lateral steering capabilities, whereas multi-element phased arrays may allow electronic 3D steering to compensate for positioning inaccuracies if the actual position is known, and the respective bone aberration correction parameters have been pre-calculated or can be re-calculated in real-time.

### 3.3.2. Stimulation Exposure and Dose

In the context of TUS for neuromodulation, *exposure* is defined as the amount of acoustic pressure the target brain volume is exposed to as sound waves pass through the tissue. In contrast, ***dose*** usually refers to the amount of energy deposited, which for ultrasound fields may translate into thermal or mechanical bioeffects at the target. Dose concepts for ultrasound are currently not well defined or understood, particularly for neuromodulation where the mechanisms are not yet fully established. It is not yet known whether neuromodulatory response is proportional to pressure, intensity or total deposited energy for example, or exactly which physical aspect of the ultrasound exposure (e.g., acoustic radiation force, particle displacements, pressure, etc.) interacts with the neural tissue. However, heating and radiation force, for example, are dependent on energy absorption in tissue, which varies with the absorption coefficient of the tissue, ultrasound frequency and intensity. The resulting bioeffects then likely also depend on additional factors such as neuron type, orientation of neural elements relative to the beam, and density and types of mechanosensitive receptors expressed; this is mirrored in the concept of effective dose for ionizing radiation.

Irrespective of these unknowns, aspects of both spatial distribution and temporal summation of the acoustic energy in the target tissue need to be considered separately and will be discussed in the following subsections, as they depend on the specific parameter selections and are also associated with different biomechanical and thermal safety considerations (section 3.4).

As for other NIBS techniques, stimulation exposure and dose are important parameters for both safety considerations and determining the scale and directionality of the neuromodulatory effect. Notably, exposure- or dose-response relationships are likely non-linear, and higher exposure or dose may not always yield stronger neuromodulatory effects. In any case, the intended tissue exposure at the sonication focus may be selected based on theoretical considerations or previously published work.

#### 3.3.2.1. *In-situ* exposure estimates

As mentioned throughout the guide, the acoustic pressure is the fundamental quantity of interest in an ultrasound field. Derived quantities such as intensity or acoustic radiation force, which may drive neuromodulation, can then be calculated from the pressure and tissue properties (under the plane wave assumption, or in combination with particle velocity in the more general case). Knowledge of exposure parameters is also fundamental to the assessment of thermal and mechanical safety. Mechanical index and quantities related to thermal effects (thermal index, temperature rise, or thermal dose) can be calculated or simulated from the *in-situ* pressure distribution for reporting and safety assessment.

Given the considerable attenuation of the acoustic pressure through absorption, scattering, and reflection by the skull bone, the actual pressure and intensity levels within the brain (aka *intracranial or in-situ pressure/intensity*) are significantly smaller compared to free-field, or water measurements. Free-field reference measurements are acquired for ultrasonic field characterization and then often used as the basis for assessment of exposure and safety in neuromodulation. The *ITRUSST working group for Standardized Reporting* has produced a consensus paper on the reporting of free-field acoustic parameters, pulse timing parameters, and in-situ exposure estimates (Martin et al, Brain Stim, 2024).

Improved estimates of in-situ exposure parameters can be made simply by using a generic, frequency-specific derating factor to take bone and tissue attenuation into account. Derating factors based on ex-vivo measurements of ultrasound transmission through skull fragments (Gimeno et al., 2019; Pichardo and Hynynen, 2010) have been determined, but they vary considerably across individual skulls as well as skull sites. More recently, a method has been developed for estimating a single attenuation factor for skull and soft tissue that varies as a function of transducer frequency and aperture (Attali et al., 2023). The work specifies the $95^{th}$ percentile transmission in a lookup table, which can be used conservatively estimate your own pressure transmission stays within recommended levels. This simply requires free-field measurements of the focal pressure in the acoustic field, which is then derated by the table specified value. Importantly, use of this method will almost certainly lead to an overestimation of ultrasound pressures and intensities in the brain, providing a conservative estimate in the context of safety limits. The Radboud FUS Initiative TUS calculator includes the lookup table by Attali et al. and can aid in determining these features, as well as expected thermal rise (https://www.socsci.ru.nl/fusinitiative/tuscalculator/). Similar levels of accuracy can be expected from acoustic simulation based on generic anatomical templates, while the use of personalized head models with individual CT-/MR-derived bone thickness and density information can in principle produce more accurate in-situ exposure estimates. Importantly, local, or even global, pressure maxima occurring outside the intended focus due to side lobes, grating lobes, or reflections can only be identified by simulating the acoustic field. Personalized simulations employ acoustic properties derived from either CT Hounsfield Units or Pseudo CT values which have significant variation across the literature, and even within studies. Subject-specific simulations are the gold standard and should be performed if possible. Yet, it should be born in mind that all simulation outputs have an associated uncertainty. The acoustic properties of bone and the effect of sub-image-resolution bone microstructure is a major source of uncertainty. It is therefore important to integrate uncertainty estimates in acoustic and thermal simulations, which allow users to assess the probability of unintentionally exceeding predefined safety limits.

### 3.3.2.2. Temporal Summation, Pulsing Pattern & Thermal Simulations

While ultrasound exposure may be defined by the spatial distribution of pressure, the resulting primary biomechanical effects and secondary neurophysiological processes may also depend on temporal summation Therefore, in addition to the pressure amplitude, the specific temporal pulsing pattern has to be determined, setting additional parameters (see section 1.4, **Table 1**) such as Pulse Duration (PD), Pulse Repetition Interval (PRI) or frequency (PRF), Pulse Train Duration (PTD) and Pulse Train Repetition Interval (PTRI), as well as pulse form (e.g., square or ramped pulse envelope) to produce the desired neuromodulatory effects. Although it might be assumed that the temporal accumulation of biomechanical effects may accrue intracellular consequences (e.g., $Ca^{2+}$-influx) non-linearly over time, the complex relationship between these stimulation parameters and the resulting neuromodulatory effects is largely unknown. Researchers new to the field may therefore want to start with single or repeated use of published sonication protocols (di Biase et al., 2019; Fomenko et al., 2018; Kim et al., 2021; Murphy et al., 2024; Pellow et al., 2024; Sarica et al., 2022c) that have safely and successfully induced neuromodulatory effects in humans or animals, as the exploratory design of novel protocols requires careful incremental adjustment of stimulation parameters within a safe range.

In addition to mechanical effects, the temporal accumulation of energy absorption by brain tissue is highly relevant to tissue heating. Importantly, PD, PRI, and PTD, together with the peak pressure determine the temperature rise in the sonicated tissue. Researchers may want to constrain thermal rise not only because of safety considerations (section 3.6), but also to disentangle the contribution of mechanical vs. thermal neuromodulatory effects, which may occur with even subtle temperature change (e.g. 1 °C)(Darrow et al., 2019). Importantly, thermal simulations, which can typically be conducted using the available acoustic simulation software packages (section 2.8), are required to estimate the temperature rise over time for the entire sonication protocol of the treatment/experiment (not just a single Pulse Train). While some models only consider conductive cooling in between successive sonications, blood and cerebrospinal fluid perfusion may cause additional heat dissipation. Terms for simulating this type of heat dissipation can be excluded to take obtain a conservative safety estimate of brain heating. It should also be kept in mind that the skull bone absorbs most of the energy and the largest temperature rises may not occur at the sonication focus, but in the skull bone or neighboring tissue subject to heat transfer. This is particularly true for cortical targets close to the skull bone and can even be the main limiting factor when designing sonication protocols.

### 3.4. Validation of target exposure

While acoustic simulations based on realistic personalized head models can be very accurate when based on correct and sufficiently detailed bone information, some uncertainty always remains, and empirical validation of the intended sonication may enhance the success of TUS for neuromodulation.

In principle, two MR-based techniques are available to empirically measure the sonication exposure, based on the mechanical and thermal effects it exerts on the brain tissue. Ideally, one may want to choose a technique that relies on the same effect as the intended neuromodulation.

### 3.4.1. MR-based acoustic radiation force imaging (MR-ARFI)

Brain tissue displacements in the micrometer range that are induced by the acoustic radiation force can be visualized by MR-based acoustic radiation force imaging (*MR-ARFI*). The imaging method employs special pulse sequences with motion encoding gradients coincident with a short ultrasound sonication.

The displacement of tissue due to the ARF pushes spins to an altered magnetic field, changing their precessional frequency. This change in precessional frequency can be seen as a shift in the signals phase, which can be used to build a spatial profile the sonication focus. The displacement measurements by MR-ARFI have been validated in both small and large animals, including in sheep where the measured MR-ARFI phase was related to the neuromodulatory effect (Mohammadjavadi et al., 2022) and is currently translated into human application.

### 3.4.2. MR thermometry

Absorption of the ultrasound field and the resulting temperature rise can be visualized with MR thermometry. An increase in tissue temperature causes hydrogen bonds between the water and oxygen to break. When this happens, the electrons shield the water protons more, reducing the magnetic field seen by the hydrogen protons in water. The reduced magnetic field due to the increased shielding results in a reduction in the proton precessional frequency and is the basis for proton resonance frequency shift (PRFS) thermometry. The frequency shift can be seen as a change in the phase of gradient echo images. While MR thermometry based on the proton resonant frequency shift can be very accurate in imaging changes in temperature, it cannot provide absolute temperature. Clinical thermal therapies using MR thermometry guidance assume a brain temperature of 37°C. However, for quantitation of small shifts in temperature that we might be interested in with TUS, the variability in baseline temperature between individuals becomes a bigger fractional source of error. An MR-spectroscopy method based on the shift of the water precessional frequency with respect to a reference peak, such as N-Acetylaspartate, can potentially provide absolute temperature. While accurate absolute temperature quantification may be of interest for the calculation of **thermal dose**, or measure of the cumulative thermal energy delivered to tissue over time, PRFS thermometry should work well for documenting the focal location and size, as well as documenting a temperature rise less than 2°C.

### 3.5. Evidence of neural target engagement

***Neural target* engagement**, as opposed to mere *target exposure*, refers to the intended *neuromodulatory* effect itself, such as TUS induced changes in neuronal spiking, membrane potential, synaptic or network plasticity, oscillatory synchronization, etc. caused by stimulation of the target structure or network. Direct evidence of neural target engagement requires measurement of brain activity through imaging or electrophysiological signals (e.g., fMRI, PET, EEG, intracranial electrodes, etc.). Other outcome measures such as TUS-TMS-EMG(-EEG), and precisely quantifiable perceptual or behavioral effects can provide indirect support of neural target engagement when using appropriate experimental designs. Different (non-invasive) measures of neural activity are available as readouts (dependent variables) and can be combined with online as well as offline TUS protocols to assess the respective acute and early-phase plasticity effects. A few of these measures are discussed here (see **Figure 23** for an overview).

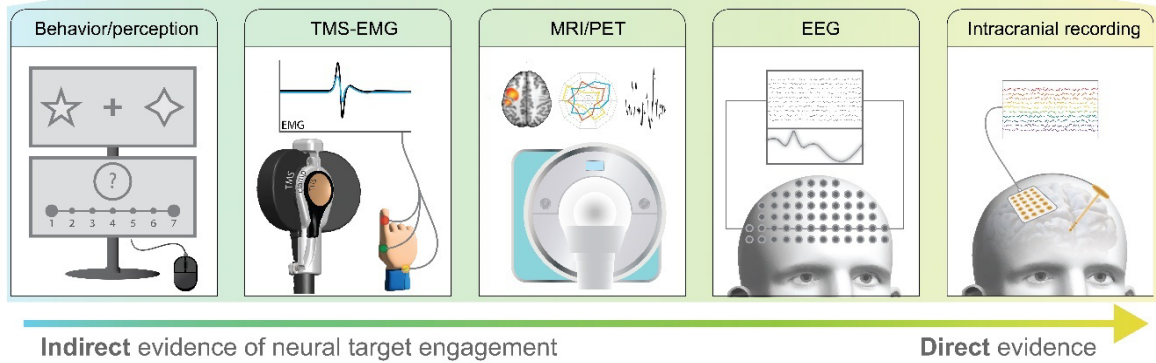

**Figure 23. Example outcome measures**. Different dependent variables can be obtained to measure the neuromodulatory effects of TUS, here show in the order (left to right) from less to more direct evidence of neural target engagement. Note that the level of directness (i.e., how many causal steps away from the neural source we measure) does not necessarily equal the accuracy or spatial resolution for assessing those effects, A well-controlled behavioral experiment with strong a-priory knowledge of the relevant functional anatomy can go a long way, while e.g., EEG measures may not provide the most precise spatial localization of effects without additional a-priory knowledge either.

### 3.5.1. Intracranial electrode recording

Brain disorders are increasingly being treated through the use of stereotactically brain implanted electrodes, often referred to as deep brain stimulation (DBS) (Lozano et al., 2019). While DBS leads vary across manufacturer and intended use, most electrodes consist of two or more platinum–iridium wires coated in a polyurethane sheathe with exposed contacts at or near the site of stimulation. Differential voltages can be placed across exposed contacts to shape electrical fields which are then pulsed at a given frequency (e.g., ~140 Hz) to alter the activity of proximal neural tissue. In some instances, these electrodes can be used to monitor the local field potential of the surrounding neurons, providing an opportunity to record the impact of neuromodulation with site specificity (Yuan et al., 2012). Unlike PET and fMRI, intracranial stereoelectroencephalography (sEEG) offers a means of directly recording neural activity rather than a proxy such as glucose uptake or blood volume. Compared to scalp EEG recordings, sEEG provides much higher spatial resolution and sensitivity. While this method may be applied in the recording of both immediate and enduring effects of TUS, this method is highly sensitive to mechanical perturbation which may lead to artifactual interpretations in signal (Collins and Mesce, 2020; Sarica et al., 2022a). For subject safety, it is important to consider the possibility of ultrasound waves being absorbed by, and heating the electrodes which could result in damage to local tissue. Several published and ongoing studies have performed TUS in combination with sEEG (Lee et al., 2022), and at least one study has directly monitored the effects of focused ultrasound fields on a commonly implanted DBS electrode (Sarica et al., 2022a). While the probe does heat considerably with a previously used protocol (1.67 C°, I$_{SPPA}$: 30 W/cm$^2$, I$_{SPTA}$: 15 W/cm$^2$, D.C. 50%) (Bobola et al., 2020; Fomenko et al., 2020), there have been no findings suggesting that this level of heating would damage tissue. Nevertheless, this work suggests the electrodes of choice should be examined with respect to temperature rise and the protocol of choice if TUS field will overlap with the electrode in the brain. Studies may also target areas largely outside of the ultrasound focus to avoid contamination by mechanical artifacts and tissue heating. On the other hand, mechanical artifacts arising from the TUS field (Sarica et al., 2022a) may be leveraged to determine the sonication focus in the brain.

### 3.5.2. Electroencephalography (EEG)

Electroencephalography (EEG) can be used for investigating both direct TUS-evoked potentials and the TUS-induced modulation of task-/sensory-evoked potentials or resting state oscillatory activity.

Limited evidence suggests that TUS can both evoke EEG potentials (Lee et al., 2016b), and modulate the amplitude and spectral content of potentials evoked by sensory stimulation (Legon et al., 2014; Nandi et al., 2023). For EEG potentials to become detectable at the human scalp, it requires the summation of synchronized local field potentials across comparably large, mainly excitatory, neuron populations with aligned dendritic trees to create a sufficiently strong dipole. Therefore, in some cases the highly focal effects of TUS may synchronize an insufficient number of neurons to result in an observable scalp EEG potential. Given the proximity of the cortex to the scalp where EEG electrodes are placed, EEG is more sensitive to cortical changes, in particular radial sources from the gyral crowns. While this suggests that TUS effects in cortical targets may be more easily detectably by EEG, the indirect cortical effects resulting from the sonication of deeper targets may still be investigated with EEG. For example, TUS of the sensory thalamus has been shown to modulate sensory evoked potentials on their way to the cortex (Legon et al., 2018a), and TUS of neuromodulatory nuclei may cause brain wide oscillatory changes. In some instances, these indirect effects may even result in larger signal changes as they involve larger neuron populations compared to the immediately affected targets. It should also be considered that TUS effects may emerge with variable delay or integration periods as suggested by TUS induced motor responses in rodents and neuronal or glial tissue culture (Lee et al., 2023; Yoo et al., 2022) which is challenging for time-locked averaging of rapid onset EEG response to TUS. In addition, when using EEG-derived outcomes, evoked potentials due to auditory and somatosensory confounds must be considered. Methods to mask, attenuate or eliminate these confounds are discussed in section 3.2. High-density electrode montages are required to attain proper source localization and check whether potentials may be related to these confounds. A major limitation for concurrent TUS-EEG is that no EEG electrodes can be mounted directly below the transducer, and caps may need to be customized with holes for the transducer. Depending on the EEG signals of interest, such montages with electrodes around the transducer may or may not be adequate to pick up the relevant sources below the transducer (while radially oriented sources may be missed, superficial tangentially oriented sources from the sulcal walls may still be detected). Also, EEG and TUS coupling gels should not touch to avoid short-circuiting EEG channels. Moreover, the PRF may be visible as an artifact in the EEG even when masking of the associated sound was successful and may thus be of technical origin. High PRFs (e.g., 1000 Hz) above the frequency range relevant for EEG analyses may be largely attenuated with a narrow low-pass filter, while lower frequency PRFs may require post-hoc removal (e.g., via ICA).

### 3.5.3. MRI and Positron Emission Tomography (PET)

A variety of MR techniques like functional MRI (fMRI) measuring the blood-oxygenation level dependent (BOLD) signal, including both task- and resting-state fMRI (rs-fMRI), arterial spin labeling (ASL) measuring blood perfusion changes, MR spectroscopy (MRS), diffusion MRI (dMRI), etc., may be used to characterize the neuromodulatory effects of TUS, and potentially the underlying metabolic and biochemical mechanisms. The temporal resolution of each MR technique must be considered in the context of the chosen TUS protocol. For instance, BOLD-fMRI and ASL (percent signal and/or volume change, respectively) could capture both acute effects when applied concurrently (Ai et al., 2018b; B. W. Badran et al., 2020; Cain et al., 2021; Martin et al., 2024b; Riis et al., 2023) and early-phase plasticity effects when applied consecutively (Badran et al., 2020; Cain et al., 2021; Yaakub et al., 2023) while MRS (Yaakub et al., 2023) and dMRI (brain microstructure change) may be restricted to the subsequent measurement of early-phase plasticity effects that outlast the stimulation. Some MR techniques may match the spatial resolution of TUS which is in the order of millimeters but others like MRS will cover a larger area. While the offline assessment of early plasticity effects of TUS with neuroimaging does mainly pose logistical constraints (Bergmann et al., 2016), the safety and reliable operation of the TUS equipment in an MR environment is paramount for concurrent TUS-MRI when imaging acute TUS effects (see section 2.6.1.). In general, it is recommended that several test pulses at the experimental intensity are applied within the scanner environment at the beginning of any experimental session. This will allow the researcher to assess subject comfort and any sensations including auditory co-stimulation, which may evoke extraneous BOLD activity, and to ensure that the transducer is stably mounted. This may be more prominent with PRFs well in the audible range which may produce a tone (i.e. 1-4 kHz). Furthermore, some considerations will be specific to each MR technique. For instance, neuronavigation may be required to ensure overlap between the TUS focus and the target region and the respective MRS voxel in addition to choosing an MRS voxel orientation that is aligned with the TUS beam. Finally, PET imaging is a powerful technique for monitoring the presence of radio-labeled molecules in a biological system. 18-fluorodeoxyglucose is commonly used to detect glucose uptake in the brain, which is directly related to neural activity. Although PET imaging has been used to validate target engagement in animal models with fairly accurate prediction of focal position, it has yet to be implemented in human TUS studies (Kim et al., 2013; Murphy et al., 2022).

### 3.5.4. TMS-EMG/-EEG

TMS-evoked MEPs or TMS-evoked EEG potentials (TEPs) can be used to measure TUS effects below the neuronal firing threshold (i.e., excitability changes as opposed to immediate excitation), which cannot be measured directly using techniques like EMG, EEG, fMRI etc. Direct TUS-evoked EMG responses have been demonstrated in small animals (Hesselink et al., 2023; King et al., 2013; Yuan et al., 2020; Zhu et al., 2023), but not in non-human primates and humans, possibly due to the relatively smaller portion of the brain exposed to TUS. Therefore, modulatory effects of TUS on motor cortical excitability that do not elicit immediate action potential firing can be probed via concurrent TMS-EMG (Fomenko et al., 2020; Legon et al., 2018b) but auditory confounds resulting from audible PRFs (such as 1000 Hz) have to be ruled out to obtain valid results (Kop et al., 2024). Though TMS itself is limited to cortical areas, TMS-evoked potentials may also be used to probe TUS effects on deeper structures, like the thalamus, which are connected to the cortex. However, even when the targets of TUS and TMS are identical, the TMS induced electric field (low focality and depth) and TUS pressure field (high focality and possibly depth) will not completely overlap and may thus interact with different neuronal populations. For instance, TMS of the primary motor cortex may have greater effects along the crown and lip region of the precentral gyrus and differentially stimulate corticospinal projections or interneurons depending on the orientation of the magnetic field (Osada et al., 2022; Siebner et al., 2022), while it is possible that TUS exerts greater effects on different (e.g., deeper) parts of the gyrus and thus different neuron populations. Further research is required to determine whether TMS-evoked MEPs depend on the same neuronal populations that are stimulated by TUS. However, since both techniques can stimulate neurons trans-synaptically, TMS-evoked MEPs may provide a readout of TUS effects even when the two techniques stimulate different neurons. Practically, when targeting the same brain area with TMS and TUS concurrently in an online paradigm, the transducer has to be placed between the scalp and TMS coil. Consequently, the increased TMS coil-to-cortex distance will influence the choice of transducer (physical dimensions and focal steering) and coupling medium. Here, the transducer and coil may be attached to each other and navigated as a single unit (cf. **Figure 23**). A limitation of this approach is that the optimal TMS site is often several millimeters away, typically anteriorly, from the target primary motor representation. Therefore, aligning TUS with the center of the TMS coil may not sufficiently expose the target region to ultrasonic stimulation. Where possible, for example in an offline paradigm where TMS-evoked MEPs are used to evaluate the offline effects of TUS in comparison of pre- and post-TUS MEPs (Zeng et al., 2022), it is preferable to navigate TUS and TMS separately to allow positioning at two different locations that optimally engage the anatomical and/or functional target. Possible electromagnetic interactions between the transducer and TMS coil, as well as methods for testing such interactions are discussed in section 2.6.2.

### 3.5.5. Behavior and Perception

Similar to other NIBS techniques, TUS may elicit or alter sensory perception and influence behavior. Similar to TMS, TUS has also been reported to induce mild or dim phosphene generation, indicating direct activation of visual cortex neurons (Gimeno et al., 2019; Lee et al., 2016b; Nandi et al., 2023), and somatosensory percepts when targeting the somatosensory cortex (Lee et al., 2016a). While awaiting replication, these initial results suggest that TUS of primary visual and somatosensory cortices may be able to elicit positive phenomena, which can be subjectively reported. Given the possibility of targeting both superficial and deep structures, TUS can be used to examine distinct contributions of different parts of a brain network, to a given behavior or perception. For instance, somatosensory cortex TUS improved tactile discrimination (Legon et al., 2014). Another example is that offline TUS to the basal ganglia, such as the subthalamic nucleus or anterior putamen, induces a sustained impaired motor response inhibition during a stop-signal task (Nakajima et al., 2022). Given the high spatial specificity, TUS will principally allow researchers to study fine-grained brain-behavior relationships, for instance, finger specificity within the primary motor and somatosensory cortices. Behavioral effects, potentially even clinically relevant ones, are surely the ultimate goal of TUS for neuromodulation. However, strong a priori knowledge is required regarding the functional anatomy that mediates the cognitive function or pathological dysfunction under investigation, to allow valid reverse inference regarding the specific neuronal target engagement, given the many elements in the chain of causation leading from neuromodulation to behavioral change (Bergmann and Hartwigsen, 2020). Nonetheless, if obtained in the context of an appropriately designed behavioral experiment with strong control conditions, a clear behavioral or perceptual change may be considered even stronger evidence for a neuromodulatory TUS effect of relevant size and ecological validity. While the relationship between stimulation related changes in neuronal activity and its behavioral or perceptual consequences is complex, TUS (similar to TMS or tES) may produce anything from a mild modulation to disruptive interference at both levels, and potentially different types of effects may be expected for online and offline approaches (Bergmann et al., 2016; Bergmann and Hartwigsen, 2020).

### 3.6. Safety considerations

### 3.6.1. Safety assessment

The ITRUSST consortium recently reached a consensus on biophysical safety in TUS (Aubry et al., 2023). Based on existing regulatory guidelines ITRUSST recommends levels that ensure minimal and nonsignificant risk for TUS. In brief, the two main bioeffects of TUS relevant for safety assessment are thermal and mechanical bioeffects. Concerning **mechanical bioeffects**, the primary consideration is the risk of acoustic cavitation caused by rapid changes in negative pressure. While cavitation can result in stable bubble oscillation (stable cavitation) which may contribute to neuromodulatory effects, it may also result in collapse (inertial cavitation) at extreme pressures. Inertial cavitation can cause very high local temperatures and mechanical forces. While these resulting forces would likely lead to changes in neural activity, they are also known to cause tissue damage and must be carefully considered. The risk of cavitation depends on peak negative pressure, ultrasound frequency, and pulse duration, and has been correlated with high mechanical index (MI). Cavitation is a threshold effect i.e., at low peak negative pressures, the risk is negligible, and only above a threshold does the risk increase with increasing dose. In diagnostic ultrasound with relatively short ultrasound pulses, cavitation can only occur at sufficient peak negative pressures and is absent at lower intracranial pressures (e.g., < 2 MPa in diagnostic ultrasound (Nightingale et al., 2015)). Only at more extreme negative pressure (e.g., > 4 MPa at 558 kHz) does the risk of acoustic cavitation become tangible (Hynynen, 1991). The current ITRUSST safety recommendations, in line with the FDA guidelines for diagnostic ultrasound (Administration, 2019), state that mechanical risks are negligible if the MI or MI$_{tc}$ does not exceed 1.9, in the absence of ultrasound contrast agents. They also highlight that it is the responsibility of the operator/manufacturer to determine which mechanical index corresponds best to the configuration: MI or MI$_{tc}$. Concerning **thermal bioeffects**, some of the mechanical energy is converted to thermal energy via absorption as ultrasonic waves pass through tissues, leading to tissue heating. ITRUSST recommends that thermal risk can be minimized to negligible levels by limiting one of three parameters: temperature rise, thermal dose, or thermal index (TI). Either of these three parameters is sufficient to constrain thermal risk, in alignment with FDA, BMUS, and AIUM guidelines for diagnostic ultrasound, with international regulations for (implantable) medical devices, and expert consensus for magnetic resonance radiofrequency exposure. First, ITRUSST considers thermal risks negligible if the thermal rise does not exceed 2°C at any time in any tissue, or the thermal dose does not exceed 0.25 CEM43, or if the exposure time does not exceed 80 min for 1.5 < TI ≤ 2.0; 40 min for 2.0 < TI ≤ 2.5; 10 min for 2.5 < TI ≤ 3.0; 160 s for 3.0 < TI ≤ 4.0; 40s for 4.0 < TI ≤ 5.0; and 10 s for 5.0 < TI ≤ 6.0. It is the responsibility of the operator/manufacturer to determine which thermal index corresponds best to the configuration: soft tissue thermal index (TIS), bone at focus thermal index (TIB), or, most likely for standard TUS configurations, the cranial thermal index (TIC). Skull bone has a higher absorption coefficient than soft tissues, and the degree of heating depends on both TUS exposure and exposure rate. Either standardized indices such as the thermal index, direct temperature measurements, or numerical simulations can be used to estimate the temperature rise for a given protocol. Note that these models need to be validated via experimental measurements on a representative set of human skulls and described in sufficient detail in publications. The reader is strongly advised to carefully study the *ITRUSST safety guidelines* for more details (Aubry et al., 2023).

### 3.6.2. Risk assessment

Prior to participation, all subjects should be screened for general contraindications to non-invasive brain stimulation. Importantly, as for human research, the risk-benefit ratio must be carefully assessed. While it may be most reasonable to simply exclude participants with any potential risk factor from basic science studies, clinical research in patient populations with a potential gain from TUS application may be less restrictive in their inclusion criteria. While an in-depth discussion of neurophysiological safety is outside the scope of this guide, we want to briefly mention a few potentially relevant risk factors. While there are no reports of seizures associated with TUS, all techniques that stimulate the brain (including sensory stimulation), come with a low theoretical possibility of inducing seizures in people who are particularly susceptible to them. Furthermore, the impact of any active or passive skull or brain implants has to be carefully assessed. On the one hand, they may impact ultrasound propagation and cause reflections and standing waves, and on the other hand, they may themselves be influenced by the ultrasound via mechanical movement or heating of e.g., stents, coils, or wires of active implants like stimulation or recording electrodes. Notably, the calcification of blood vessels may interact with ultrasound and should be considered if the focal target overlaps with severe calcifications observed in scans. Particular attention may also be required when anatomical anomalies that are sensitive to mechanical stress, such as aneurysms and arteriovenous malformations, overlap with the focal beam. Since such pathologies can be asymptomatic, they can also occur in participants expected to be healthy. While there are no empirical safety data suggesting that calcifications, aneurysms, or arteriovenous malformations actually represent a substantial risk for the low-intensity application of TUS, the prospective evaluation of pre-treatment MR images offers a possibility to screen for vascular anomalies and exclude respective participants following individual risk assessment.

### 3.6.3. Potential side effects and adverse events

As for any other medical intervention, including the use of drugs or medical devices, undesirable effects can occur following brain stimulation (Rossi et al., 2021), which are referred to as **adverse events (AE)**, **serious adverse events (SAE)**, or **side effects (SE)** depending on their severity, context, and causal relationship with the intervention. According to ISO14155:2020 (*Clinical investigation of medical devices for human subjects — Good clinical practice*), in the context of clinical investigations, AEs refer to any untoward medical occurrence, unintended disease or injury, or untoward clinical signs (including abnormal laboratory findings) in subjects, users or other persons. SAEs refer to AEs that lead to either death or a serious deterioration in health that resulted in a life-threatening illness or injury, a permanent impairment, in-patient hospitalization or prolongation of existing hospitalization, or in a medical intervention to prevent the former. Importantly, AEs and SAEs are independent of whether or not related to the investigational medical device and whether anticipated or unanticipated, whereas SEs are unintended and usually undesirable effects that that have been determined to result from the normal use of the medical device (here, when TUS is applied according to the international safety guidelines) but are secondary to a main or therapeutic effect, typically expected, and usually accepted following risk-benefit evaluation of the intervention.

For TUS, undesired auditory or localized tactile sensations and scalp heating can be clearly caused by TUS application and are thus considered side effects, whereas the causal relationship remains unclear for other mild adverse effects. Please note that participation in TUS experiments, whether receiving verum or sham stimulation, can be accompanied by side effects of participation in general, such as fatigue, muscle pain, problems with attention, and anxiety. The most common symptoms in this domain, perceived as unrelated to the intervention, included sleepiness and neck pain (Legon et al., 2020). Researchers and participants attributed these symptoms to factors external to the TUS treatment itself, such as the general effort and tediousness of participating in the experiment. Specifically, they may be caused by prolonged attentional focus, restriction of movement, or psychological/physical strain. Importantly, no SAEs have been reported during TUS neuromodulation trials in humans (Sarica et al., 2022b).

Although the number of laboratories and clinical settings administering TUS has increased rapidly in recent years, there are no standardized and agreed procedures for reporting SEs, AEs, SAEs to the international community. Given the early translational stage of this technique, it is recommended to use dedicated side-effect questionnaires to assess and monitor side effects and adverse events that occur throughout the course of a TUS study. Evaluating side effects and adverse events related to TUS is of paramount importance to provide an evaluation of TUS safety for human neuromodulation. Additionally, it will allow the comparison of symptomology to other forms of non-invasive brain stimulation or relate the frequency of adverse events to TUS variables or other forms of clinical features involved in the intervention. Standardized forms and/or neurological assessments should be used to monitor side effects and adverse events where applicable. Several types of assessments exist, focusing either on immediate side effects during or right after the TUS intervention or on long-term side effects following days, weeks, months, and years after the intervention. As TUS is in its infancy, researchers are currently focused on immediate, short-, and medium-term events while long-term assessments remain limited. Beside standard items relating to non-specific side-effects previously identified in NIBS studies (e.g., headaches, neck pain, and sleepiness), researchers may want to add study-specific items pertaining to the anatomical on- and off-target function. However, one should also consider typical responses biases, such as inflation of false positive responses. These may arise simply from the application of too many test items and a tendency of participants to exhibit compliance. This may be ameliorated by a tiered approach in which more detailed items are presented conditionally following positive responses on more general items. Moreover, a comparison to sham stimulation is always necessary to correctly interpret side-effect frequencies against their normal occurrence during the course of a NIBS study.

## 4. Conclusion

It is our hope that this extensive, but non-exhaustive practical guide to TUS for neuromodulation will enable researchers to start their TUS research, plan and conduct studies according to the highest methodological standards. However, our understanding of TUS effects is rapidly evolving and some of the methodological recommendations will adapt future insights. We suggest referring to the ITRUSST website (https://itrusst.com) for further updates of this guide following its initial publication.

**Conflict of Interest Statement**
K.R.M. hold equity in, and receives a salary from Attune Neurosciences Inc.
K.R.M. receives royalties from a patent application assigned to Stanford University.
H. R.S. has received honoraria as speaker and consultant from Lundbeck AS, Denmark, and as editor (Neuroimage Clinical) from Elsevier Publishers, Amsterdam, The Netherlands.
H.R.S. has received royalties as book editor from Springer Publishers, Stuttgart, Germany, Oxford University Press, Oxford, UK, and from Gyldendal Publishers, Copenhagen, Denmark.
All other authors have no conflict of interest relevant to this paper.

**Acknowledgements**

We thank the members of the ITRUSST consortium as well as Andrew Thomas (BrainBox Ltd., UK), Jean-Francois Aubry, Samuel Pichardo, Martin Monti, and Axel Thielscher for their feedback on the manuscript.

T.O.B. received funding supporting this work from the Boehringer Ingelheim Foundation (grant on "Methods Excellence in Neurostimulation"), the European Innovation Council (EIC Pathfinder project CITRUS, Grant Agreement No. 101071008), the Ministry of Science and Health of the State of Rhineland-Palatinate, Germany ("ACCESS" grant), the Leibniz Association (ScienceCampus "NanoBrain"), and the DFG (Grant No. 468645090).

T.O. is supported by JSPS KAKENHI (21K07255) and Brain Science Foundation.

W.A.N received funding supports from the French National Research Agency (ANR-16-TERC-0017, ANR-21-CE19-0007, ANR-21-CE19-0030), the French Laboratories of Excellence (LabEx) DevWeCan and Cortex and the American Focused Ultrasound Foundation (LabTAU, Center of Excellence of the FUSF)

H.R.S. received support by a Grand Solutions grant "Precision Brain-Circuit Therapy - Precision-BCT)" from Innovation Funds Denmark to Hartwig R. Siebner (grant nr. 9068-00025B) and a collaborative project grant "ADAptive and Precise Targeting of cortex-basal ganglia circuits in Parkinson´s Disease - ADAPT-PD" from Lundbeckfonden to Hartwig R. Siebner (grant nr. R336-2020-1035).

Y.U. received grants (KAKENHI grant numbers 23H00459) from Japan Society for the Promotion of Science and grants from the Association of Radio Industries and Business and Tokyo Metropolitan Institute of Medical Science.

E.M. is supported by UKRI Future Leaders Fellowship [grant number MR/T019166/1], and in part by the Wellcome/EPSRC Centre for Interventional and Surgical Sciences (WEISS) (203145Z/16/Z), and i part by the EIC (Pathfinder project CITRUS, Grant Agreement No. 101071008).

# Supplementary information on

# A Practical Guide to Transcranial Ultrasonic Stimulation from the IFCN-endorsed ITRUSST Consortium

**Table S1. List of commercially available TUS system (alphabetical order).** Note that this list may not be exhaustive and does not reflect any recommendation or preference of the authors or ITRUSST.

| Companies | Products | Features | Links |
|---|---|---|---|
| Blatek Industries, Inc. | Custom built transducers | | http://www.blatek.com/medical.html |
| Brainbox Ltd. & Sonic Concepts Inc. | NeruoFUS LT, NeuroFUS Pro | Frequency: 0.25, 0.5, 1.0 or 2.5 MHz, steerable multi-element array | https://brainbox-neuro.com/products/neurofus or https://neurofus.com/ |
| BrainSonix Corp. | BrainSonix, BxPulsar | Mechanically steered single element focused transducer, MRI compatible | http://www.brainsonix.com/ |
| MRInstruments | MR-guided TRUST™ | Layered 2 stack 128 steerable array, 500 kHz operating frequency | https://mrinstruments.com/focused-org/ |
| NaviFUS | NaviFUS system | Multi-element array | http://www.navifus.com/ |
| Neurosona | NS-US100 | Image-guided tFUS system | http://www.neurosona.com/technology/focused_ultrasound.php (Website in Korean) |
| Stortz Medical AG | Neurolith, | Shockwave protocols, High intensity, electromagnetic plasma based FUS generation | https://www.storzmedical.com/en/disciplines/neurology/neurolith |

**Table S2. List of commercially available transducer fabrication companies (alphabetical order).**
Note that this list may not be exhaustive and does not reflect any recommendation or preference of the authors or ITRUSST.

| Companies | Links |
|---|---|
| Daxsonics | https://www.daxsonics.com/ |
| IBMT Fraunhofer | https://www.ibmt.fraunhofer.de/de/ibmt-kernkompetenzen/ibmt-ultraschall.html |
| Imasonic | https://www.imasonic.com/en/home/ |
| Olympus | https://www.olympus-ims.com/en/ultrasonic-transducers/ |
| Sonele | https://www.sonele.com/ |
| Sonic Concepts Inc. | https://sonicconcepts.com/transducer-selection-guide/ <br> https://brainbox-neuro.com/products/neurofus <br> https://neurofus.com/ |
| Transducer Works | https://transducerworks.com/ |
| Vermon | https://www.vermon.com/index.php |

**Table S3. Hydrophone and hydrophone scanning tank manufacturers (alphabetical order).** Note that this list may not be exhaustive and does not reflect any recommendation or preference of the authors or ITRUSST.

| Manufacturer | Link |
|---|---|
| Focused Ultrasound Foundation Open Source | https://osf.io/ys7jr/ |
| Gampt Ultrasonic Solutions | https://www.gampt.de/en/product/3d-sound-field-scanner/ |
| Onda Corporation | https://www.ondacorp.com/scanning-tank/ |
| Precision Acoustics | https://www.acoustics.co.uk/product/ums3-scanning-tank/ |
| Sonic Concepts | https://sonicconcepts.com/hydrophone-transducers/#High_Intensity_Hydrophones |

**Table S4. List of software packages for simulating acoustic pressure fields and/or temperature rise (alphabetical order).** Note that this list may not be exhaustive and does not reflect any recommendation or preference of the authors or ITRUSST.

| Library | Language | Short description | Link |
|---|---|---|---|
| BabelBrain | Python | An open-source, stand-alone tool for the prediction of low-intensity transcranial ultrasound. Works in tandem with neuronavigation and visualization software to assist in the planning of TUS procedures in humans. Supports major GPU incl. NVidia, AMD and Apple for accelerated computing[96]. | https://github.com/ProteusMRIgHIFU/BabelBrain |
| COMSOL Multiphysics | Standalone environment | Advanced user software for acoustic simulation with an emphasis on material geometry, material physics, and material properties. Includes a variety of built-in materials. | https://www.comsol.com/ |
| Field II | MATLAB, python, .exe | Field II allows for simulation of ultrasound transducer fields and imaging through linear acoustics based on the Tupholme-Stepanishen method for calculating pulsed ultrasound fields. | https://field-ii.dk/ |
| FOCUS | MATLAB | FOCUS is a cross-platform ultrasound simulation software tool for rapid and accurate calculation of pressure fields generated by single transducers or phased arrays. | https://www.egr.msu.edu/~fultras-web/ |
| k-Plan | Standalone environment | k-Plan is a commercial client for running k-wave simulations of pressure fields and temperature rise remotely via a cloud service, providing a study management system and a graphical user interface with an integrated workflow to load template or individual medical images, select transducers models, define transducer positions, and specify sonication parameters. | https://brainbox-neuro.com/products/k-plan |
| k-Wave | MATLAB | k-Wave is an open-source MATLAB toolbox designed for the time-domain simulation of propagating acoustic waves in 1D, 2D, or 3D. Initial value problems are provided for new researchers seeking to gain practical knowledge of acoustic simulations. | http://www.k-wave.org/ |
| j-Wave | Python | A JAX based implementation of k-wave which specializes in the use of graphics processing units, and differentiable implementation, allowing for large simulation speed increases. | https://github.com/ucl-bug/jwave |
| TUSX | MATLAB | A wrapper developed for k-Wave to create accessibility to new researchers. | https://www.tusx.org/ |

**Table S5. List of commercially available frameless stereotactic neuronavigation systems and robotic navigation systems (alphabetical order).** Note that this list may not be exhaustive and does not reflect any recommendation or preference of the authors or ITRUSST.

| Company | Products | Features | Links |
|---|---|---|---|
| ANT neuro | Visor | Unknown whether ultrasound transducer navigation is implemented already | https://www.ant-neuro.com/products/visor2 |
| Axilum Robotics | TMS-Cobot | Adapters available to mount Sonic Concepts transducers and control the robot via compatible neuronavigation systems | https://www.axilumrobotics.com |
| Localite GmbH | TMS-Navigator | Models of some transducers from Sonic Concepts and BrainSonix already included; compatible with Axilum Robotics TMS-Cobot | https://www.localite.de/en/home/ |
| Rogue Research | BrainSight | Models of some transducers from Sonic Concepts already included; compatible with Axilum Robotics TMS-Cobot | https://www.rogue-research.com/tms/brainsight-tms/ |